\newif\ifShowKeys
\ShowKeystrue

\documentclass[11pt,a4paper]{article} 
\usepackage[height=8.8in,width=6.45in]{geometry}

\usepackage[hidelinks]{hyperref}

\usepackage{tocloft}

\usepackage{amsmath, amssymb,amsthm}
\usepackage{amsthm}
\numberwithin{equation}{section}

\usepackage{bm,environ,mathrsfs,array,arydshln}
\usepackage{booktabs,float,slashed,hyperref}
\usepackage[mathcal]{euscript}
\usepackage{tensor} 						
\usepackage{mathabx}
\usepackage{simpler-wick}

\usepackage{aurical}
\usepackage[T1]{fontenc}

\usepackage[nodayofweek]{date time}
\usepackage{graphicx,epsfig,epic}
\usepackage{tikz} 
\usetikzlibrary{decorations.pathreplacing,decorations.markings,snakes}
\tikzset{middlearrow/.style={decoration={markings, mark= at position 0.5 with {\arrow{#1}} ,
}, postaction={decorate}}}

\usepackage{framed}						
\definecolor{shadecolor}{rgb}{0.95,0.95,0.97}

\usepackage{comment}

\allowdisplaybreaks

\newcommand{\bs}{\begin{shaded}}
\newcommand{\es}{\end{shaded}}
\def\ba#1\ea{\begin{align}#1\end{align}}		
\newcommand{\be}{\begin{equation}}
\newcommand{\ee}{\end{equation}}
\newcommand{\mc}{\mathcal }
\newcommand{\mb}{\mathbb }
\newcommand{\la}{\label}

\newcommand{\lp}{\notag \\ & }

\DeclareMathOperator{\tr}{\text{tr}}
\DeclareMathOperator{\sech}{\text{sech}}

\newcommand{\cf}{\textit{cf.} }
\newcommand{\ie}{\textit{i.e.} }
\newcommand{\eg}{\textit{e.g.} }

\makeatletter
\DeclareFontFamily{OMX}{MnSymbolE}{}
\DeclareSymbolFont{MnLargeSymbols}{OMX}{MnSymbolE}{m}{n}
\SetSymbolFont{MnLargeSymbols}{bold}{OMX}{MnSymbolE}{b}{n}
\DeclareFontShape{OMX}{MnSymbolE}{m}{n}{
<-6>  MnSymbolE5
   <6-7>  MnSymbolE6
   <7-8>  MnSymbolE7
   <8-9>  MnSymbolE8
   <9-10> MnSymbolE9
  <10-12> MnSymbolE10
  <12->   MnSymbolE12
}{}
\DeclareFontShape{OMX}{MnSymbolE}{b}{n}{
<-6>  MnSymbolE-Bold5
   <6-7>  MnSymbolE-Bold6
   <7-8>  MnSymbolE-Bold7
   <8-9>  MnSymbolE-Bold8
   <9-10> MnSymbolE-Bold9
  <10-12> MnSymbolE-Bold10
  <12->   MnSymbolE-Bold12
}{}

\let\llangle\@undefined
\let\rrangle\@undefined
\DeclareMathDelimiter{\llangle}{\mathopen}%
 {MnLargeSymbols}{'164}{MnLargeSymbols}{'164}
\DeclareMathDelimiter{\rrangle}{\mathclose}%
 {MnLargeSymbols}{'171}{MnLargeSymbols}{'171}
\makeatother

\catcode`,\active

\catcode`\,12

\def\XXint#1#2#3{{\setbox0=\hbox{$#1{#2#3}{\int}$}
     \vcenter{\hbox{$#2#3$}}\kern-.5\wd0}}

\newcommand{\gs}{g_{\text{s}}}
\newcommand{\gcs}{g_{\rm CS}}
\newcommand{\vev}[1]{\langle  #1 \rangle}

\newcommand{\J}{\mathcal{J}}

\def \OO  {{\cal O}} 

\def\ci{\cite}
\def \ed {
\bibliography{BT-Biblio}
\bibliographystyle{JHEP}
\end{document}}
\def\be{\begin{equation}}
\def\ee{\end{equation}}
\def\bea{\begin{eqnarray}}
\def\eea{\end{eqnarray}}

\def \ha {{1 \ov 2}}\def \iffa {\iffalse} 
\def \sql {{\sqrt{\l}}\ }
\def \del{\partial}
\def \a {\alpha}
\def \aa {{\a'}}
\def\ov{\over}
\def \ci {\cite}

\def \foot {\footnote}

\def\la{\label}\def\foot{\footnote}\newcommand{\rf}[1]{(\ref{#1})}

\def \no {\nonumber}
\def \adss {${\rm AdS}_5 \times S^5\ $}
\def \a {\alpha } 
\def  \gym {g_{_{\rm YM}}}
\def \str {s} \def \l {\lambda} \def \sql {\sqrt{\l}}\def \RR  {{\rm R}} \def \str {{\rm s}}
\def \abjm  {AdS$_4\times CP^3$ }
\def \N {{\cal N}}
\def \W {{\cal W}}
\def \kkappa  {\varkappa}

\def \gcs {g_{_{\rm CS}}}
\def \te {\textstyle}
\def \tet {\textstyle}

\begin{document}

\begin{titlepage}


\ \hfill{\small  Imperial-TP-AT-2020-06 }

\begin{tabbing}
\hspace*{11.5cm} \=  \kill 
\end{tabbing}

\vspace*{15mm}
\begin{center}
{\LARGE  
On the  structure  of  non-planar     strong coupling corrections  }\vskip 5pt {\LARGE  to correlators  of   BPS Wilson loops 
and chiral primary  operators} 

\vspace*{10mm}

{\Large M. Beccaria${}^{\,a}$  and   A.A. Tseytlin${}^{\,b,}$\footnote{\ Also at the Institute of Theoretical and Mathematical Physics, MSU and Lebedev Institute, Moscow. 
}
}

\vspace*{4mm}
	
${}^a$ Universit\`a del Salento, Dipartimento di Matematica e Fisica \textit{Ennio De Giorgi},\\ 
		and I.N.F.N. - sezione di Lecce, Via Arnesano, I-73100 Lecce, Italy
			\vskip 0.3cm
${}^b$ Blackett Laboratory, Imperial College London SW7 2AZ, U.K.
			\vskip 0.3cm
			
\vskip 0.2cm
	{\small
		E-mail:
		\texttt{matteo.beccaria@le.infn.it, \ \ tseytlin@imperial.ac.uk}
	}
\vspace*{0.8cm}
\end{center}

\begin{abstract}
\noindent
Starting with  some  known localization (matrix model)  representations for  correlators involving   1/2 BPS   circular Wilson  loop 
  $\cal W$  in  ${\cal N}=4$ SYM theory we  work out   their $1/N$ expansions  in  the limit of large 't Hooft coupling $\lambda$.  Motivated by a possibility of   eventual   matching  to higher  genus corrections in dual string  theory  we 
 follow   arXiv:2007.08512   and express the result  in terms of  the   string  coupling $g_{\rm s}  \sim g^2_{\rm YM} \sim \lambda/N$ and  string  tension $T\sim \sqrt \lambda$.  Keeping only  the leading  in  $1/T$ term  
  at  each order in $g_{\rm s}  $ we observe that   while the  expansion of $\langle  {\cal W} \rangle$ 
   is a series in  $g^2_{\rm s} /T$, 
   the  correlator of  the Wilson loop   with   chiral primary operators ${\cal O}_J $  
     has expansion in powers of $g^2_{\rm s}/T^2$. Like  in the case of $\langle  {\cal W} \rangle$ where  these leading terms are known to   resum into an exponential of  a  ``one-handle''   contribution $\sim g^2_{\rm s}  /T$, 
     the  leading strong coupling  terms in   $\langle  {\cal W}\, {\cal O}_J  \rangle$   sum up to a simple square root  function of $g^2_{\rm s}/T^2$.  Analogous expansions     in   powers of $g^2_{\rm s}/T$   are found   for correlators of several 
     coincident Wilson  loops   and  they   again have   a  simple resummed form.  We also  find   similar expansions for  correlators of coincident    1/2  BPS Wilson loops in the ABJM theory. 
 \end{abstract}

\iffa 
Starting with  some 
  known localization (matrix model)  representations for  correlators involving  $1\over 2$-BPS   circular Wilson  loop
   $\cal W$
in the ${\cal N}=4$ SYM theory we  work out   their $1/N$ expansions  in  the limit of large 't Hooft coupling $\lambda$.  Motivated by a possibility of   eventual   matching  to higher  genus corrections in dual string  theory  we 
   follow   arXiv:2007.08512   and      express the results  as   series 
    in  the   string  coupling $g_{\rm s}  \sim g^2_{\rm YM} \sim \lambda/N$ and inverse string  tension $1/T\sim 1/\sqrt \lambda$. 
Keeping only  the leading  in  $1/T$ term   at  each order in $g_{\rm s}  $ we observe that 
 while the  expansion of $\langle  {\cal W} \rangle$  is a series in  the ratio  $g^2_{\rm s}  /T$, 
   the  correlator of  the Wilson loop   with  
   chiral primary operators ${\cal O}_J $  
     has expansion in powers of $g^2_{\rm s}/T^2$. Like  in the case of $\langle  {\cal W} \rangle$ where  these leading terms were known to   resum into an exponential of  a  ``one-handle''   contribution $\sim g^2_{\rm s}  /T$, 
     the  leading strong coupling  terms in   $\langle  {\cal W}\, {\cal O}_J  \rangle$   sum up to a simple square root  function of $g^2_{\rm s}/T^2$.  Analogous expansions 
      in   powers of $g^2_{\rm s}/T$   are found   for correlators of several 
     coincident Wilson  loops   and  they   again have   a  simple resummed form. 
      We also  find   similar expansions for  correlators of coincident  $1\over 2$-BPS Wilson loops in the ABJM theory. 
    \end{abstract}
\fi 

\

\

%

\vskip 0.5cm
	{
	}
\end{titlepage}

\tableofcontents
\vspace{1cm}
\setcounter{footnote}{0}

\section{Introduction  and summary} 

An important  direction is  to extend  checks of AdS/CFT 
correspondence   to   subleading  orders in  $1/N$  expansion  on the gauge theory side 
 or   higher genus corrections on the  dual string theory side. 
 One of the  simplest  observables  to consider  is  the expectation  value $\langle \mc W \rangle $ of 
 $1\ov2$-BPS  circular Wilson loop   for which  the exact  in $N$ 
 expressions  are  available   in  both  $SU(N)$   $\mc N=4$ SYM
  \ci{Erickson:2000af,Drukker:2000rr,Pestun:2007rz,Zarembo:2016bbk}   and  $U(N)_{k}\times U(N)_{-k}$   ABJM  \ci{Kapustin:2009kz,Marino:2009jd,Drukker:2010nc}  theories.  
 It was recently observed  in \ci{Giombi:2020mhz}  that   the expressions for 
  $\langle \mc W\rangle$   expanded first  in $1/N$ and then 
 in  large 't Hooft  coupling $\l$  have   a universal form  when   written in terms of the corresponding  string coupling $g_\str $ and  
 string tension $T= {\RR^2\ov 2 \pi \a'}$ defined as 
 \ci{Maldacena:1997re,Aharony:2008ug}
  \ba \la{1.1}
&{\rm SYM}:\qquad  g_{\str}  = \frac{g^2_{\rm YM}}{4\pi}=\frac{\l}{4\pi N}  \ , \qquad 
     \l = g^2_{\rm YM} N \ , \qquad   T
     = {\sql \ov 2 \pi} \ , \\
& {\rm ABJM}:\qquad   g_{\str}  = {   \sqrt \pi\,(2 \l)^{5/4}\ov  N}  \ , \qquad 
     \l = {N \ov k}  \ , \qquad \ \ \ \ \   T = {\sqrt{2 \l} \ov 2}  \ .  \la{1.2}
\ea
Explicitly,\foot{\la{f0}  Following \ci{Giombi:2020mhz}   in this paper we define $\W=\tr [ P \exp (...)] $   without the 
 $1/N$ prefactor.}
 \ba
\no
\langle \mc W \rangle =& \ e^{2\pi\,T}  \sum^\infty_{p=0}\text{c}_{p}\Big(\frac{g_\str }{\sqrt T}\Big)^{2p-1}\,\Big[1+\mc O(T^{-1})\Big]\\
= &\  e^{2\pi\,T}\,  {\sqrt T \ov g_\str} \Big\{ \text{c}_{0} \big[ 1 + \OO (T^{-1}) \big] 
  + \text{c}_{1}  { g^2_\str \ov T } \big[ 1 + \OO (T^{-1}) \big]
+  \text{c}_{2}    \Big({g^2_\str \ov T } \Big)^2 \big[ 1 + \OO (T^{-1}) \big] + ... \Big\} \ . \la{1.3}
\ea
Indeed, this is what one finds by   expanding  (in large $N$ and then in  large $\l$) 
 the  exact result  in, e.g., SYM theory \ci{Drukker:2000rr}\foot{
 Here  $L^{(k)}_n$ is the generalized Laguerre polynomial.
 This  expression is for the $U(N)$  group while  for $SU(N)$ one  gets   an additional factor
 $e^{-\frac{\lambda}{8N^2}}$  so that  $ \langle \mc W\rangle$  is 1 for $N=1$.}
\ba
\la{1.4}
\langle \mc W\rangle &= e^{\frac{\lambda}{8N}}\,L_{N-1}^{(1)}\big(-\frac{\lambda}{4N}\big) 
=  N\,e^{\sqrt\lambda}\sum_{p=0}^{\infty}\frac{\sqrt 2}{96^{p}\sqrt\pi p!}\frac{\lambda^{\frac{6p-3}{4}}}{N^{2p}} 
\Big[1+\mc O\big(\frac{1}{\sqrt\lambda}\big)\Big]\ ,
\ea
\ie we get  (\ref{1.3}) with $c_{0}=\frac{1}{2\pi}$, $c_{1}=\frac{1}{24}$, {\em etc.}
The universal   structure  of  \rf{1.3}  is a manifestation of the fact  that  the two   gauge  theories are  expected to be dual to   similar 
superstring theories in  \adss  and ${\rm AdS}_4 \times CP^3$   where  $\langle \mc W \rangle$    should be given by a  string path integral over  surfaces  
ending 
on a circle at the boundary of AdS.  $e^{2\pi\,T} $  in \rf{1.3} 
  is the semiclassical factor corresponding  
 to an AdS$_2$ minimal surface 
\ci{Berenstein:1998ij,Drukker:1999zq,Drukker:2000ep}.
 The expansion is done first  in small  string coupling $g_\str $  (i.e.  large $N$ for fixed $T\sim \sql$) 
  and then  in   $1/T$   at each order in $g_\str $. The power of   string coupling is 
  the  Euler number $\chi = 1- 2p$   of a disc  with $p=0,1,2, ...$ handles. 

 A  non-trivial  feature of \rf{1.3} is that the  leading power  of the inverse  string tension $1/T$  
 at  each order in $g_\str $  is  precisely 
 $-\ha \chi = p-\ha$, i.e. it   is  correlated   with  the  power of $g_\str $. 
  A string theory explanation of this fact was  suggested in \ci{Giombi:2020mhz}  by   showing that dependence of the string  partition  function 
 on the AdS$_n$  radius $\RR$ (and thus on the  string tension)  is controlled   by the Euler   number  of the surface. 
 
 Another remarkable fact about  the SYM  result \rf{1.4}  is that  the leading large $T$ terms  in \rf{1.3} exponentiate  \ci{Drukker:2000rr}  (according to \rf{1.4}, the coefficients $\text{c}_{p} $  in \rf{1.3}  are given by 
   $\text{c}_{p} = \frac{1}{2\pi p!}\big(\frac{\pi}{12}\big)^{p}$)
   \iffa
   \foot{\la{f1} One   may try 
    understand this exponentiation as a result of resummation of  thin  far separated handles; 
 an  attempt \ci{Drukker:2000rr}  to explain  the  negative power of $T$ in the $ {\gs^{2}}/{T}$ argument in the exponent  in \rf{1.5}  based on  supergravity approximation as in \ci{Berenstein:1998ij}
does not  appear to work  directly requiring subtle  cancellations  presumably implied by  supersymmetry
 (see Appendix \ref{app:KK} below).}
 \fi
   \ba
\la{1.5}
\langle \mc W\rangle = 
 { 1 \ov 2 \pi} \frac{\sqrt T}{\gs}\, e^{2\pi\,T \  +\    \frac{\pi}{12}\frac{\gs^{2}}{T} }\,\Big[1+\mc O(T^{-1})\Big]\ . 
\ea
Surprisingly,  in the ABJM theory  the   coefficient  of the  first subleading  correction is the  same  $\pi \ov 12$ as in  the SYM case 
  \cite{Drukker:2010nc} 
\be
\la{1.6}
\langle \mc W \rangle = \Big(\frac{N}{4\pi\,\lambda}+\frac{\pi\,\lambda}{6N}+\cdots\Big)\,e^{\pi\,\sqrt{2\lambda}} = 
{1  \ov \sqrt{2\pi}} \frac{\sqrt T}{\gs}\,e^{2\pi\,T}
\Big[1+\frac{\pi}{12}\,\frac{\gs^{2}}{T}+\OO(g^4_s)\Big]\ . 
\ee
However, the coefficients  of higher order terms  (that   can be found from \cite{Klemm:2012ii}) 
  turn out to be  different  than   in the SYM case \rf{1.4}, i.e.   here the exponentiation 
   does not   happen.\foot{This   may be  surprising    given that    such an exponentiation  
     may be  expected in   the   large tension (``thin handle'')   approximation on  the string theory  side \ci{Drukker:2000rr,Giombi:2020mhz}  
     and the fact that  the 
      dual   string theories in \adss   and \abjm  are similar.}
  Instead, we will  find   (see  Appendix ~\ref{ABJM:simple:genus})  that  in the ABJM case the leading strong-coupling terms in 
  \rf{1.3}    can be resummed  as 
  \be  \la{1.7}
  \langle \mc W \rangle = {1  \ov 2  \sin [ \sqrt\frac{\pi}{2}\,\frac{\gs}{\sqrt T}]   } \, e^{2\pi\,T}  \Big[1+\mc O(T^{-1})\Big] \ , \ee
   where    $\sqrt\frac{\pi}{2}\,\frac{\gs}{\sqrt T}= 2\pi\,\frac{\lambda}{N}= {2 \pi \ov k }$
  (see \rf{1.2}). 
  
  \
  
  Our aim below  will be to extract similar predictions about  the structure  of 
  small $g_\str$, large $T$    string theory 
  corrections  and  their possible resummation 
     for other  closely related observables   for which  the  exact   gauge theory  results can be  found 
    from   matrix model  representations  following from localization
   (in some cases generalizing   partial  results in the literature). 
   
  Namely, we shall consider correlators of  $\ha$-BPS  Wilson loop  with   chiral primary operators  (CPO)
    and also   correlators  of   several  coincident   Wilson loops  (mostly in the SYM  theory). 
   Like  in the case of $ \langle \mc W \rangle$ in \rf{1.3}  we 
     will   observe  certain universal   patterns in  their  expansion in small $g_\str $ and  large $T$
     that should be related  to supersymmetry of these  observables.
  This may hopefully  aid future investigations  on the dual string theory side.

  Let us summarize  our  main results.

  \subsection{Correlators  of  $\frac{1}{2}$-BPS Wilson loop with  chiral  primary  operators}
  
In section \ref{2}  we shall consider the  SYM correlator  of a  circular Wilson loop   with  a chiral primary 
operator  $\mc O_{J} = \tr \varphi^{J}$.\foot{ 
As is well known, in $\mc N=4$ SYM  one  can construct Maldacena-Wilson loops with various amounts of supersymmetry 
 \cite{Zarembo:2002an}, \eg the $\frac{1}{4}$-BPS circular loop \cite{Drukker:2006ga}
and  $\frac{1}{8}$-BPS loops  \cite{Drukker:2007yx,Drukker:2007dw,Drukker:2007qr}.
Correlators of  these  loops 
 and local operators were  considered in 
\cite{Semenoff:2001xp,Pestun:2002mr,Semenoff:2006am}.
Correlators of  $\frac{1}{8}$-BPS circular loop and various chiral primaries have been computed
by localization  in 
\cite{Giombi:2009ds,Giombi:2012ep,Bassetto:2009rt,Bassetto:2009ms,Bonini:2014vta}.
Correlators  involving   Wilson loops in higher representations were   discussed in 
\cite{Giombi:2006de,Gomis:2008qa}. 
In the planar limit at strong coupling  the results   were  successfully compared with AdS/CFT predictions 
\cite{Berenstein:1998ij,Giombi:2006de,Gomis:2008qa}.
~Beyond the planar limit and for $J>3$ 
the definition of the $\mc N=4$ SYM   BPS  operators dual to 
single-particle string (supergravity)  states requires the addition  to $\tr \varphi^J$ of multi-trace terms 
(see \cite{Aprile:2020uxk} and references therein). 
We have verified by explicit calculations that this  does not change the  qualitative  structure of the 
$1/N$   expansions discussed below. }
The  correlator $\langle \mc W\,\mc O_{J}\rangle $   
was originally  discussed in  \cite{Berenstein:1998ij}  at the 
 leading order   in  strong coupling  in  connection with the Wilson loop
OPE expansion. 
In the planar limit  this correlator was  computed exactly in $\l$  in \ci{Semenoff:2001xp}:
$\langle \mc W\,\mc O_{J}\rangle \sim  I_J (\sql) $  ($I_J$ is the Bessel   function). 

We have  extended the   computation  to  non-planar corrections;  expanded in small $g_\str$ and then in large $T$ 
as in \rf{1.3}  the result reads\foot{Here we ignore 
 the   R-symmetry   factor $Y$  depending on the choice of  the  CPO    and the scalar coupling in $\W$ 
  \ci{Semenoff:2001xp}  
and the  factor of dependence on the operator   insertion point 
 which is  fixed   by conformal invariance
(see  section 2).} 
\ba
\la{1.8}
\frac{\vev{\mc W\,\mc O_{J}}}{\vev{\mc W}} = c_{J}\,T\,\Big\{1 &+\frac{a_{0}^{(1)}(J)}{T}+\frac{a_{0}^{(2)}(J)}{T^{2}}+\cdots   +\frac{\gs^{2}}{T^{2}}\,\Big[a_{1}^{(0)}(J)+\frac{a_{1}^{(1)}(J)}{T}+\cdots\Big] \lp
+\frac{\gs^{4}}{T^{4}} \,\Big[a_{2}^{(0)}(J)+\frac{a_{2}^{(1)}(J)}{T}+\cdots\Big]+\OO\Big( \frac{\gs^{6}}{T^{6}}\Big)\Big\}\ , 
\ea
where   dots stand for terms subleading in $1/T$. 
On the string theory  side,   the  overall factor of $T$  should   come from the semiclassical value of  the vertex 
operator  dual to $\OO_J$   evaluated on the AdS$_2$ minimal surface. 
The coefficient $c_J$   is fixed by  normalization  of  $\mc O_{J}$  and 
 $a_{i}^{(j)}(J)$ are  polynomials in $J$, cf. \rf{2.28} 
 (for example, 
$ a_{0}^{(1)}= - {1 \ov 4 \pi} (J^2-1)$  as in  \ci{Semenoff:2001xp}). 
   Compared to the series in  $\frac{\gs^2}{T} \sim \frac{\lambda^{3/2}}{N^2}$ in  \rf{1.3}  here the 
natural expansion parameter turns out to be 
$x\equiv \frac{\gs^2}{T^2} \sim \frac{\lambda}{N^2}$.

Remarkably, it is possible  to explicitly sum up  all  leading   large $T$  terms in \rf{1.8} 
as 
\be
\la{1.9}
\frac{\vev{\mc W\,\mc O_{J}}}{\vev{\mc W}} =   c_{J}\,T\,F_{J}\Big(\frac{\gs^{2}}{T^{2}}\Big)  \, \Big[ 1 +   \OO(T^{-1})\Big]\ 
,\qquad\qquad   F_{J}(x) = \frac{2}{J\sqrt x}\sinh\Big(J\,\text{arcsinh}\frac{\sqrt x}{2}\Big).
\ee
Here $F_J$ is a finite polynomial for  odd $J$ and  $\sqrt{1+\tfrac{1}{4} {\gs^{2}\ov T^2}} $  times a polynomial  for even $J$. 
For example,  in  the $J=2$  case    
 one finds  simply 
\be
\la{1.10}
\frac{\vev{\mc W\,\mc O_{2}}}{\vev{\mc W}} = \pi\,\sqrt{T^{2}+\tfrac{1}{4} \gs^{2}}\,\Big[1+\mc O\Big(T^{-1}\Big)\Big].
\ee
The  same expression   applies  also  to the  correlator of $\mc W$ with the 
 dimension 4 dilaton operator $\mc O_{\rm dil}$    which is a  supersymmetry descendant of 
$\OO_2$.\foot{\la{f4} For higher  $J$, the   generalized   dilaton  operator  $\mc O_{{\rm dil}, J'}$  with non-zero $R$-charge  $J'$ 
  and dimension $\Delta= 4 + J'$   is a  supersymmetry descendant  of  $ \mc O_{J}$ 
with $J= 2 + J'$ and thus  ${\vev{\mc W\,\mc O_{{\rm dil}, J-2}}}/{\vev{\mc W}}$ is the same as \rf{1.8}.}
For  any  $g_\str $ and $T$  the    expectation value $\frac{\vev{\mc W\,\mc O_{\rm dil }}}{\vev{\mc W}} $
can be found  directly  from $\vev{\mc W}$ in \rf{1.3},\rf{1.4}     by differentiating over $\l$ 
  so that  using \rf{1.1} we  have (see  \ci{Giombi:2020mhz} and refs. there) 
\be
\la{1.11}
\frac{\vev{\mc W\,\mc O_{2}}}{\vev{\mc W}} = \frac{\vev{\mc W\,\mc O_{\rm dil}}}{\vev{\mc W}}
= \l  { \del \ov \del \l}  
\log \langle \mc W \rangle= \Big(  \ha  T { \del \ov \del T}   +   g_\str {\del \ov \del g_\str }  \Big)  \log \langle \mc W \rangle  \ .\ee
The  small $\gs$, large $T$ expansion of $\log\vev{\mc W}$  following  from  (\ref{1.4}) is found to be  
\ba
\la{x1}
\log\vev{\mc W} = 2\pi T  &- \ha  \log\Big(4\pi^2 { g^2_\str \ov T}  
\Big) -\frac{3}{16\pi T}+\cdots+\frac{\pi}{12}\frac{\gs^{2}}{T}\,\Big(
1-\frac{3}{4\pi T}+\cdots\Big)\lp
-\frac{\pi}{320}\frac{\gs^{4}}{T^{3}}\,\Big(1-\frac{1945}{768\pi T}+\cdots\Big)+\mc O\Big({\gs^{6}\ov T^5}\Big)\ ,
\ea
and therefore
\ba
\la{x2} \Big(  \ha  T { \del \ov \del T}   +   g_\str {\del \ov \del g_\str }  \Big)  \log \langle \mc W \rangle  
= \pi\,T\,\Big[1 & -\frac{3}{4\pi T}+\cdots+\frac{\gs^{2}}{8\,T^{2}}\Big(1-\frac{1}{2\pi T}+\cdots\Big)\lp-\frac{\gs^{4}}{128\,T^{4}}
\Big(1-\frac{389}{192\pi T}+\cdots\Big)
+ \mc O\Big({\gs^{6}\ov T^6}\Big)
\Big] \ , 
\ea
in agreement with \rf{1.8}. 
The  reason  why  the  large $T$    expansion \rf{1.8}  
  has   a different structure than  \rf{1.3} and  thus also why   the resummed  expressions in  \rf{1.5}  and in \rf{1.10}
  are not directly related by \rf{1.11}  is that  subleading in $1/T$ terms  at each order in $g_\str $ in $  \langle \mc W \rangle$ in \rf{1.3}  contribute to 
  $\log \langle \mc W \rangle$ 
 and, as a result,    reorganize   its   large $T$   expansion  (see \rf{2.39}--\rf{2.41} for details).

There is  still  an interesting  connection   between  the resummed  expressions for $\langle \mc W \rangle$ in \rf{1.5}  and 
the correlator  ${\vev{\mc W\,\mc O_{2}}}$
 in  \rf{1.9},\rf{1.10}:    both  can be given a  ``D3-brane'' interpretation  \ci{Drukker:2005kx,Giombi:2006de}.  
  To recall, 
for   a circular Wilson loop in $k$-symmetric  $SU(N)$ representation    in the limit  of 
  large  $k,\, N$  and $\l$ with $\kkappa= { k \sql \ov 4 N}  =   {k\, g_\str  \ov 2\,  T}=$fixed   one expects that  $ \langle \mc W \rangle$ 
should  be  given by  $\exp ( - S_{\rm D3})$   where  $S_{\rm D3}$ is the  D3-brane 
 action on  the  corresponding  classical  solution  \ci{Drukker:2005kx}. 
For  $1 \ll k \ll  N $  this   should apply also to the  Wilson loop 
  in the $k$-fundamental representation
  described by a   minimal  surface  ending on a multiply wrapped circle; here   one finds \ci{Drukker:2005kx}: \    $ S_{\rm D3} =N f(\kkappa) = 
 - k \sql   - { k^3 \l^{3/2} \ov 96  N^2}  + \OO ({ k^5 \l^{5/2} \ov   N^4}) $.  
  Extrapolating  this  to  the 
 $k=1$ case   corresponds  to the   resummation of the expansion in 
 \rf{1.3},\rf{1.4}  for fixed $ {  \sql \ov N} \sim {g_\str \ov T}$
 (i.e. 
      when $g_\str  \sim T $  formally  are  both large, cf.  \ci{Drukker:2005kx,Okuyama:2006ir}).
      Then       
 $ S_{\rm D3} = - 2 \pi T  -  { \pi   \ov 12 }\,   { g^2_\str \ov T}   + \OO(   { g^4_\str \ov T^3}   )$, 
 reproducing the  exponential  factor in \rf{1.5}. 
 Similar D3-brane interpretation is possible also  in the case of the    correlator  ${\vev{\mc W\,\mc O_{J}}}$ 
  \cite{Giombi:2006de}.\foot{We thank S. Giombi for pointing this out to us.}
 Indeed, the resummed expression \rf{1.9}  is in perfect agreement with the result  found in  \cite{Giombi:2006de} 
 in the fixed $\kkappa={k\, g_\str  \ov 2\,  T}$  limit (both from the  derivative of the D3-brane   action over  the corresponding  graviton source and 
 from the matrix model  in the case of $k$-fundamental representation)  after  formally  interpolating
  to  the $k=1$ case.

In section \ref{3} 
 we  shall   also consider  the correlator  $\vev{\mc W\,\mc O_{J_1}\,\mc O_{J_2}}$ 
  with two  chiral primary operators.  
 In the two special cases 
 (a)  $J_1=J_2$ and (b) $J_1=2, J_2=2J$ 
it  is  possible to  
reduce their  computation 
 to correlators in the Gaussian 1-matrix model. 
 The structure of the resulting  $1/N$  strong coupling expansion 
 is found to be similar to \rf{1.8} 
\ba
\la{1.14}
&\frac{\vev{\mc W\,\mc O_{J_1}\,\mc O_{J_2}}}{\vev{\mc W}} = d_{J_1J_2}\,T^2\,\Big\{1+\frac{b_{0}^{(1)}(J_1,J_2)}{T}+\frac{b_{0}^{(2)}(J_1,J_2)}{T^{2}}+\cdots \\ 
&\quad+\frac{\gs^{2}}{T^{2}}\,\Big[b_{1}^{(0)}(J_1,J_2)+\frac{b_{1}^{(1)}(J_1,J_2)}{T}+\cdots\Big]
+\frac{\gs^{4}}{T^{4}}\,\Big[b_{2}^{(0)}(J_1,J_2)+\frac{b_{2}^{(1)}(J_1,J_2)}{T}+\cdots\Big]+ \OO\Big(   { g^6_\str \ov T^6}\Big)\Big\},\no 
\ea
where $b_{i}^{(j)}$ are  polynomials in $J_1,J_2$  (see \rf{3.19},\rf{332},\rf{3322}). 

\subsection{Correlators  of coincident  Wilson loops} 
\def \TT {{\rm T}}

Another class of tractable examples  that we shall consider in section \ref{4}
are the expectation values of  coincident  circular  Wilson loops $\vev{\mc W^{n}}$ in SYM theory.\foot{ 
Correlators   of separated  loops  were  considered   in 
 \cite{Gross:1998gk,Zarembo:1999bu,Correa:2018lyl,Correa:2018pfn};
  supersymmetric configurations with oppositely oriented loops  were discussed in 
  \cite{Dorn:2018srz,Giombi:2009ms};  for various 
matrix model calculations, see  \cite{Sysoeva:2018xig,CanazasGaray:2018cpk,Okuyama:2018aij,CanazasGaray:2019mgq,Muck:2019hnz}.
}
\iffa 
Notice also that the multi-trace $\mc W^{n} = (\tr U)^{n}$ is completely different from the $n$-wound simple loop, \ie $\tr(U^{n})$. The multiply wound case 
is obtained from the standard simple loop by the rescaling $\lambda\to n^{2}\,\lambda$. This means $T\to n\,T$ and $\gs\to n^{2}\,\gs$, so instead of (\ref{1.5})
we get $W_{1}\to W_{1}\,n^{-3/2}$ (including also the rescaled AdS area) and $H\to n^{3}H$.
\fi 
The  $n=2$  case   
in the planar limit was  discussed, in particular,  in \cite{Drukker:2000rr,Arutyunov:2001hs}.
Extending  calculation to  subleading orders in $1/N$ 
in   large $\l$  limit and  rewriting the resulting expansion in terms of $g_\str $ and $T$  as in \rf{1.3} 
we have found that 
\ba
\la{1.15}
 \frac{\vev{\mc W^{2}}}{\vev{\mc W}^{2}} = &
\sum_{p=0}^{\infty}\frac{\pi^p }{(2p-1)!!}\Big( {g_\str \ov T}\Big)^p\Big[1 +\mc O(T^{-1})\Big] \no\\
= &1+e^{\frac{\xi}{2}}\sqrt\frac{\pi\,\xi}{2}\,\text{erf}\Big(
\sqrt\frac{\xi}{2}\Big)+\mc O(T^{-1}),  
\ \ \ \ \ \qquad 
 \xi \equiv  \pi\frac{g^2_\str}{T}\ .
\ea
 The analogous   expression for  $n=3$ is 
\be
\la{1.16}
\frac{\vev{\mc W^{3}}}{\vev{\mc W}^{3}} = 1+3e^{\frac{\xi}{2}}\sqrt\frac{\pi\,\xi}{2}\,\text{erf}\Big(
\sqrt\frac{\xi}{2}\Big)
+\frac{4\pi}{3\sqrt 3}\,\xi\,e^{2\xi}\,\Big[1-12\,\TT\Big(\sqrt{3\xi}, \frac{1}{\sqrt 3}\Big)\Big]+\mc O(T^{-1}),
\ee
where $\TT(h,a)$ 
 is the Owen T-function (see \rf{ttt}). 
 For  general  $n$   we found   similar expansion  
 (see \rf{443}  and Appendix \ref{anew}) 
\ba
\la{1.17}
\frac{\vev{\mc W^{n}}}{\vev{\mc W}^{n}}=  
 1&+\frac{n\,(n-1)}{2}\,\xi+\frac{\,n(n-1)(3n-5)\,(n+2)}{24}\,\xi^{2}\notag \\ &+\frac{\,n(n-1)(15n^{4}+30n^{3}-75n^{2}-610 n+1064)}{720}\,\xi^{3}+...  \ . 
\ea
Thus like  for $\vev{\W}$ in \rf{1.3}  here we get  again series in $\xi= \pi  \frac{g^2_\str}{T} $
 while   for the correlators  
 with  chiral primary operators \rf{1.8},\rf{1.14} 
the  expansion  was   in powers of $ x= {g^2_s \ov T^2}$.

In   section \ref{s4.3}  we shall derive a   similar expansion  
for   the   correlator  $\mc W^{(1,-1)}$ of  coincident   Wilson loops   in  fundamental and  anti-fundamental representations. It turns out that   in contrast to \rf{1.15},\rf{1.17} 
  $\mc W^{(1,-1)}$ has   trivial connected part, i.e. 
 to all orders in $\xi$   (and to leading order  in $1/T$) 
\be
\la{1.18}
\langle{\mc W^{(1,-1)}}\rangle = \vev{\mc W}^{2}\,\Big[1+\mc O(T^{-1})\Big]\simeq  \vev{\mc W}^{2}\ .
\ee
It  would    be interesting to  explain   this fact 
 from the string theory point of view.

\subsection{Comments  on  correlators  in   ABJM}

Obtaining the above  results  in the $\N=4$ SYM theory case  is facilitated   by  a relative   simplicity  of the 
associated  Gaussian matrix model. 
In the ABJM theory   the   computations of  similar  correlators involving 
 $\frac{1}{2}$-BPS circular  Wilson loop \cite{Drukker:2009hy}
  are  substantially more involved.  

The  structure of the   strong-coupling expansion of the  correlators  
  with  chiral primary operators 
is expected   to be similar to \rf{1.8}.\foot{For a discussion of  single-trace  
CPO  in ABJM theory see, e.g.,    \cite{Aharony:2008ug,Papathanasiou:2009zm}.}
This is suggested by the  observation \ci{Giombi:2020mhz}  that 
the expansion \rf{1.3}  of $ \vev{\mc W}$  looks the same  in the SYM and ABJM theories
and that, in particular,  for the dilaton operator  the correlators 
  $\vev{\mc W\,\mc O_{\rm dil }}$  and  $ \vev{\mc W}$  should be 
 again related  as in the    last equality   in  \rf{1.11}.  
 \iffa  
 \foot{Note that in  view of \rf{1.2} 
 here     $\l  { \del \ov \del \l}    =  \ha  T { \del \ov \del T}   + {5 \ov 4}   g_\str {\del \ov \del g_\str }  
 $     so that   the insertion of the dilaton   operator on the  ABJM  theory side does not appear to 
 correspond to applying  $\l  { \del \ov \del \l}$.}
 \fi
 Indeed, the dilaton vertex operator 
 has the same  structure  in both \adss   and  AdS$_{4}\times CP^{3}$   string theories
 and thus the  derivative over the zero-momentum dilaton should be related  to the  string partition function 
 in the same  way as in \adss  case in \ci{Giombi:2020mhz}, i.e.   as in  \rf{1.11}.


Below  in Appendix \ref{5}  we shall discuss   the computation of  correlators  of coincident  Wilson loops 
 $\vev{{\mc W}^{n}}$  in ABJM   theory. 
In particular,  for  the $n=2,3$ 
we  will  find   
\ba
\vev{\mc W^{2}} = \vev{\mc W}^{2}  
\,\Big[1+0\times \frac{\gs^{2}}{T}+0\times \Big(\frac{\gs^{2}}{T}\Big)^{2}+\cdots 
\Big], \qquad 
\vev{\mc W^{3}} =\vev{\mc W}^{3}  
\, \Big[1+0\times \frac{\gs^{2}}{T}+0\times \Big(\frac{\gs^{2}}{T}\Big)^{2}+\cdots 
 \Big],  \la{1.19}
\ea
 suggesting  the  conjecture  that, up to subleading $1/T$ terms  at each order in the genus expansion,    here
 the connected part of the correlator  $\vev{\mc W^{n}} $  vanishes, 
  i.e.  $\vev{\mc W^{n}} \simeq \vev{\mc W}^{n}$. 
This is in  contrast to  the non-trivial relation  \rf{1.17} found  for $\vev{\mc W^{n}} $     in the SYM case
(but is similar to the  behaviour of $ \langle{\mc W^{(1,-1)}}\rangle $ in   \rf{1.18}). 


 \subsection{Structure of the paper}
 
 In section \ref{2} we   compute the  $1/N$ expansion \rf{1.8} of 
 the SYM correlator $\vev{\mc W\,  \mc O_{J}}$ of the $\frac{1}{2}$-BPS Wilson loop with a chiral primary 
 operator  starting with its matrix model representation  implied by localization. 
 In section \ref{3} we repeat  the same analysis  for  the correlator  $\vev{\mc W\, \mc O_{J_{1}}(x_1) \mc O_{J_{2}}(x_2) }$
 assuming  a  special (supersymmetric)   choice of  insertion points $x_1$ and $x_2$ 
  that allows a matrix model calculation confirming 
  that its  strong coupling expansion has the form \rf{1.14}.
  
   In section \ref{4}  we consider  correlators of 
 coincident  BPS Wilson loops. We establish   the structure of the expansions  in \rf{1.15},\rf{1.16}  and prove 
 their exact form  by exploiting the Toda integrability
 structure of the  underlying Gaussian matrix model.
  In section \ref{s4.3} we consider the   correlator  of  Wilson loops  in  the fundamental  and  in the anti-fundamental representation
 where special features are expected 
 due  to supersymmetry. 
  Indeed,  in this case one  finds \rf{1.8}, i.e.   there are no leading order corrections to $\mc W^{(1,-1)}$ 
 beyond those in $\vev{\mc W^{2}}$.

In Appendix \ref{app:KK}  we discuss   an  attempt \ci{Drukker:2000rr}  to explain  the  negative power of $T$ in the $ {\gs^{2}}/{T}$  term in \rf{1.3}  by    
 assuming  that for large $T$  one  can use  supergravity approximation as in \ci{Berenstein:1998ij}. As we explain,
  this 
argument  may  work  only if there are non-trivial  cancellations of the dominant large $T$ terms 
  that should    be  implied by  supersymmetry.
 Appendix \ref{app:B}  contains some technical  details  of  the $1/N$ expansion 
of    $\vev{\mc W}$.  In 
Appendix \ref{sec:semi}  we consider the  $1/N$   expansion of the 
correlator $\vev{\W\, \mc O_{J}}$ in the 
 string semiclassical limit $J\sim \sql\gg 1$.
In Appendix \ref{anew} we work out  the $1/N$ expansion of $\vev{\mc W^n}$ deriving  the expansion \rf{1.17}.

The Appendices \ref{ABJM:simple:genus} and \ref{5} are devoted to the 
correlators $\vev{\mc W^n}$ of    $\frac{1}{2}$-BPS circular Wilson loop in the ABJM theory.
In Appendix \ref{ABJM:simple:genus} we comment on the single Wilson loop case 
  case by  reviewing the  known  matrix model results  
 pointing out that here the expansion  has again the same structure as in the SYM case in  \rf{1.3} 
 and deriving the representation  \rf{1.7}.  
 Appendix \ref{5} discusses   correlators of 
 $n=2,3$  coincident Wilson loops where  we 
 use the  topological expansion  of the algebraic curve characterizing the ABJM matrix model
 to first  derive  the exact expressions valid for all couplings,  and then  expand at strong coupling 
 demonstrating the  validity of \rf{1.19}.

\iffa 
\item The relation (\ref{1.17}) shows that exponentiation does not hold for $n\neq 1$, at least not in terms of a simple linear
function of $\gs^{2}/T$. Instead ``simple factorization'' holds for (\ref{1.18}), where it is just inherited from squaring $\vev{\mc W}$.  
Is there an explanation in terms of the dual minimal surfaces and/or supersymmetry ? 
\item Apart from ``simple exponentiation'', a separate issue is the presence of ``irreducible'' corrections in $\vev{\mc W^{n}}/\vev{\mc W}^{n}$.
In this regard, what happens in ABJM ? What is the high genera extension of (\ref{1.21}) ? Is it true that in that case we have at all orders the relation 
$\vev{\mc W^{2}}/\vev{\mc W}^{2} = 1+\mc O(T^{-1})$ ? And if yes, why ?
\fi

\section{
Expansion  of    
 $ \vev{\mc W\, \mc O_J} $  
\la{2} }

In this section we will  compute the $1/N$ expansion of the $\N=4$ SYM correlator of $\frac{1}{2}$-BPS circular Wilson loop   
with  chiral primary operators
$\mc O_{J}$. 
As was shown in \cite{Semenoff:2001xp},
in the  leading planar   approximation  the 
expression  for   this   correlator  
   is proportional to the  Bessel function,   $\langle \mc W\,\mc O_{J}\rangle \sim  I_J (\sql)$. This result
was 
 obtained by summing all planar rainbow Feynman graphs under the assumption that radiative corrections from planar graphs with internal vertices 
cancel to all orders in perturbation theory. This result was   later confirmed    in the framework of 
 supersymmetric localization where $\vev{\mc W\,\mc O_{J}}$
was  computed   using  a suitable hermitian 2-matrix model \cite{Giombi:2009ds}. 

Below  we shall   first   obtain    the finite $N$ localization result  for this correlator     using 
 a simplified equivalent 1-matrix model suggested by  similar computations  in the 
   $\mc N=2$ superconformal models \cite{Fucito:2015ofa,Billo:2018oog}.  We  shall  then  derive    the 
    $1/N$ expansion  of this    correlator  (up  to the $1/N^{6}$  order)  
      with  the  coefficients 
     being 
    $J$-dependent combinations of Bessel functions   of $   \sqrt  \l$. 
    Finally, we will extract the leading large $\l$ behaviour of these  coefficients. 

In general, the $\frac{1}{2}$-BPS  Wilson loop  depends \cite{Maldacena:1998im} on a  unit  6-vector 
$n_i$  
defining   the coupling to the SYM  scalars $n_{i}\Phi_{i}$. 
 The chiral primary operator 
 may be  chosen as   $\mc O_{J} = \tr\big(u_{i}\Phi_{i}(x)\big)^{J}$
where $u$ is a complex null 6-vector $u^{2}_i=0$.
The dependence of the correlator $\vev{\mc W\,\mc O_{J}}$ 
 on $n$ and $u$  factorizes \cite{Semenoff:2001xp}, i.e.  is  contained only in the  overall
 factor $Y(n, u) = (n_i u_i)^{J}$.
We shall  choose the  6-vector $n_i$  in $\W$   along the 1-direction  and the 
vector $u_i$  to be non-zero only  in (1,2) directions,   so that 
\ba  &\la{1w}
\mc W = \tr P \exp\Big\{g_{_{\rm YM}}\int_{C} d\sigma\,[i\,A_{\mu}(x)\,\dot x^{\mu}(\sigma)+\, R\,\Phi_{1}(x)]\Big\},
\\ \la{2w}
&\mc O_{J} = \tr \big[\varphi (x)\big]^J \ ,\qquad  \qquad \varphi = \tfrac{1}{\sqrt 2}(\Phi_{1}+i\Phi_{2}),
\ea
where $C$ is a circle of radius $R$ (that can be set to 1 as we assume   below).  
Then $ Y(n, u) = 2^{-J/2}$; we will not  explicitly indicate  this factor in $\vev{\mc W\, \mc O_J(x)}$  
as it   can be absorbed  into  normalization of $\mc O_{J}$   discussed below. 
Note also that   $\varphi=\varphi^{I}T^{I}$ where $T^{I}$ are $U(N)$ generators.

  Let us  also assume that  the unit-radius circular loop in 4-space $(x_1,x_2,x_3,x_4) $ 
    lies   in the $(x_1,x_2)$ plane (with the center at the origin) 
   and  define   the  ``transverse  distance''
\be
\la{2.2}
d_{\perp}(x)  = \tfrac{1}{2}\big[(r^{2}+h^{2}-1)^{2}+4h^{2}\big]^{1/2},\qquad \qquad  
r^{2}\equiv x_{1}^{2}+x_{2}^{2}, \qquad  h^{2}\equiv x_{3}^{2}+x_{4}^{2}\ .
\ee
Conformal symmetry implies   that \cite{Berenstein:1998ij,Alday:2011pf,Giombi:2018hsx}
\be
\la{2.3}
\vev{\mc W\, \mc O_J(x)} = \frac{1 }{\big[2d_{\perp}(x) \big]^{J}} \vev{\mc W\,\mc O_{J}(0)}  \  .
\ee
%
In what follows   we shall  thus assume  that  $\vev{\mc W\, \mc O_J}$  stands 
 for the $x$-independent part of \rf{2.3}, i.e. its value at $x=0$.

\subsection{Matrix model formulation \la{s2.1}}

\def \gym  {g_{_{\rm YM}}}

We we will  use a hermitian  1-matrix model formulation 
that computes expectation values like \rf{2.3}    using 
 a Gaussian hermitian 1-matrix model with the variable $a=a^{I}T^{I}$ as
 \foot{
This is the approach that can be used for chiral correlators in $\mc N=2$ superconformal theories \cite{Fucito:2015ofa,Billo:2018oog}.
Other equivalent approaches are available in the $\mc N=4$ case, such 
  as the complex matrix model formulation \cite{Kristjansen:2002bb}
or the associated normal matrix model version applicable  for suitable chiral observables \cite{Okuyama:2006jc}.
Nevertheless, our choice will be more convenient for most of our purposes.}
\be
\la{2.5}
\vev{\mc O} = \int Da\,\mc O(a)\,e^{-\tr a^{2}}\ ,\ \qquad \qquad  Da = \prod_{I=1}^{N^{2}}\frac{da^{I}}{\sqrt{2\pi}} \ . 
\ee
The explicit  map from the   gauge theory operator to the matrix model one  is  
\be
\la{2.6}
\mc O_{J} = \tr \big[\varphi(0)\big]^{J} \  
{\longrightarrow}\  \Big(\frac{g^2}{8\pi^{2}}\Big)^{J/2}\,\mathsf{O}_{J},\qquad\qquad  \mathsf{O}_{J} = :\tr a^{J}:\ , \qquad \qquad  g\equiv g_{_{\rm YM}} \ , 
\ee
where the coupling factor comes from the scaling needed to have the simple normalization  in the 
exponent  $e^{-\tr a^{2}}$ in \rf{2.5}.
Normal ordering in  $\mathsf{O}_{J}$  (\ref{2.6}) amounts to subtraction of  all self-contractions.
It is required as  the   correlators   involving   the  standard (flat 4-space)   chiral 
operator do not have self-contractions (there is no $\varphi\varphi$ propagator)   so that 
 on the matrix model  side  (derived from gauge theory formulated on  4-sphere) 
 these self-contractions should   be explicitly removed  (see, e.g.,  \cite{Gerchkovitz:2016gxx}).\foot{\la{x3}
Subtraction of self-contractions is in turn equivalent to the requirement of orthogonality
to all lower dimensional operators \cite{Billo:2017glv}. Denoting by $\{\Omega_{\alpha}\}$ the (single or multi-trace) operators with dimension 
strictly less than $\dim \mc O$, one has 
\be
\no 
:\mc{O}: = \mc{O}-\sum_{\alpha, \beta}\vev{\mc O\,\Omega_{\alpha}}\,(C^{-1})_{\alpha\beta}\,\Omega_{\beta},\qquad C_{\alpha\beta}=\vev{\Omega_{\alpha}\Omega_{\beta}}.
\ee
}
The matrix model   counterpart of the   BPS  Wilson loop operator  is  simply (with no $1/N$ normalization) 
\be\la{266}
\mc W \   
{\longrightarrow} \ \tr \, e^{\, \frac{g}{\sqrt 2}\,a}  \ . 
\ee
The correlator  
 in (\ref{2.5})  is computed by  Wick contractions 
with the free propagator $\vev{a^{I}a^{J}} = \delta^{IJ}$.
In the following, we will need generic multi-trace correlation functions of the form 
\be
t_{n_{1}, \dots, n_{\ell}} = \vev{\tr a^{n_{1}}\cdots \tr a^{n_{\ell}}}.
\ee
They may be computed by repeated application of the $U(N)$ fusion/fission  identities
\be
\la{2.10}
\tr (T^{I}\, A \, T^{I} \, B) = \frac{1}{2}\tr A\, \tr B\ ,\qquad\qquad  \tr(T^{I}A)\,\tr(T^{I}B) = \frac{1}{2}\tr(AB)\ ,
\ee
leading to the  recursion relations  \cite{Billo:2017glv}
\ba
t_{n} &= \frac{1}{2}\sum_{m=0}^{n-2}t_{m,n-m-2}\ ,\qquad\qquad 
t_{n, n'} = \frac{1}{2}\sum_{m=0}^{n-2}t_{m,n-m-2,n'}+\frac{n'}{2}t_{n+n'-2}\ ,\qquad \textit{etc.}
\ea

\subsection{Differential relations \la{s2.2}}

Using  the methods of \cite{Billo:2018oog} we can  compute  the one-point correlation 
functions \rf{2.3}  in presence of the Wilson loop. 
Remarkably,
   they can be found  directly from 
the knowledge of $\vev{\mc W}$    since it is possible to show that   for all $J$ one has 
 $\vev{\mc W\,\mathsf{O}_{J}} = \mathscr{D}_{J}(g, \partial_{g})\,\vev{\mc W}$,  
where $\mathscr D_{J}$ is a linear differential operator of order $J-1$. 
This follows from the matrix model 
representation of $\vev{\mc W\,\mathsf{O}_{J}}$ and is ultimately related to the supersymmetry.   
 For example,  
the  $J=2$  CPO  correlator   is related  to the  correlator  with the dilaton  operator  and the latter  may be  found  by differentiation  over 
the coupling  $g$ or $\l= g^2 N $  as in \rf{1.11}.  
 Explicitly,   in the $J=2$  case  one finds 
\ba
\la{2.12}
\vev{\mc W\,\mathsf{O}_{2}} = \vev{\mc W\,:\tr a^{2}:}  &= \sum_{k=0}^{\infty}\frac{g^{k}}{2^{\frac{k}{2}}\,k!}\,\Big(t_{k,2}-\frac{N^{2}}{2}t_{k}\Big)
= \sum_{k=0}^{\infty}\frac{g^{k}}{2^{\frac{k}{2}}\,k!}\,\Big(\frac{k+N^{2}}{2}t_{k}-\frac{N^{2}}{2}t_{k}\Big)\notag \\
&= \frac{1}{2}\sum_{k=0}^{\infty}\frac{g^{k}}{2^{\frac{k}{2}}\,k!}\,k\,t_{k} = \frac{1}{2}g \,\partial_{g}\vev{\mc W} \ .  
\ea
The  $J=4$  case is slightly more complicated
\ba
\vev{\mc W\,\mathsf{O}_{4}} = \vev{\mc W\,:\tr a^{4}:} &= \sum_{k=0}^{\infty}\frac{g^{k}}{2^{\frac{k}{2}}\,k!}
\vev{ \tr a^{k}\,\Big(\tr a^{4}-(\tr a)^{2}-2N\,\tr a^{2}+\frac{N^{3}}{2}+\frac{N}{4}\Big)} \notag \\
&= \sum_{k=0}^{\infty}\frac{g^{k}}{2^{\frac{k}{2}}\,k!}\Big(t_{k,4}-t_{k,1,1}-2N\,t_{k,2}+(\frac{N^{3}}{2}+\frac{N}{4})\,t_{k}\Big) \ , 
\ea
where we used the explicit form of $:\tr a^{4}:$ obtained by resolving the mixing with dimension $<4$ operators. From the relations (\ref{2.10}), we find (doing Wick contractions)
\ba
\la{2.14}
t_{k,2} &= \frac{N^{2}+k}{2}t_{k}, \qquad\qquad  t_{k,1,1} = \frac{N}{2}t_{k}+\frac{k(k-1)}{4}t_{k-2},\notag \\
t_{k,4} &= N\,t_{k,2}+\frac{1}{2}\,t_{k,1,1}+\frac{k}{2}\,t_{k+2} = \frac{k}{2}t_{k+2}+\frac{N(2N^{2}+1+2k)}{4}t_{k}+\frac{k(k-1)}{8}t_{k-2}.
\ea
Hence, 
\ba
\vev{\mc W\,\mathsf{O}_{4}} &= \sum_{k=0}^{\infty}\frac{g^{k}}{2^{\frac{k}{2}}\,k!}\Big(\frac{k}{2}t_{k+2}-\frac{k N}{2}t_{k}-\frac{k(k-1)}{8}t_{k-2}\Big)= 
\sum_{k=0}^{\infty}\frac{g^{k}}{2^{\frac{k}{2}}\,k!}\Big(-\frac{g^{2}}{16}-\frac{kN}{2}+\frac{k(k-1)(k-2)}{g^{2}}\Big)\, t_{k}
\notag \\
&= \Big(-\frac{g^{2}}{16}-\frac{N}{2}\,g\,\partial_{g}+g\,\partial_{g}^{3}\Big)\,\vev{\mc W}. \la{213} 
\ea
A completely similar calculation for $\mathsf{O}_{6}$ and $\mathsf{O}_{8}$ gives
\ba
\la{2.16}
\vev{\mc W\,\mathsf{O}_{6}} &= \Big[
2g\partial_{g}^{5}+\frac{3}{8}(N^{2}+1)\,g\partial_{g}-\frac{3}{4}(g\partial_{g})^{2}-2N\, g\partial_{g}^{3}+\frac{3N}{32}g^{2}
\Big]\,\vev{\mc W},
\\
\vev{\mc W\,\mathsf{O}_{8}} =& \Big[
4g\partial_{g}^{7}
-6N g\partial_{g}^{5}
-\frac{1}{64} g^2 (1+6 N^2)
+\frac{5}{32} g^3 \partial_{g}
-\frac{1}{8}  N (17+2 N^2) g\partial_{g}
\lp
+\frac{15}{8}  N (g\partial_{g})^{2}
+\frac{5 (4+N^2))}{2} g\partial_{g}^{3}
-\frac{15}{4 g} \partial_{g}(g^{3}\partial_{g}^{3})
\Big]\,\vev{\mc W}. \la{217}
\ea
The   above differential relations \rf{2.12},\rf{213},\rf{2.16}  written in terms of $\lambda=N\,g^{2}$ read 
\ba
\vev{\mc W\,\mathsf{O}_{2}} =& \lambda\partial_{\lambda}\vev{\mc W}, \qquad \qquad 
\vev{\mc W\,\mathsf{O}_{4}} = \Big(8N\lambda^{2}\partial^{3}_{\lambda}+12N\lambda\partial^{2}_{\lambda}-N\lambda\partial_{\lambda}
-\frac{\lambda}{16N}\Big)\,\vev{\mc W},  \la{216} \\
\vev{\mc W\,\mathsf{O}_{6}} = &\Big[
64N^{2}\lambda^{3}\partial_{\lambda}^{5}+320N^{2}\lambda^{2}\partial^{4}_{\lambda}+16N^{2}(\lambda-15)\,\lambda\,\partial_{\lambda}^{3}-3\lambda(8N^{2}+\lambda)\partial_{\lambda}^{2}\lp
\qquad +\frac{3}{4}(N^{2}-3)\lambda
\partial_{\lambda}+\frac{3\lambda}{32}
\Big]\,\vev{\mc W}\ . 
\ea
Similar  representations are found for higher even  $J$   and also for odd $J$, e.g., 
\ba
& \vev{\mc W\,\mathsf{O}_{1}}  = \frac{1}{2}\sqrt{\textstyle \frac{\lambda}{2N}} \, \vev{\mc W} \ ,  
\qquad \qquad   \vev{\mc W\,\mathsf{O}_{3}} = -\frac{1}{4}\sqrt{\tfrac{\lambda }{2N}}\,N\big(1-8\partial_\l-16\partial_\l^{2}\big) \vev{\mc W} \ ,\la{2188} \\
&  \vev{\mc W\,\mathsf{O}_{5}}   = \frac{1}{8}\sqrt{\tfrac{\lambda }{2N}}\Big[ N^{2}-1-6 (\lambda 
+4 N^2) \partial_{\l}-48 (-4+\lambda ) N^2 \partial_{\l}^{2}+768 \lambda  
N^2 \partial_{\l}^{3}+256 \lambda ^2 N^2 \partial_{\l}^{4}\Big]\vev{\mc W} , \notag \\
& \vev{\mc W\,\mathsf{O}_{7}} = \frac{1}{32N}\sqrt{\textstyle \frac{\lambda}{2N}}\Big[
\lambda -2 N^4 +16 N^2 (-5+3 \lambda +6 N^2) \partial_{\l}+32 N^2 (-35 \lambda -60 N^2+6 \lambda  N^2) \partial_{\l}^{2}\notag \\
& -640 N^2 (
\lambda ^2-24 N^2+12 \lambda  N^2) \partial_{\l}^{3}-2560 
(-36+\lambda ) \lambda  N^4 \partial_{\l}^{4}+61440 \lambda ^2 N^4 \partial_{\l}^{5}+8192 \lambda ^3 N^4 \partial_{\l}^{6}
\Big]    \vev{\mc W}. \notag \ea

\def \te {\textstile }

\subsection{$1/N$ and strong coupling expansion \la{s2.3}}

From the large $N$ expansion of $\vev{\mc W}$ in (\ref{B.2}) we can  then compute the 
corresponding expansion of the  ratios  $\vev{\mc W\,\mathsf{O}_{J}}/\vev{\mc W}$.
The strong coupling regime 
we are interested in is  defined by 
first expanding in  large $N$  for  fixed $\lambda$ and then expanding the coefficient of each $1/N$ term 
at large $\lambda$.
We find  for $J=2,4,6$ (the expressions for $J=8$   and odd $J$ are  similar) 
\ba
& \Big. \frac{\vev{\mc W\,\mathsf{O}_{2}}}{\vev{\mc W}}\Big|_{N\gg 1,\, \lambda \gg 1} = \frac{\sqrt\lambda}{2}\,\Big[
1-\frac{3}{2\,\sqrt\lambda}+\cdots
+\frac{1}{N^{2}}\Big(
\frac{\lambda }{32}-\frac{\sqrt{\lambda 
}}{32}+\cdots\Big)\notag \\
& \qquad \qquad  +\frac{1}{N^{4}}\Big(
-\frac{\lambda ^2}{2048}+\frac{\lambda ^{3/2}}{512}+\cdots\Big)
+\frac{1}{N^{6}}\Big(
\frac{\lambda ^3}{65536}-\frac{\lambda ^{5/2}}{8192}+\cdots\Big)
+\mc O\Big(\frac{1}{N^{8}}\Big)
\Big], \la{218}
\\
& \Big. \frac{\vev{\mc W\,\mathsf{O}_{4}}}{\vev{\mc W}}\Big|_{N\gg 1,\, \lambda \gg 1} = \frac{N\,\sqrt\lambda}{2}\Big[
1-\frac{15}{2\sqrt\lambda}+\cdots
+\frac{1}{N^{2}}\,\Big(\frac{5\lambda}{32}-\frac{15\sqrt\lambda}{32}+\cdots\Big)\notag \\
& \qquad\qquad  +\frac{1}{N^{4}}\,\Big(\frac{7\lambda^{2}}{2048}-\frac{\lambda^{3/2}}{512}+\cdots\Big)
+\frac{1}{N^{6}}\,\Big(-\frac{3\lambda^{3}}{65536}+\frac{\lambda^{5/2}}{8192}+\cdots\Big)
+\mc O\Big(\frac{1}{N^{8}}\Big)
\Big], \la{219}
\\
& \Big. \frac{\vev{\mc W\,\mathsf{O}_{6}}}{\vev{\mc W}} \Big|_{N\gg 1,\, \lambda \gg 1} = \frac{3N^{2}\,\sqrt\lambda}{8}\Big[
1-\frac{35}{2\sqrt\lambda}+\cdots
+\frac{1}{N^{2}}\,\Big(\frac{35\lambda}{96}-\frac{105\sqrt\lambda}{32}+\cdots\Big)\notag \\
& \qquad \qquad +\frac{1}{N^{4}}\,\Big(\frac{63\lambda^{2}}{2048}-\frac{63\lambda^{3/2}}{512}+\cdots\Big)
+\frac{1}{N^{6}}\,\Big(\frac{33\lambda^{3}}{65536}-\frac{\lambda^{5/2}}{8192}+\cdots\Big)
+\mc O\Big(\frac{1}{N^{8}}\Big)
\Big] \ . \la{220}
\ea
The leading (planar) terms in the square brackets are $1-\frac{J^{2}-1}{2\sqrt\lambda}+\cdots$ in agreement with the 
expansion of $I_{J}(\sqrt\lambda)$  in
\cite{Semenoff:2001xp}.\footnote{This planar $1/N^{0}$ part in the square brackets has the  full $\lambda$
dependence given 
 by the Bessel function ratio $I_{J}(\sqrt\lambda)/I_{1}(\sqrt\lambda)$ up to a power of $\lambda$ fixed by the choice of normalization
 of the operator. 
Notice  that we are considering  the $U(N)$ gauge theory.
Beyond the  planar level, results in the $SU(N)$ gauge theory differ in the subleading terms in the $1/N$ expansion 
due to  an additional factor $\exp\big(-\frac{\lambda}{8N^{2}}\big)$, and also
due  to $1/N$ modifications in the fusion/fission relations for  the $SU(N)$ generators compared to (\ref{2.10}).}
To determine the higher order $J$-dependent terms in the expansion of  $\vev{\mc W\,\mathsf{O}_{J}}$ up to some fixed order in $1/N$ it is
convenient to use the  
representation derived in \cite{Okuyama:2006jc}
\be
\la{222}
\vev{\mc W\, \mathsf{O}_{J}} = \frac{2^{1-J/2}N^{1+J/2}}{\sqrt\lambda}\,e^{\frac{\lambda}{8N}}\oint\frac{dw}{2\pi i}
w^{J}\,e^{\frac{\sqrt\lambda}{2}w}\,\Big(1+\frac{\sqrt\lambda}{2N w}\Big)^{N}\,
\Big[\Big(1+\frac{\sqrt\lambda}{2N w}\Big)^{J}-1\Big]. 
\ee
Expanding at large $N$ gives
\be
\vev{\mc W\, \mathsf{O}_{J}} = J\,(N/2)^{J/2}\,\oint\frac{dw}{2\pi i}
w^{J-1}\exp\Big(\frac{\sqrt\lambda}{2w}+\frac{w\sqrt\lambda}{2}\Big)\,\Big[
1+\frac{2(J-1)w\sqrt\lambda+\lambda(w^{2}-1)}{8w^{2}N}+\cdots\Big].
\ee
Using the identity
\be
I_{J}(\sqrt\lambda) = \oint\frac{dw}{2\pi i}
w^{J-1}\exp\Big(\frac{\sqrt\lambda}{2w}+\frac{w\sqrt\lambda}{2}\Big),
\ee
this gives 
(including all required additional terms in (\ref{2.24}))
\be
\vev{\mc W\, \mathsf{O}_{J}} = J\,(N/2)^{J/2}\Big[
I_{J}+\frac{1}{N^{2}}\Big(
\frac{(J+1)(J-2)}{96}\lambda I_{J}+\frac{\lambda+2(J^{2}-1)(J-2)}{96}\sqrt\lambda I_{J-1}
\Big)+\cdots
\Big],
\ee
where we notice that the $1/N$ correction due to the first term in square brackets in (\ref{2.24}) happens to cancel 
due to the Bessel function identity $2(J-1) \sqrt\lambda\,I_{J-1}(\sqrt\lambda)+\lambda(I_{J}(\sqrt\lambda)-I_{J-2}(\sqrt\lambda))=0$.
Extending this procedure to determine all terms up to order $1/N^{6}$, we find 
\ba
\la{2.24}
\vev{\mc W\,\mathsf{O}_{J}} &= (N/2)^{J/2}\, 
J\,\Big[A^{(0)}_{J}(\lambda)+\frac{1}{N^{2}}\,A^{(1)}_{J}(\lambda)+\frac{1}{N^{4}}\,A^{(2)}_{J}(\lambda) +\frac{1}{N^{6}}\,A^{(3)}_{J}(\lambda) + \cdots\Big],
\ea
where $A_{J}$ are expressed in terms of the modified Bessel functions $I_{n}\equiv I_{n}(\sqrt\lambda)$  
\ba
& A^{(0)}_{J}(\lambda) = I_{J},\qquad \qquad 
A^{(1)}_{J}(\lambda) =
\frac{(J+1)(J-2)}{96}\,\lambda\,I_{J}+\frac{\lambda+2(J^{2}-1)(J-2)}{96}\sqrt\lambda\,I_{J-1},\la{225}  \\
& A^{(2)}_{J}(\lambda) = 
\frac{(J-3) (-40-18 J+19 J^2+3 J^3)}{92160}\,\lambda^{2}\,I_{J}   \lp\qquad\qquad\qquad
+\Big[\frac{(J-3) (J-2) (-16-8 J+9 J^2+3 J^3)}{23040}+\frac{-12-5 J+5 J^2}{46080}\,\lambda\Big]\,\lambda^{3/2}\,I_{J-1} \lp\
\qquad\qquad\qquad
+\Big[\frac{(J-4) (J-3) (J-2) (J+1) (-6-J+3 J^2)}{23040}+\frac{1}{18432}\,\lambda^{2}\Big]\,\lambda\,I_{J-2}, \la{226} \\
& A^{(3)}_{J}(\lambda) = \frac{4800+8248 J-1254 J^{2}-1891 J^{3}+447 J^{4}+27 J^{5}-9 J^{6}}{92897280}\,\lambda^{3}\,I_{J} \lp\quad
+\Big[\frac{1}{8064}+\frac{828+196 J-413 J^{2}+21 J^{4}}{61931520}\,\lambda\Big]\,\lambda^{5/2}\,I_{J-1}   \la{224}\\ \quad
&+\Big[\frac{(J-5) (J-4) (J-3) (288+228 J-184 J^2-125 J^3+24 J^4+9 
J^5)}{15482880}\lp
+\frac{(J+1) (-8592+2788 J+1604 J^2-733 J^3+36 J^4+9 J^5)}{61931520}\,\lambda-\frac{14+5 J-5 J^2}{8847360}\,\lambda^{2}\Big]\,\lambda^{2}\,I_{J-2}\lp
+\Big[\frac{(J-6) (J-5) (J-4) (J-3) (J+1) (80+34 J-57 J^2-18 J^3+9 
J^4)}{23224320} 
+\frac{\lambda^{3}}{5308416}\Big]\,\lambda^{3/2}\,I_{J-3} \no 
\ea
It is then   straightforward  to expand the  coefficients $A^{(n)}_J (\lambda)$ at large $\l$ for any $J\ge 2$.\foot{
\la{fo14} The  expressions \rf{222} and \rf{225}--\rf{224}  apply  for $J\ge 2$.
  The case of   $J=0$ is trivial 
 while for $J=1$  we  get  $\vev{\mc W\,\mathsf{O}_{1}} = \frac{g}{2\sqrt 2}\,\vev{\mc W}$ 
by  contraction of $\tr a$ with $\tr \exp(\frac{g}{\sqrt 2}a)$. The $J\ge 2$ 
restriction can be understood at the planar level  by noting that the  recursion  relation 
leading to the $I_{J}$ term is based on a recursion over the number of 
scalar propagator endpoints and this has a regular structure only for $J\ge 2$ (\cf section 2.2 in 
\cite{Semenoff:2001xp}).}

\subsection
{String theory interpretation}

Let us now  rewrite the above expansions
 in terms of the  string   coupling and tension in \rf{1.1},  
 setting  $N = \pi\,T^{2}/\gs$ and $\lambda=(2\pi T)^{2}$. 
 Let us also   choose a particular  normalization of the chiral primary operator. 
One possibility could be to  impose  as in \cite{Semenoff:2001xp}  the condition that the  two-point function  
should be unit-normalized. 
However,  this choice does not appear to be  natural in  the string theory context.\foot{For   example,  the 
 string  dilaton   vertex  has a factor  of $T\sim \sqrt\lambda$   and no $\gs\sim 1/N$ factors (see, e.g., 
 \cite{Giombi:2020mhz}).
 Its  gauge theory counterpart   is the SYM Lagrangian  
$\frac{1}{g_{\rm YM}^2}  \tr F^2_{mn}+...$ and its 2-point function scales as $N^2$.}
Below we shall assume that the operators $\mc O_{J}  $ 
that should   correspond to the string vertex operators should be  normalized  relative to the matrix model operator 
$\mathsf{O}_{J}$ in \rf{2.6}  as\foot{Below we  shall use the label  $\mc O_{J} $  for the CPO as in \rf{2.6} 
even though its  normalization will be  different.}
\be\la{2255}
\mc O_{J} = \Big({\gs\ov T^2} \Big)^{\frac{J}{2}-1}\,\mathsf{O}_{J} =  \Big({\pi \ov N} \Big)^{\frac{J}{2}-1}\, \mathsf{O}_{J} \ .
\ee
Since   at  strong coupling   the correlators in \rf{218}--\rf{220}  scale  as  $ N^{{J\ov 2}-1} \sqrt \l $
we will then have  at the  leading  planar order
\be\la{2311}
\frac{\vev{\mc W\, \mc O_{J}}}{\vev{\mc W}}\ \sim \  \sqrt \l  \, N^{{J\ov 2}-1} \,   \Big({\pi \ov N} \Big)^{\frac{J}{2}-1} \sim T \ , 
\ee
in agreement with  the canonical  normalization of the  corresponding 
string vertex operator. 

Including  subleading  corrections and using  (\ref{B.2}), we  then obtain from (\ref{2.24})  the following 
expression  for general value of $J$
\ba
\la{2.28}
 \Big. \frac{\vev{\mc W\,\mathcal{O}_{J}}}{\vev{\mc W}}\Big|_{\gs\ll 1, \, T\gg 1} = &
c_{J}\,T
\,\Big\{
1 -\frac{J^{2}-1}{4\pi\,T}
+\cdots
+\frac{J^{2}-1}{24}\,\frac{\gs^{2}}{T^{2}}\,\Big[
1-\frac{J^{2}-4J+6}{4\pi\,T}
+\cdots
\Big]\lp
+\frac{J^{2}-1}{1920}\,\frac{\gs^{4}}{T^{4}}\,\Big[
J^{2}-9-\frac{J^{4}-8J^{3}+16J^{2}+32J-120}{4\pi\,T}
+\cdots
\Big]\\
&+\frac{J^{2}-1}{322560}\,\frac{\gs^{6}}{T^{6}}\,\Big[
(J^{2}-9)(J^{2}-25)\lp
-\frac{5040-768 J-944 J^2+240 J^3+22 J^4-12 J^5+J^6}{4 \pi  T}
+\cdots
\Big]
+\mc O\Big(\frac{\gs^{8}}{T^{8}}\Big)\Big\},\notag
\ea 
where dots stand for  terms subleading at  large  $T$ and the  value of the   overall  coefficient 
\be
\la{2.29}
c_{J} = J\,\Big(\frac{\pi}{2}\Big)^{J/2}
\ee 
 reflects  our choice of normalization of $\mc O_{J}$ in \rf{2255}. 

\subsubsection{Resummation of  leading   strong coupling  
 terms}
 \la{s2.4.1}

Separating  the   leading  $(\gs/T)^{2n}$ terms in the brackets in  (\ref{2.28})  we get 
\ba
& \frac{\vev{\mc W\,\mathcal{O}_{J}}}{\vev{\mc W}} = 
c_{J}\,T\, 
\Big[ \,F_{J}\Big(\frac{\gs^{2}}{T^{2}}\Big)\,  +\cdots\Big],\la{229} \\ \la{2.30}
F_{J}(x) &= 1+\frac{J^{2}-1}{24}\,x+\frac{(J^{2}-1)(J^{2}-9)}{1920}\,x^{2}+
\frac{(J^{2}-1)(J^{2}-9)(J^{2}-25)}{322560}\,x^{3}+\cdots\ , 
\ea 
where  
dots  in \rf{229}   stand for the terms  which are   subleading  in $1/T$   at each order in $g_\str$.
Thus, formally,   keeping only  $F_J$ part of \rf{229}
 is the same as   keeping only the terms  that are non-vanishing 
at $T\to\infty$ for  fixed $x= \frac{\gs^2}{T^2}= \frac{\l}{4 N^2 }$.
The pattern of the leading coefficients 
in \rf{2.30}  suggests the all-order  conjecture
\be\la{231} 
F_{J}(x) = \sum_{n=0}^{\infty}\frac{\prod_{k=1}^{n}\big[J^{2}-(2k-1)^{2}\big]}{4^{n}(2n+1)!}\,x^{n} = \frac{2}{J\,\sqrt x}\sinh\Big(J\,\text{arcsinh}\frac{\sqrt x}{2}\Big).
\ee
As was mentioned   in  section 1.1  in the Introduction, this resummed expression 
 agrees with the  semiclassical   D3-brane calculation in \cite{Giombi:2006de} 
generalizing the  computation of $\vev{\mc W}$ in  \cite{Drukker:2005kx} to the case of 
correlators with chiral primary operators.  

To explain  the reason  this  agreement, let us  recall that the semiclassical D3-brane probe 
description applies   to the  
expectation value of  the  circular Wilson loop in the $k$-symmetric representation 
and in the limit where $\kkappa= \frac{k\sql}{4N}$ is fixed  for  large $\lambda$ and $N$. 
Remarkably, at large $N$ and large $\lambda$   the result  for  the $k$-symmetric Wilson loop 
is the  same as for the  simpler  $k$-fundamental Wilson loop 
\cite{Okuyama:2006jc,Hartnoll:2006is,Kawamoto:2008gp}
for which  the dependence on $k$ is  obtained from the $k=1$   case  by simply 
    rescaling  $\lambda\to k^{2}\lambda$.
Hence,  in the above large $N,\l$  limit with fixed $\frac{\sql}{N}\sim \frac{\gs}{T}$ 
 the  semiclassical D3-brane  description  should  also reproduce the 
result for   the Wilson loop in the 
fundamental   ($k=1$)  representation,
  but this  limit  is equivalent to the  one  we considered 
when we  neglected the subleading $1/T$  terms  in  the full expansion (\ref{2.28}). 
This  leading
contribution \rf{229},\rf{231} may be  obtained also  by  directly  from  the matrix model saddle point at fixed 
$\frac{\sql}{N}$ \cite{Giombi:2006de}.

For odd $J$ 
 the function $F_{J}(x)$ in \rf{231} reduces to a polynomial in $x$,   while  for even $J$
 the series expansion in $x$ does not truncate -- in this case 
$F_{J}(x)$ turns out to be $\sqrt{1+x/4}$ times a polynomial in $x$. 
Indeed,  from the definition of the  Chebyshev polynomials
\be
\TT_{n}(\cos\theta) = \cos(n\,\theta),\qquad {\rm U}_{n}(\cos\theta)\sin\theta = \sin\big((n+1)\theta\big)\ ,
\ee
we obtain
\be
\sinh(J\,\text{arcsinh}\, t)  = \begin{cases}
i\,(-1)^{\frac{J}{2}}\,\sqrt{1+t^{2}}\,{\rm U}_{J-1}(i\,t),  & \quad J\ \text{even} \\
i\,(-1)^{\frac{J+1}{2}}\,\TT_{J}(i\,t), & \quad J\ \text{odd}
\end{cases}
\ee 
For even $J$, the overall factor $\sqrt{1+t^{2}}=(1+x/4)^{1/2} = (1+\frac{\gs^{2}}{4T^{2}})^{1/2}$  (that  has  an imaginary branch point) 
shows that the $\frac{\gs}{T}$ expansion has a finite radius of
convergence. Explicitly, one finds for $F_J$  in \rf{231} 
\ba
F_{2}(x) &= \frac{1}{2}\sqrt{4+x}\ , \qquad
F_{3}(x) = 1+\frac{x}{3}\ , \qquad
F_{4}(x) = \frac{1}{4}(2+x)\,\sqrt{4+x}\ ,\la{235} \\
F_{5}(x) &= 1+x+\frac{x^{2}}{5}\ ,\qquad F_{6}(x) = \frac{1}{6}(1+x)(3+x)\,\sqrt{4+x}\ .
\ea
Starting with the resummed  expression \rf{229},\rf{231} 
we can  formally   consider the limit  when  the parameter   $x= \frac{\gs^2}{T^2}$  that was fixed  in the resummation is 
now taken to be   
 large.  Using that 
$F_{J}(x) \stackrel{x\to\infty}{=} {J}^{-1}\,(\sqrt x\,)^{{J-1}}+\cdots, $  and \rf{2.29}
we  then get\foot{The limit $ \gs/T \gg 1$  assumed   here is of course formal as in the original expansion we assumed  that both 
$\gs$   and $1/T$ are small.}
\be
\frac{\vev{\mc W\,\mathcal{O}_{J}}}{\vev{\mc W}} \simeq  
\Big(\frac{\pi}{2}\Big)^{J/2} T\,\Big(\frac{\gs}{T}\Big)^{J-1}+\cdots \ . 
\ee
One can also consider the limit of large $J$.  The result depends on the  assumption  about growth of $J$ relative to $T$. 
To be able to ignore the $1/T$ corrections in  square brackets in (\ref{2.28})  and thus 
 use  the resummed expression in \rf{229},\rf{231} 
$J^2  $  should    grow  slower than $T$ (i.e. $J \ll \lambda^{1/4}$). 
Then $c_J F_J(x) \sim \exp ( J \text{arcsinh}\frac{\sqrt x}{2})$. 
Another interesting limit corresponds to  the semiclassical   large charge 
expansion  in  the  dual string  theory  when  $J\sim T \gg 1$.
 In this  case  the $1/T$ corrections in (\ref{2.28})
are not  negligible  and \rf{229},\rf{231} cannot be used.   This limit will be  discussed in  Appendix \ref{sec:semi} below.

\subsubsection{Comparison of  expansions  of  $\vev{\mc W}$ and $\vev{\mc W\,\mc O_{2}}/ \vev{\mc W}$}

Let us recall that 
the chiral primary  operator $\mc O_{J}\sim \tr \varphi^{J}$  with dimension $\Delta=J$ 
 belongs to the same   short   supermultiplet as the  R-charge  generalization of the dilaton operator $\mc O_{{\rm dil}, J'}\sim \tr (\varphi^{J'}  F^2_{mn} ) + ...$  of dimension $\Delta = 4 + J'$  with $J'= J-2$. 
The   standard $\Delta =4$  dilaton operator is  the  supersymmetry   descendant of the 
$J=2$   CPO  and thus   their   correlators with the 
BPS Wilson loop should   be   directly related. 
  Indeed,  like the  dilaton correlator,  
the CPO correlator  can be obtained   from $ \vev{\mc W}$   by the  differentiation over the  coupling  using 
 (\ref{2.12}),\rf{216},\rf{2255}   (cf.  \rf{1.11})\footnote{For  general $J$, the supersymmetry  relation between the  Wilson loop 
 correlators with 
  CPO   $\mc O_{J}$ 
      and  with  the dilaton operator  $\mc O_{{\rm dil}, J-2}$ 
imply that  $\vev{\mc W\,  \mc O_{{\rm dil}, J-2} }$     can also   be obtained   from $\vev{\W}$  by the differential 
relations  like  \rf{216}--\rf{2188}.} 
\be
\la{2.37}
\frac{\vev{\mc W\,\mc O_{2}}}{\vev{\mc W}}= \lambda\,\partial_{\lambda}\log \vev{\mc W} \ . 
\ee
According to \rf{229},\rf{235},\rf{2.29}  the result  of the resummation of the 
strong coupling  expansion  for the $J=2$   case is simply 
\be
\la{2.38}
\frac{\vev{\mc W\,\mathcal{O}_{2}}}{\vev{\mc W}} \simeq c_{2}\,T\,\sqrt{1+\frac{\gs^{2}}{4T^{2}}} = \pi\,\sqrt{T^{2}+\frac{1}{4}\gs^{2}} \ , 
\ee
or, in gauge theory notation, $\frac{1}{2}\,\sqrt{\lambda+\frac{\lambda^{2}}{16\pi^{2}\,N^{2}}}$. 
The leading strong coupling term here agrees  with \rf{2.37}    since   $\vev{\mc W} \sim e^{\sql}$   and thus 
 $\lambda\partial_{\lambda}\log\vev{\mc W} = \frac{1}{2}\sqrt\lambda+\cdots$.
 However, 
 the resummed   expression  \rf{1.5}  for  $\vev{\mc W} $  does not lead to \rf{2.38}  if substituted  into \rf{2.37}.
As already discussed in the Introduction, the reason why the two resummations   are not  directly related is that 
subleading in $1/T$ terms  in   $\vev{\mc W} $  cannot  be in general  ignored  in $\log  \vev{\mc W} $ in \rf{2.37} (see \rf{x1}--\rf{x2}).
In more detail, 
the  structure of the  expansion of $\vev{\mc W}$   is 
\ba
\la{2.39}
\vev{\mc W} &= e^{2\pi T}\Big(\frac{\gs^{2}}{T}\Big)^{-1/2}\,\sum_{p=0}^{\infty}\frac{1}{2\pi p!}\Big(\frac{\pi}{12}\Big)^{p}
\Big(\frac{\gs^{2}}{T}\Big)^{p}\,
\Big(1+\frac{a_{p}^{(1)}}{T}+\frac{a_{p}^{(2)}}{T^{2}}+\cdots\Big),
\ea
where  the  values of the  coefficients $a_{p}^{(n)}$  may be extracted from (\ref{1.4}) 
 \cite{Drukker:2000rr}. Then including the subleading terms   we have 
\ba
\la{2.41}
 \frac{\vev{\mc W\,\mc O_{2}}}{\vev{\mc W}} = \lambda\,\partial_{\lambda}\log \vev{\mc W} 
= \pi\,T\,\Big[
1 &-\frac{3}{4\pi T}-\frac{a_{0}^{(1)}}{2\pi T^{2}}+\cdots+\frac{\gs^{2}}{T}\,\Big(\frac{1}{8T}
+\frac{a_{1}^{(1)}-a_{0}^{(1)}}{12\,T^{2}}+\cdots\Big)\lp
+\Big(\frac{\gs^{2}}{T}\Big)^{2}\,\Big(
 \,\frac{5 \pi  \big(a_{0}^{(1)}-2a_{1}^{(1)}+a_{2}^{(1)}\big)}{576 T^{2}}+\cdots\Big)+\cdots\,
\Big].
\ea
The  resummation of $\vev{\mc W}$  leading to the $g^2_\str\ov T$  exponent in 
 (\ref{1.5}) amounts to dropping 
 all  subleading  $a_{p}^{(n)}$ corrections in \rf{2.39}  but they actually contribute  to the leading order terms in 
 \rf{2.41}   starting with the order  $\big(\frac{\gs^{2}}{T}\big)^{2}$.
 Using that $a_{p}^{(1)} = -\frac{3(12p^{2}+8p+5)}{80\pi}$  
 one finds  indeed the agreement with the result of the direct computation  of the order $\frac{\gs^{4}}{T^4}$
 term in  the $J=2$   CPO correlator in  the brackets in \rf{218},\rf{2.28}   which  corresponds to the 
 $g^4_\str$ term in the expansion of the square root in \rf{2.38}. 
 
\iffa 
 \footnote{
Of course, from the exact expression of $\vev{\mc W}$ we can improve with respect to (\ref{1.5}).
From instance, one has, \cf eq.(B.7) of ,
$a_{p}^{(0)} = -\frac{3(12p^{2}+8p+5)}{80\pi}$,
and plugging this into (\ref{2.39}) gives  
\ba
\notag
\frac{\vev{\mc W\,\mc O_{2}}}{\vev{\mc W}} &= c_{2}T\,\Big[
1-\frac{3}{4\pi T}+\frac{3}{32\pi^{2}T^{2}}+\cdots+\frac{\gs^{2}}{T}\,\Big(\frac{1}{8T}
-\frac{1}{16\pi T^{3}}+\cdots\Big)
+\Big(\frac{\gs^{2}}{T}\Big)^{2}\,\Big(-\frac{1}{128T^{4}}+\cdots\Big)+\cdots\,
\Big],
\ea
where the omitted terms depend on $a_{p}^{(1)}$. The predicted terms are again in agreement with (\ref{2.28}).
}
\fi 

Similar  remarks apply  to higher $J$ cases of  the resummed  expression   for the 
 correlator \rf{229},\rf{231} (understood using analytic continuation in $J$) 
 when  applying  the differential relations like \rf{2.16}--\rf{2188} 
and comparing to the  resummed  expression for  $\vev{\mc W}$. 

\def \rJ {{\rm J}}

\section{Expansion of $\vev{\mc W\,\mc O_{J_{1}}\,\mc O_{J_{2}}}$ 
 \la{3}}

One  may  also   consider   a  correlation function of a circular Wilson loop with two  scalar 
 chiral primary 
operators at generic positions  $x_1$, $x_2$. 
Such correlator is fixed by conformal invariance up to a function of  $N$ and  $\lambda$
and two scalar combinations u and v 
of the positions invariant under the conformal transformations preserving the circle \cite{Buchbinder:2012vr}. 
Explicitly, for 
 $\vev{\mc W\,\mc O_{1}(x_{1})\,\mc O_{2}(x_{2})}$  
where $\mc O_{1}$ and $\mc O_{2}$ are  scalar primary 
operators of dimensions $\Delta_{1}$,$\Delta_{2}$ 
 at points $x_{1}, x_{2}\in \mathbb R^{4}$ and $\mc W$ is the circular $\frac{1}{2}$-BPS loop 
of unit radius the conformal symmetry  implies  that\foot{One can conformally map $\mathbb R^{4}\to AdS_{2}\times S^{2}$ so that the circle is mapped to the boundary 
of $AdS_{2}$. Then   $ \vev{\mc W\,\mc O_{1}(x_{1})\,\mc O_{2}(x_{2})}/\vev{\W}$   
is invariant under the 6 isometries of $AdS_{2}\times S_{2}$ (corresponding  to 
6 conformal transformations that preserve the circle in $\mathbb R^{4}$). It is expressed in terms of two functions
(u  and v)  of the $AdS_{2}$
and $S^{2}$ geodesic distances between the operators (see  \cite{Buchbinder:2012vr} for details).}
\be
\la{3.1}
\frac{\vev{\mc W\,\mc O_{1}(x_{1})\,\mc O_{2}(x_{2})}}{\vev{\mc W}} = \frac{{\cal F}({\rm u}, {\rm v}; N, \lambda)}{d_{\perp}^{\Delta_{1}}(x_1)\, d_{\perp}^{\Delta_{2}}(x_2) }\ ,
\ee
where $d_{\perp}(x) $ for a point $x\in \mathbb R^{4}$ was
 defined in (\ref{2.2}). 
Fixing particular values of  $x_1$,$x_2$  and  thus of 
u and v  one may then study  the   $1/N$ expansion of the resulting   function.

It turns out that  for   special supersymmetric   configurations 
 correlators  of  certain BPS  Wilson loops  with  local operators 
  may  be computed to  all orders by localization by reducing  them   to 
 correlators in a  multi-matrix model \cite{Giombi:2012ep}.  
 Examples   include 
special $\frac{1}{8}$-BPS Wilson loop   which is  a  contour on  a  2-sphere   $S^{2}\subset \mathbb R^{4}$.

In  the   general  $1\ov 8$-BPS case,  one considers   \cite{Giombi:2012ep}  the  operators 
 ${\mc O }_J (x) = \tr \big[ x_n \Phi_n(x) + i \Phi_4(x)\big]^J $  (for   $x^2_n=1$, \ $n=1,2,3$)
 and the Wilson loop  for  a contour  on   $S^{2}\subset \mathbb R^{4}$
 with the scalar coupling being $\int \epsilon_{nkl} \Phi_n x_k d x_l $  (cf. \rf{1w}). 
 The  special $1\ov 2$-BPS case  we are interested in here  corresponds to 
 placing  the operators 
at  the   poles of the 2-sphere   and the unit-circle  Wilson loop at  its 
 equator. 
This  results in the  following choice  of $x_1$ and $x_2$ 
\be 
x_{1}=(0,0,1,0)\ , \  \quad  \ x_{2}=(0,0,-1,0)\ , \ \ \qquad  
{\rm u}=-{\rm v}=1\ , \ \ \qquad  \ 
d_{\perp}(x_1)=d_{\perp}(x_2)=\tfrac{1}{2}  \ . \la{ddd} 
\ee
Then  the correlator in \rf{3.1}    becomes  explicitly 
\be  \vev{\mc W\, \tr\big[(\Phi_3 + i \Phi_4)^{J_1}\big] \,  \tr\big[ (-\Phi_3 + i \Phi_4)^{J_2} \big]   } \ ,  \la{333} \ee 
 and the Wilson loop  scalar  coupling becomes the same   as in \rf{1w}   with $\Phi_1\to \Phi_3$  (and $R=1$).
 For general $x_1,x_2$  the correlator \rf{333} has the  structure \rf{3.1}  but  its  value  can be 
  computed by localization at specific  positions in \rf{ddd}.

In detail, it can be  computed using  a 3-matrix model with the following  
action depending on the hermitian matrices $X_{1}$, $X_{2}$, $X_{3}$  \cite{Giombi:2012ep}\foot{
We  specialize  
the expression   in \cite{Giombi:2012ep} 
  to the case of \rf{ddd}.}
\ba\la{32}
S &= \frac{8\pi^{2}}{g^{2}}\tr\Big(X_{1}^{2}-\frac{1}{4\pi^{2}} X_{2}^{2}+X_{3}^{2}-\frac{i}{\pi}X_{1}X_{2}+\frac{i}{\pi}X_{2}X_{3}\Big),\qquad \ \ g\equiv  g_{_{\rm YM}}.
%
\ea
The connected part of  the  correlator \rf{3.1}   is related to a particular  matrix model 
correlator  which admits the following $1/N$ expansion 
\be\la{34} 
\vev{\tr X_{1}^{J_{1}}\tr e^{X_{2}}\tr X_{3}^{J_{2}}}_{\rm conn} \equiv Q_{J_{1},J_{2}}(\lambda; N) =    \frac{Q_{J_{1},J_{2}}^{(1)}(\lambda)}{N}+ \frac{Q_{J_{1},J_{2}}^{(2)}(\lambda)}{N^{2}}+\cdots\ . 
\ee
For the coefficient $Q_{J_{1},J_{2}}^{(1)}(\lambda) $ of the  leading planar  contribution  one finds \cite{Giombi:2012ep}  
\ba
\la{3.5}
&Q_{J_{1},J_{2}}^{(1)}(\lambda)   = J_{1}J_{2}\,\Big(\frac{i\sql}{4\pi}\Big)^{J_{1}} \Big(-\frac{i\sql}{4\pi}\Big)^{J_{2}} 
 \Big[\sum_{k=1}^{\min(J_{1},J_{2})}(J_{1}+J_{2}-2k)\,I_{J_{1}+J_{2}-2k}(\sql)\lp\qquad \qquad \qquad \qquad \qquad \qquad \qquad \qquad \qquad 
+\sum_{k=1}^{\infty}(J_{1}+J_{2}+2k-2)\,I_{J_{1}+J_{2}+2k-2}(\sql)\Big].
\ea
The 3-matrix model representation \rf{32},\rf{34}
can be translated into a  Gaussian 1-matrix model one  similar to the   one 
 considering in the previous section (cf. \rf{2.5},\rf{2.6}). Indeed,  
after  the change of variables
\be
X_{1}\to A+i C\ ,\qquad X_{2}\to 2\pi C\ ,\qquad X_{3}\to B-i C\ ,
\ee
 the correlator in \rf{34}   becomes 
\be
\la{3.12}
\vev{\tr (A+i C)^{J_1}  \tr(B-i C)^{J_2} \tr e^{2\pi C} },
\ee
computed in  the matrix model with  the  
decoupled Gaussian action $S\sim A^{2}+B^{2}+C^{2}$. 
Integrating out the $A$ and $B$  matrices  amounts to subtracting from $\tr (A+i C)^{J_1}$ and $\tr(B-iC)^{J_2}$ their self contractions, resulting in 
the normal ordering 
  discussed in section \ref{s2.1}.\footnote{The 
 relation between the 2-matrix model and the 1-matrix model with explicit normal ordering
follows also from the equivalence between the 2-matrix model and the complex matrix model of \cite{Kristjansen:2002bb}
(see, for instance,  Appendix C of \cite{Okuyama:2006jc}).}
We then end up with the following  correlator in the 1-matrix model for $C$
\be\la{38}
\vev{:\tr C^{J_{1}}:\ :\tr C^{J_{2}}:\ \tr e^{2\pi C}}\ .
\ee
  Below   we shall  consider  two examples  of  the 
 correlators \rf{3.1}. The first has $J_{1}=J_{2}=J$ and the second $J_{1}=2$ and $J_{2}=2J$ ($J$ is integer).
  We shall  use them to  
illustrate the general features of the  strong coupling  limit   of the coefficients  of the $1/N$ expansion  of 
   (\ref{3.1}).

\subsection{$J_{1}=J_{2}=J$  \la{s3.1}}

In this case   the explicit  form of the relation   between  
the matrix model correlator and the  function of $\l,N$ in 
 \rf{3.1},\rf{ddd}  is 
\ba\la{71}
Q_{J}(\lambda; N) =\vev{\tr X_{1}^{J}\tr e^{X_{2}}\tr X_{3}^{J}}_{\rm conn} &=
 \Big(\frac{\lambda}{8\pi^{2}N}\Big)^{J}\,\Big[\vev{ \mc W\,\mathsf O_{J}\mathsf O_{J}}-\vev{\mc W}\, \vev{\mathsf O_{J}\mathsf O_{J}}\Big]\ , 
  \notag \\
 &= \frac{\pi^{2}}{N^{2}}\Big(\frac{\lambda}{8\pi^{3}}\Big)^{J}\,\Big[\vev{ \mc W\,\mc O_{J}\mc O_{J}}-\vev{\mc W}\, \vev{\mc O_{J}\mc O_{J}}\Big],
\ea
where $\mathsf O_{J}$  are the matrix model operators  
  the notation 
of  section \ref{s2.1} (cf. \rf{2.6}), i.e.   $\mathsf O_{J} =  :\tr a^ J: $ after 
renaming  $C\to a$.
We used that  $\vev{\mathsf O_{J}}=0$.\footnote{Recall that   for  any 3 operators 
$\vev{O_1 O_2 O_3  }_{\rm conn} = \vev{O_1 O_2 O_3 }-\vev{O_1}\vev{O_2 O_3 }-\vev{O_2}\vev{O_1O_3}-\vev{O_3}\vev{O_1O_2}+2\vev{O_1}\vev{O_2}\vev{O_3}$.}
 The operators  $\mc O_{J}$   in \rf{71}   are  assumed  to be 
 normalized  as  in (\ref{2255}).
Let us   consider explicitly  the  $J=2$ case  when 
 \ba\la{316}
 \mathsf{O}_{2}^{2} &= \mathsf{O}_{2,2}+2\mathsf{O}_{2}+\frac{N^{2}}{2}, \qquad   \vev{\mathsf{O}_{2}^{2}} = \frac{N^{2}}{2}, \qquad  \mathsf{O}_{J} = :\tr a^ J: \ , \quad  \mathsf{O}_{2,2} = :(\tr a^{2})^{2}:\ , \\
 \vev{\,\mc W\,\mathsf{O}_{2}^{2}} &= \vev{\mc W\,\mathsf{O}_{2,2}}+2\,\vev{\mc W\,\mathsf{O}_{2}}+\frac{N^{2}}{2}\vev{\mc W}
 = \lambda^{2}\partial_{\lambda}^{2}\vev{\mc W}+2\,\lambda\partial_{\lambda}\vev{\mc W}+\frac{N^{2}}{2}\vev{\mc W}\ .
 \ea
 Here we used the relation
  $ \vev{\mc W\,:(\tr a^{2})^{2}:} = \frac{1}{4}(g^{2}\partial_{g}^{2}-g \partial_{g})\,\vev{\mc W} = \lambda^{2}\partial_{\lambda}^{2}\vev{\mc W} $
 that may be proved  using the  same  method  as in  section \ref{s2.2}. 
 As a result, we get  the  following differential relation  for the $J=2$ case of \rf{71} 
 \be
 \la{3.16}
 Q_{2}(\lambda; N) = \Big(\frac{\lambda}{8\pi^{2}N}\Big)^{2}\,(\lambda^{2}\partial^{2}_{\lambda}+2\lambda\partial_{\lambda})\, \vev{\mc W}\ .
 \ee
 Using the $1/N$ expansion of $\vev{\W}$ in  (\ref{B.2}) we  find    
 \be
 Q_{2}(\lambda; N) = \frac{1}{N}\,\Big(\frac{\lambda}{16\pi^{2}}\Big)^{2}\,\Big[2\sqrt\lambda\,I_{1}(\sqrt\lambda)
 +\frac{1}{N^{2}}\frac{\lambda^{3/2}}{48}\Big(\sqrt\lambda\,I_{0}(\sqrt\lambda)+4I_{1}(\sqrt\lambda)\Big)+\mc O\Big(\frac{1}{N^{4}}\Big)\Big],
 \ee
 Similar  calculation can be repeated for  higher $J$ and  leads to   
 \ba 
 Q_{J}(\lambda;N ) =& { 1\ov N} \Big[ Q_{J}^{(1)}(\lambda)  +{1 \ov N^2}  Q_{J}^{(2)}(\lambda)  + \OO\big({1\ov N^4}\big)\Big] \ , \qquad \qquad 
Q_{J}^{(1)}(\lambda) = \frac{J^{2}}{2}\,\Big(\frac{\lambda}{16\pi^{2}}\Big)^{J}\,\sqrt\lambda\,I_{1}(\sqrt\lambda), \notag \\
Q_{J}^{(2)}(\lambda) =&\frac{J^{2}}{192}\,\Big(\frac{\lambda}{16\pi^{2}}\Big)^{J}\,\sqrt\lambda\,\Big\{
\big[4 (J^{2}-1)(J-2)+\lambda\big]\,\sqrt\lambda\,I_{0}(\sqrt\lambda)\lp\qquad \qquad \qquad \qquad \qquad 
+\big[4(J^{2}-1)(J-2)^{2}+2(J^{2}-2)\,\lambda\big]\,I_{1}(\sqrt\lambda)\Big\}.
 \ea
 Dividing  \rf{71} over  ${\vev{\mc W}} $  leads to  (cf. \rf{3.1})  
 \ba
 \la{3.19}
 {\vev{ \mc W\,\mc O_{J}\mc  O_{J}}\ov \vev{\W}}  -\vev{\mc O_{J}\mc O_{J}}
 & = N^{2}\pi^{J-2}\,\Big[
 \frac{J^{2}\lambda}{4N^{2}}+\frac{J^{2}(J^{2}-1)}{N^{4}}\,\Big(
 \frac{\,\lambda^{2}}{192}+\frac{(J-2)\,\lambda^{3/2}}{96}+ ... 
 \Big)
 +\mc O\Big(\frac{1}{N^{6}}\Big)
 \Big]
 \notag \\
 & = 
 \pi^{J}\, J^2\,T^{2}\,\Big[1+\frac{J^{2}-1}{12}\frac{\gs^{2}}{T^{2}}\,\Big(1+\frac{J-2}{\pi T}+\cdots\Big)+\cdots\Big].
 \ea

\subsection{$J_{1}=2$, $J_{2}=2J$ \la{s3.2}}

\iffa 
In this case the specialization of (\ref{3.5}) is  
 \ba
&Q_{J_{1},J_{2}}^{(1)}(\lambda)   = 2 (2J)\,\Big(\frac{i\sqrt\lambda}{4\pi}\Big)^{2}\Big(-\frac{i\sqrt\lambda}{4\pi}\Big)^{2J}\times\notag \\
&\times\Big[\sum_{k=1}^{2}(2J+2-2k)\,I_{2J+2-2k}(\sqrt\lambda)+\sum_{k=1}^{\infty}(2J+2k)\,I_{2J+2k}(\sqrt\lambda)\Big] \notag \\
&= 4J\,(-1)^{J+1}\Big(\frac{\lambda}{16\pi^{2}}\Big)^{J+1}\,\Big[\frac{\sqrt\lambda}{2}I_{1}(\sqrt\lambda)-\sum_{k=1}^{J-2}2k\,I_{2k}(\sqrt\lambda)\Big]
\ea
The first cases are $J=2, 3$ for which the planar results read
\ba
\la{3.21}
Q_{2,4}^{(1)}(\lambda)   &= -4\,\Big(\frac{\lambda}{16\pi^{2}}\Big)^{3}\,\sqrt\lambda\,I_{1}(\sqrt\lambda), \notag \\
Q_{2,6}^{(1)}(\lambda)   &= 12\,\Big(\frac{\lambda}{16\pi^{2}}\Big)^{4}\,\Big[\frac{\sqrt\lambda}{2}I_{1}(\sqrt\lambda)-2\,I_{2}(\sqrt\lambda)\Big].
\ea
\fi
 In this case 
the   1-matrix model representations  for the correlators \rf{34}   with  $J=2,3$ are\footnote{The sign is $i^{J_{1}}(-i)^{J_{2}}$ from (\ref{3.12}).}
\ba
&Q_{2,4}(\lambda; N) = -\Big(\frac{\lambda}{8\pi^{2}N}\Big)^{3}\,\vev{\mc W\,\mathsf{O}_{2}\,\mathsf{O}_{4}} = -\frac{\pi^{2}}{N^{2}}\,\Big(\frac{\lambda}{8\pi^{3}}\Big)^{3}\,\vev{\mc W\,\mc{O}_{2}\,\mc{O}_{4}}, \la{339} \\
& Q_{2,6}(\lambda; N) = \Big(\frac{\lambda}{8\pi^{2}N}\Big)^{4}\,\vev{\mc W\,\mathsf{O}_{2}\,\mathsf{O}_{6}} = \frac{\pi^{2}}{N^{2}}\Big(\frac{\lambda}{8\pi^{3}}\Big)^{4}\,\vev{\mc W\,\mc {O}_{2}\,\mc {O}_{6}}, \la{319}
\\
&\mathsf{O}_{4} = :\tr a^{4}: =\textstyle   \tr a^{4}-2N\,\tr a^{2}-(\tr a)^{2}+\frac{1}{4}N(1+2N^{2})\ , \notag \\
&\mathsf{O}_{6} = :\tr a^{6}: =\textstyle   \tr a^{6}-3N\,\tr a^{4}+\frac{15}{4}\,(N^{2}+1)\,\tr a^{2}-\frac{5}{8}N^{2}(2+N^{2})\lp\qquad\qquad\qquad\quad\textstyle 
+\frac{15}{4}N\,(\tr a)^{2}-3 \tr a\,\tr a^{3}-\frac{3}{2}(\tr a^{2})^{2}.
\ea
The {exact} differential relations  for  \rf{339},\rf{319} are   found to be 
\ba
\vev{\mc W\,\mathsf{O}_{2}\,\mathsf{O_{4}}} &= \Big[
-\frac{\lambda}{16 N}-\frac{\lambda  (\lambda +16 N^2)}{16 N}\,\partial_{\lambda}
-(-36+\lambda ) \lambda  N \,\partial^{2}_{\lambda}+44 
\lambda ^2 N\,\partial_{\lambda}^{3}+8 \lambda ^3 N\,\partial_{\lambda}^{4}
\Big]\,\vev{\mc W}\ , \notag \\
\vev{\mc W\,\mathsf{O}_{2}\,\mathsf{O_{6}}} &= \Big[
\frac{3}{32} (-80+\lambda ) \lambda  \partial_{\lambda}+\frac{3}{4} \lambda 
 (-17 \lambda -80 N^2+\lambda  N^2) \partial_{\lambda}^{2}-\lambda  (3 \lambda 
^2-960 N^2+80 \lambda  N^2) \partial_{\lambda}^{3}\lp\qquad 
-16 (-115+\lambda ) 
\lambda ^2 N^2 \partial_{\lambda}^{4}+704 \lambda ^3 N^2 \partial_{\lambda}^{5}+
64 \lambda ^4 N^2 \partial_{\lambda}^{6}
\Big]\,\vev{\mc W}\ .
\ea
As a result, using \rf{B.2} we get 
\ba
&Q_{2,4}(\lambda; N) = -\frac{4}{N}\Big(\frac{\lambda}{16\pi^{2}}\Big)^{3}\,\Big\{
\sqrt{\lambda } I_1(\sqrt{\lambda })+\frac{1}{N^{2}}\Big[\frac{1}{96} \lambda  (24+
\lambda ) I_0(\sqrt{\lambda })+\frac{1}{6} \lambda ^{3/2} 
I_1(\sqrt{\lambda })\Big]\lp
\qquad +\frac{1}{N^{4}}\,\Big[\frac{\lambda ^2 (32+13 \lambda ) 
I_0(\sqrt{\lambda })}{7680}+\frac{\lambda ^{3/2} (-768+672 \lambda +5 
\lambda ^2) I_1(\sqrt{\lambda })}{92160}\Big]+\mc O\Big(\frac{1}{N^{6}}\Big)
\Big\}, \la{323} \\
&Q_{2,6}(\lambda; N) = \frac{12}{N}\Big(\frac{\lambda}{16\pi^{2}}\Big)^{4}\,\Big\{
-2I_{0}(\sqrt\lambda)+\frac{(8+\lambda)I_{1}(\sqrt\lambda)}{2\sqrt\lambda}
+\frac{1}{N^{2}}\Big[\frac{  \lambda  (24+\lambda ) I_ 0(\sqrt{\lambda}) }{192} 
\lp \qquad +\frac{\sqrt{\lambda } (-3+2 \lambda ) I_ 1(\sqrt{\lambda 
})}{12} \Big]
+\frac{1}{N^{4}}\Big[\frac{\lambda ^2 (160+7 \lambda ) I_ 0(\sqrt{\lambda})}{3840}
+\frac{\lambda ^{5/2} (3408+5 \lambda ) I_ 1(\sqrt{\lambda})}{184320}\Big]
+\mc O\Big(\frac{1}{N^{6}}\Big)
\Big\}.\no 
\ea
Taking the ratio of \rf{319} and  $\vev{\mc W}$ in \rf{B.2} 
and expanding at strong coupling  gives\foot{The absence of $1/T$ corrections  at leading  planar order in \rf{332} is due to cancellation of  the planar  $I_1(\sql)$ term in 
$Q_{2,4}(\lambda; N)$ in \rf{323}  and in $\vev{\W}$ in \rf{B.2}.}
\ba
\frac{\vev{\mc W\,\mc{O}_{2}\,\mc{O}_{4}}}{\vev{\mc W}} = &
\pi^3 T^2 \,\Big[1+
\frac{3}{4}\frac{\gs^{2}}{T^{2}}\Big(1+\frac{2}{3\pi T}+\cdots\Big)
+\frac{1}{8}\frac{\gs^{4}}{T^{4}}\,\Big(1+\frac{1}{2\pi T}+\cdots\Big)\lp\qquad\quad \quad
+ \frac{\gs^{6}}{T^{6}}\Big( -\frac{1}{256\pi T}+\cdots\Big)+
\cdots
\Big], \la{332} \\
\frac{\vev{\mc W\,\mc O_{2}\,\mc O_{6}}}{\vev{\mc W}}  = &
\frac{3\pi^4}{4}\,T^{2} \,\Big[
1-\frac{2}{\pi T}+\cdots
+\frac{19}{12}\frac{\gs^{2}}{T^{2}}\,\Big(1+\frac{3}{19\pi T}+\cdots\Big)
+\frac{2}{3}\frac{\gs^{4}}{T^{4}}\,\Big(1+\frac{143}{128\pi T}+\cdots\Big)\lp
\qquad \qquad\qquad\qquad \qquad  +\frac{1}{12}\frac{\gs^{6}}{T^{6}}\,\Big(1+\frac{119}{128\pi T}+\cdots\Big)
+\cdots
\Big] .\la{3322}
\ea
Similar expansions may be found for other values of $J$. 

Thus from  in \rf{3.19}  and \rf{332},\rf{3322} we  conclude  that   
 the  small $g_\str$, large $T$  expansion  of the   correlators \rf{3.1}  goes   
 in powers of $\frac{\gs^2}{T^2}$,   up to  subleading  $1/T$ corrections.  
  This is the same  pattern as  was found  in the case of  $\frac{\vev{\mc W\,\mc O_{J}}}{\vev{\mc W}}$ 
  in \rf{2.28}.

\section{Correlators of coincident  circular Wilson loops \la{4}}

As was mentioned in the Introduction, we can  also study the 
 $1/N$ expansion  for other observables, like  correlators of   several  circular 
Wilson loops $\vev{\mc W^n}$.
Such correlators  were  previously discussed  in particular in  the planar approximation  
in the $n=2$ case with  two circular loops in parallel planes  separated  by some distance;
  at strong coupling  one finds a transitional behaviour \cite{Gross:1998gk}   at certain critical distance 
 when the associated minimal surface  reduces to independent  
surfaces   attached to  separate  loops \cite{Zarembo:1999bu,Correa:2018lyl,Correa:2018pfn}. 

Here  we will  consider the limiting  case 
when  the loops have the same radii and are  coincident.  In this case   the  correlator $\vev{\mc W^n}$
can be found exactly using  the matrix model methods  \cite{Drukker:2000rr,Okuyama:2018aij,Sysoeva:2018xig}.
\iffa \foot{
The $n=2$ example  was used  to   check  
the  consistency of the localization approach  with weak-coupling perturbation theory \cite{Arutyunov:2001hs}.} 
\fi 
Our aim    below will be to work out 
  the large $N$, large $\lambda$ expansion of such   
 correlators.

\subsection{$\vev{\mc W^2}$ for  loops in fundamental representation}

The coincident Wilson  loops may be  considered   in generic representations 
(see, e.g.,  \cite{Sysoeva:2018xig,Aguilera-Damia:2017znn}). 
Let us consider  the case  of two loops  in the fundamental representation.\footnote{Let us  note that a   discussion of similar  correlator 
  in planar limit  at strong coupling  (i.e. using semiclassical   string theory)
  was  in section 6 of  \cite{Giombi:2009ms}  where the 
coincident $1\ov4$-BPS ``latitudes''  were considered;  the present example of  
$1\ov2$-BPS  circular  loops is  a special   case.}
The relevant $1/N$  expansions   may be written  in terms of matrix model correlators  as 
\ba
\vev{ \mc W } &= \vev{\tr\exp\Big(\sqrt{\tfrac{\lambda}{2N}}\, a\Big)} = 
\sum_{n=0}^{\infty}\frac{1}{(2n)!}\Big(\frac{\lambda}{2N}\Big)^{n}\vev{\tr a^{2n}}, \la{411} \\
\vev{ \mc W^{2} } &= 
\sum_{n,m=0}^{\infty}\frac{1}{n!\,m!}\Big(\frac{\lambda}{2N}\Big)^{\frac{n+m}{2}}\vev{\tr a^{n}\,\tr a^{m}}\ . \la{4.1}
\ea
The  expression  for \rf{411} is given by  (\ref{1.4}). A similar exact result for \rf{4.1} 
 was  found  in \cite{Drukker:2000rr,Kawamoto:2008gp,Okuyama:2018aij} (here $L^{(i)}_j$  are  the generalized  
 Laguerre polynomials  and  $L_i= L^{(0)}_i$)
\ba
\la{4.2}
\langle \mc W^{2}\rangle
 =  & 
e^{\frac{\lambda}{2N}}\,L_{N-1}^{(1)}\Big(-\frac{\lambda}{N}\Big)
+
2e^{\frac{\lambda}{4N}}\,\sum_{i=0}^{N-1}\sum_{j=0}^{i-1}\Big\{
 L_{i}\Big(-\frac{\lambda}{4N}\Big)\,L_{j}\Big(-\frac{\lambda}{4N}\Big) \lp \qquad \qquad\qquad \qquad  \qquad \qquad \qquad \qquad \qquad
-\frac{j!}{i!}\Big(\frac{\lambda}{4N}\Big)^{i-j}\,\Big[L_{j}^{(i-j)}\Big(-\frac{\lambda}{4N}\Big)\Big]^{2}
\Big\}.
\ea
This expression can be checked by directly evaluating $\langle \mc W^{2}\rangle$ at weak coupling and finite $N$
 using  the Gaussian matrix model, which gives  
\ba
\la{4.3}
\vev{\mc W^{2}} =& N^2 \Big[
1+\tfrac{1}{4}\textstyle  \big(1+\frac{1}{N^2}\big) \lambda +\tfrac{1}{192} 
\big(5+\frac{19}{N^2}\big) \lambda ^2+\tfrac{(24+65 N^2+7 N^4) }{4608 N^4}\lambda 
^3 \notag \\
&\qquad +\tfrac{(554+385 N^2+21 N^4) }{368640 
N^4}\lambda ^4 + \tfrac{(320+1239 N^2+350 N^4+11 N^6) }{7372800  N^6}\lambda ^5 +\OO(\l^6)\Big] \ . 
\ea
While \rf{4.2}   is exact, it is non-trivial  to extract the exact $\l$ dependence 
of its  coefficients  in the $1/N$ expansion  so some indirect approach may be required. 

The first non-planar   contribution to   the $1/N$ expansion of  \rf{4.2}  was  computed  exactly 
 in  $\lambda$
in \cite{Akemann:2001st} (and was checked in \cite{Arutyunov:2001hs} by  the standard weak coupling perturbation theory) 
\ba
\la{4.4}
\vev{\mc W^{2}} &= N^2 \Big\{ \frac{4}{\lambda}\big[I_{1}(\sqrt\lambda)\big]^{2}+\frac{\sqrt\lambda}{2N^{2}}\Big[
I_{0}(\sqrt\lambda)\,I_{1}(\sqrt\lambda)+\frac{1}{6}\,I_{1}(\sqrt\lambda)\,I_{2}(\sqrt\lambda)
\Big]+\mc O\Big({1\ov N^{4}}\Big)\Big\}.
\ea
Expanding  \rf{4.4}  at large $\lambda$  gives (cf. \rf{1.5})
\be
\la{4.5}
\vev{\mc W^{2}} = N^{2}\,e^{2\sqrt\lambda}\, 
\Big(\frac{2}{\pi}\lambda^{-3/2}+\frac{7}{24\pi N^2}+\cdots\Big) = 
W_{1}^{2}\, \Big(1+\frac{7\pi}{6}\,\frac{\gs^{2}}{T}+\cdots\cdots\Big) \ , \qquad 
 W_1= \frac{\sqrt T}{{2\pi}\gs}\,e^{2\pi\,T} \ . 
\ee
In  general,   writing the $1/N$ expansion as 
\be\la{47}
\vev{\mc W^{2}} = \sum_{p=0}^{\infty}\frac{1}{N^{2p-2}}\, \vev{\mc W^{2}}_{p}\ , 
\ee  
the   above previously  known   expressions \rf{4.4}   for  the 
$p=0,1$  terms  may  be written in terms of the $_{1}F_{2}$ hypergeometric function  as 
\ba
\la{4.7}
\vev{\mc W^{2}}_{0} &= {}_1F_2\Big(\tfrac{3}{2};2,3;\lambda \Big), \qquad \qquad 
\vev{\mc W^{2}}_{1} &= \tfrac{1}{4} \lambda  \, {}_1F_2\Big(\tfrac{3}{2};2,3;\lambda \Big)+\tfrac{7}{192} 
\lambda ^2 \, {}_1F_2\Big(\tfrac{5}{2};3,4;\lambda \Big).
\ea
Extending  the weak-coupling  expansion (\ref{4.3}) up to  $\mc O(\lambda^{17})$ order 
one   can   come up   with  
similar  results   for the $p=2,3,4$ terms in \rf{47}
\ba
\vev{\mc W^{2}}_{2} = &
\tfrac{1}{192} \lambda ^3 \, _1F_2\Big(\tfrac{5}{2};4,5;\lambda \Big)+\tfrac{157 
}{184320}\lambda ^4 \, _1F_2\Big(\tfrac{7}{2};5,6;\lambda \Big)+\tfrac{679 
}{22118400}\lambda ^5 \, _1F_2\Big(\tfrac{9}{2};6,7;\lambda \Big)\lp
+\tfrac{37}{141557760} \lambda ^6 \, _1F_2\Big(\tfrac{11}{2};7,8;\lambda \Big),\la{4.9} \\
\vev{\mc W^{2}}_{3} =& \tfrac{1}{23040}\lambda ^5 \, _1F_2\Big(\tfrac{7}{2};6,7;\lambda 
\Big)+\tfrac{7883}{1238630400} \lambda ^6 \, _1F_2\Big(\tfrac{9}{2};7,8;\lambda 
\Big)+\tfrac{7073}{26011238400} \lambda ^7 \, 
_1F_2\Big(\tfrac{11}{2};8,9;\lambda \Big)\lp
+\tfrac{176671}{39953262182400} \lambda 
^8 \, _1F_2\Big(\tfrac{13}{2};9,10;\lambda \Big)+\tfrac{38753 
}{1369826131968000}\lambda ^9 \, _1F_2\Big(\tfrac{15}{2};10,11;\lambda 
\Big)\lp
+\tfrac{11531}{197254963003392000} \lambda ^{10} \, 
_1F_2\Big(\tfrac{17}{2};11,12;\lambda \Big). \la{49}
\ea
These  expressions can be written also  in terms of Bessel functions; for \rf{4.9} one finds (cf. \rf{4.4})
\be
\la{4.10}
\vev{\mc W^{2}}_{2} =\tfrac{37 \lambda ^2}{2304} \big[I_0(\sqrt{\lambda })\big]^2 -\tfrac{\sqrt\lambda(24 +131 \lambda ) 
}{2880}I_0(\sqrt{\lambda }) I_1(\sqrt{\lambda 
}) +\tfrac{(192+332 \lambda +185 \lambda ^2) }{11520}\big[I_1(\sqrt{\lambda 
})\big]^2 .
\ee
This agrees with the result  in \cite{Okuyama:2018aij} found using 
 the topological recursion.  
From the point of view of  computational 
efficiency, our procedure based on the hypergeometric representation of the connected 
part of the $\vev{\mc W^{2}}$ correlator  has an advantage that it can be 
easily coded and extended to  higher  order terms in $1/N$  expansion in    \rf{47}. 
Continuing  to order $p=6$ in \rf{47},  
expanding for  large $\lambda$   and dropping subleading $1/T$ terms  
 we get the   following   generalization   of  
 (\ref{4.5})  
\ba
\la{4.11}
\vev{\mc W^{2}} &\simeq W_{1}^{2}\, \Big(1+\tfrac{7}{6}\,\xi+\tfrac{37}{72}\xi^{2}
+\tfrac{887}{6480}\,\xi^{3}+\tfrac{28379}{1088640}\,\xi^{4}+\tfrac{5045}{1306368}\,\xi^{5}+
\tfrac{1210793}{2586608640}\,\xi^{6}
+\cdots\cdots\Big),  \\
\frac{\vev{\mc W^{2}}}{\vev{\mc W}^{2}} &\simeq 1+\xi+\frac{\xi^{2}}{3}+\frac{\xi^{3}}{15}
+\frac{\xi^{4}}{105}+\frac{\xi^{5}}{945}+\frac{\xi^{6}}{10395}+\cdots\ , \qquad \qquad  \xi \equiv  \frac{\pi\gs^{2}}{T} \ . 
\la{413}
\ea
This suggests a natural  all-order  conjecture  for the resummed 
leading-order strong-coupling terms   (cf. \rf{1.5},\rf{1.9}) 
\be
\la{4.13}
\frac{\vev{\mc W^{2}}}{\vev{\mc W}^{2}}  \simeq 
1 + \sum_{p=1}^{\infty}\frac{\xi^{p}}{(2p-1)!!} = 1+e^{\frac{\xi}{2}}\sqrt\frac{\pi\,\xi}{2}\,\text{erf}\Big(
\sqrt\frac{\xi}{2}\Big).
\ee
We  prove  (\ref{4.13}) using the  Toda integrability structure of the
underlying  Gaussian matrix model in the  next subsection.

Let us note   that  one can   easily   find  also  the correlation function 
of $\W^2 $     with   $J=2$  chiral primary  operator. 
Indeed, the insertion of $\mc{O}_{2}$ is equivalent to $\lambda\partial_{\lambda}$ in presence of 
any power of $\mc W$ in the correlator  (cf.  \rf{2.12},\rf{216}). 
Then  
from (\ref{4.4})
one  finds  
\be\la{415}
\frac{\vev{\mc W^{2}\,\mc{O}_{2}}}{\vev{\mc W^{2}}} = \lambda\partial_{\lambda }\log {\vev{\mc W^{2}}}  = 2\pi\,T\,\Big[
1-\frac{3}{4\pi T}+\cdots+\frac{7}{8}\frac{\gs^{2}}{T^{2}}\,\Big(1+\frac{1}{14\pi T}+\cdots\Big)\Big] \  , 
\ee
which   has a similar   structure to the  one of the previously found correlator  in \rf{2.28}
\be\la{416}
\frac{\vev{\mc W\,\mc {O}_{2}}}{\vev{\mc W}} = \pi\,T\,\Big[
1-\frac{3}{4\pi T}+\cdots+\frac{1}{8}\frac{\gs^{2}}{T^{2}}\,\Big(1-\frac{1}{2\pi T}+\cdots\Big)\Big].
\ee

\subsection{Resummation of the $\gs^{2}/T$ expansion using Toda integrability structure}
\la{s4.2}

In the Gaussian matrix model case, the Toda integrability  structure 
\cite{Gerasimov:1990is,Morozov:1994hh,Morozov:1995pb,Mironov:2005qn} is a useful alternative to the topological recursion. Let us  now show how to  use it  
to prove  the relation (\ref{4.13})
to all orders in $\xi =\pi \gs^{2}/T$. From (\ref{4.1}) it follows that we need to find 
the exponential generating functions (here $x,y$ are  free parameters)
\ba
e_{N}(x) &= \vev{\tr e^{x a}} = \sum_{n=0}^{\infty}\frac{x^{n}}{n!}\,\vev{\tr a^{n}}\,, \quad 
e_{N}(x, y) = \vev{\tr e^{x a}\, \tr e^{y a}}_{\rm conn} = \sum_{n,m=0}^{\infty}\frac{x^{n}\,y^{m}}{n!\, m!}\,\vev{\tr a^{n}\, \tr a^{m}}_{\rm conn}\,\no
\\
\vev{\mc W} &= e_{N}\Big(\sqrt{\tfrac{\lambda}{2N}}\Big),\qquad\qquad \qquad 
\vev{\mc W^{2}}_{\rm conn} =  e_{N}\Big(\sqrt{\tfrac{\lambda}{2N}}, \sqrt{\tfrac{\lambda}{2N}}\Big).\la{417}
\ea
The Toda hierarchy analysis of \cite{Morozov:2009uy} shows that\foot{
Note that our normalization of $a$ is different by $\sqrt 2$ from the one in  \cite{Morozov:2009uy}.}
\ba
\la{4.18}
& e_{N+1}(x)+e_{N-1}(x) = 2e_{N}(x)+\frac{x^{2}}{2N}e_{N}(x),  \\
& e_{N+1}(x,y)+e_{N-1}(x,y) = 2e_{N}(x,y)+\frac{(x+y)^{2}}{2N}e_{N}(x,y)-\frac{x^{2}y^{2}}{4N^{2}}e_{N}(x)e_{N}(y)\ .\la{418} 
\ea
The first recursion is solved by 
\be
e_{N}(x) = e^{\frac{x^{2}}{4}}\,L_{N-1}^{(1)}\big(-\tfrac{1}{2}x^{2}\big)\ ,
\ee
reproducing the expression for  $ \vev{\mc W}$ in \rf{1.4}.


\subsubsection{$1/N$ expansion from Toda recursion and proof of (\ref{4.13}) \la{4.2.1}}

Using \rf{4.18},\rf{418}  one can generate the $1/N$ expansion of $\vev{\mc W^{2}}$.
Let us first show    how this is  done  for $\vev{\mc W}$.
At large $N$ we have 
\be
\la{4.20}
\vev{\mc W} \equiv  w(N, \lambda) = e_{N}\Big(\sqrt{\tfrac{\lambda}{2N}}\Big) =N\, w_{0}(\lambda)+\frac{1}{N}\,w_{1}(\lambda)+\frac{1}{N^{3}}\,w_{2}(\lambda) + \cdots.
\ee
The first recursion  (\ref{4.18}) gives 
\be
\la{4.21}
w\Big(N+1, \lambda\frac{N+1}{N}\Big)+w\Big(N-1,\lambda\frac{N-1}{N}\Big) = 2\,\Big(1+\frac{\lambda}{8\,N^{2}}\Big)\,w(N, \lambda).
\ee
Making the following   ansatz for the  large $\l$ expansion of the $1/N$ coefficients as 
(dots stand for subleading  terms at large $\l$)  
\be
\la{4.22}
w(N,\lambda) =e^{\sqrt\lambda}  \sum_{n=0}^{\infty} C_{n}\,(\lambda^{-3/4}N)^{1-2n}\,+...\ , \qquad 
C_{0} = \frac{1}{48\sqrt{2\pi}} \  
\ee
  and  plugging it  in the recursion (\ref{4.21})   gives 
\be
\la{4.23}
w(N,\lambda) = e^{\sqrt\lambda}\,F(\lambda^{-3/4}\,N)+...\ .  
\ee
Setting  $z=\lambda^{-3/4}\,N$ and taking $N\to \infty$   gives 
\be
\la{4.24}
F'(z)+\Big(1-\frac{1}{48z^{2}}\Big)\,F(z)=0\quad\to\quad F(z) = C_{0}\,z\,e^{\frac{1}{96z^{2}}} \ . 
\ee
Thus  it reproduces the 
resummed expression in  (\ref{1.5}). The derivation  of the  ``D3-brane''   limit in this approach  is 
presented for completeness in Appendix~\ref{app:Dbrane-Toda}.


For the case of  $\vev{\mc W^{2}} $ in \rf{417}  we define similarly 
\ba
\vev{\mc W^{2}}_{\rm conn}  & \equiv \sigma(N,\lambda) = e_{N}\Big(\sqrt{\textstyle  \frac{\lambda}{2N}}, \sqrt{\textstyle \frac{\lambda}{2N}}\Big) = N^{2}\,\sigma_{0}(\lambda)+\sigma_{1}(\lambda)+\frac{1}{N^{2}}\sigma_{2}(\lambda)+\cdots, \la{426}
\\ 
\la{4.26}
\vev{\mc W^{2}} &= \sigma(N,\lambda)+\big[w(N,\lambda)\big]^{2} = e^{2\sqrt\lambda}\,\sum_{n=0}^{\infty} S_{n}\,(\lambda^{-3/4}N)^{2-2n}+ ... \ . 
\ea
Eq.  (\ref{4.11}) gives the   ``initial data''   values 
$S_{0}, ...,S_6 =\big\{ \tfrac{2}{\pi },\ \tfrac{7}{24 \pi },\ \tfrac{37}{2304 \pi },\ \tfrac{887}{1658880 \pi}, \ \tfrac{28379}{2229534720 \pi }\big\}$. 
The recursion relation  in (\ref{418}) reads 
\be
\la{4.28}
\sigma\Big(N+1, \lambda\frac{N+1}{N}\Big) +\sigma\Big(N-1, \lambda\frac{N-1}{N}\Big) = \Big(2+\frac{\lambda}{N^{2}}\Big)\, \sigma(N,\lambda)-\frac{\lambda^{2}}{16N^{4}}\, \big[w(N,\lambda)\big]^{2}.
\ee
 Making, like in    (\ref{4.23}), the  strong-coupling ansatz   (cf.  (\ref{4.24})) 
\be
\la{4.29}
\sigma(N,\lambda) = e^{2\sqrt\lambda}\,G(\lambda^{-3/4}N)+... \ , 
\ee
 and taking  large $N$ limit this gives the differential equation for  $G(z)$, $z= \lambda^{-3/4}N$
\be
\la{4.30}
G'(z)+\frac{1-6z^{2}}{6x^{3}}\,G(z)+\frac{1}{38864\,\pi\,z}\,e^{\frac{1}{48z^{2}}}=0\ .
\ee
Its   general solution is 
\be
\la{4.31}
G(z) = c\,z\,e^{\frac{1}{12z^{2}}}+\frac{1}{18432\,\sqrt\pi}\,z\,e^{\frac{1}{12z^{2}}}\,\text{erf}\Big(\frac{1}{4z}\Big) \ , 
\ee
where  the integration constant $c$  should be set to zero  to match the  leading terms  in \rf{4.11}. 
As a result, we find from \rf{4.24} and \rf{4.31}\footnote{As in  similar relations above, here 
 ``$\simeq$'' stands again for the procedure of first making the $1/N$ expansion
and then  keeping the  leading  large $\lambda$  term at each order in $1/N$. We will understand this notation in the rest of the paper.} 
\be
\vev{\mc W^{2}} \simeq e^{2\sqrt\lambda}(G+F^{2}) = \vev{\mc W}^{2}\,\Big[1+\sqrt\frac{\pi}{2}\,e^{\xi/2}\,\sqrt\xi\,\text{erf}\Big(\sqrt\frac{\xi}{2}\Big)\Big] \ , \ \ \ \ \qquad  \xi= {1\ov 8 z^2} =  \pi {g^2_\str\ov T} \ . \la{432} 
\ee
This proves our conjecture in \rf{4.13}.

\subsubsection{Case of $\vev{\W^3}$}
\la{s4.2.2}

Similar approach can be applied   also for   higher correlators  $\vev{\W^n}$.
For $n=3$  we need the  generating functions with 3  arguments 
\ba
e_{N}(x, y, z) &= \vev{\tr e^{x a}\, \tr e^{y a}\, \tr e^{z a}}_{\rm conn},\qquad 
\vev{\mc W^{3}}_{\rm conn} =e_{N}\Big(\sqrt{\textstyle \frac{\lambda}{2N}}, \sqrt{\textstyle \frac{\lambda}{2N}},  \sqrt{\textstyle \frac{\lambda}{2N}}\Big) = \,t(N,\lambda).\la{433}
\ea
The Toda recursion relation here reads
\ba
& e_{N+1}(x,y,z)+e_{N-1}(x,y,z) = 2e_{N}(x,y,z)+\frac{(x+y+z)^{2}}{2N}e_{N}(x,y,z)
-\frac{(x+y)^{2}\,z^{2}}{4N^{2}}e_{N}(x,y)e_{N}(z)\notag \\
&\qquad 
-\frac{(x+z)^{2}y^{2}}{4N^{2}}e_{N}(x,z)e_{N}(y)
-\frac{(y+z)^{2}x^{2}}{4N^{2}}e_{N}(y,z)e_{N}(x)
+\frac{x^{2}y^{2}z^{2}}{4N^{3}}e_{N}(x)\,e_{N}(y)\,e_{N}(z).
\ea
Writing it in  terms of the functions $t$ in  \rf{433},   $w$ in \rf{4.20} and $\sigma$ in \rf{426}  we get 
\ba
 t\Big(N+1,\lambda\frac{N+1}{N}\Big)+t\Big(N-1,\lambda\frac{N-1}{N}\Big) =  &\Big( 2+\frac{9\lambda}{4N^{2}}\Big)\,t(N,\lambda)
-\frac{3\lambda^{2}}{4N^{4}}\,w(N,\lambda)\,\sigma(N,\lambda)\notag \\
&+\frac{\lambda^{3}}{32N^{6}}\,\big[w(N,\lambda)\big]^{3}\ . 
\ea
Making an ansatz  
\be
\la{4.37}
t(N,\lambda) = e^{3\sqrt\lambda}\,U(\lambda^{-3/4}N)+... 
\ee
and using \rf{4.23},\rf{4.24}  and \rf{4.29},\rf{4.31}   gives\footnote{The  fact that the resulting differential equation is 
1st order  and separable  holds for any $\vev{\mc W^{n}}$ due to the universal finite difference form of
the Toda recursion.}
\be
U'(z)-\frac{16z^{2}-9}{16z^{3}}\,U(z)+\frac{1}{884736\sqrt 2\,\pi z}e^{\frac{3}{32z^{2}}}\text{erf}\Big(\frac{1}{4z}\Big)=0,
\ee
Solving for $U(x)$ and  using that \be
\vev{\mc W^{3}} = e^{3\sqrt\lambda}\Big[t+3(G+F^{2})F-2F^{3}\Big],
\ee
gives  the analog of  \rf{4.13},\rf{432}   ($\xi=\pi\gs^{2}/T$)
\be
\la{4.40}
\frac{\vev{\mc W^{3}}}{\vev{\mc W}^{3}} \simeq 1+3\,\sqrt\frac{\pi}{2}\,e^{\xi/2}\,\sqrt\xi\,\text{erf}\Big(\sqrt\frac{\xi}{2}\Big)
-\frac{4\pi}{3\sqrt 3}\,\xi\,e^{2\xi}\,\Big[-1+12\,\TT\Big(\sqrt{3\xi}, \frac{1}{\sqrt 3}\Big)\Big],
\ee
where $\TT(h,a)$ is the Owen T-function  
\be\la{ttt}
\TT(h,a) = \frac{1}{2\pi}\,\int_{0}^{a}dx\,\frac{e^{-\frac{h^{2}}{2}\,(1+x^{2})}}{1+x^{2}}
= \frac{\text{arctan}(a)}{2\pi}-\frac{1}{2\pi}\sum_{n=0}^{\infty}\frac{(-1)^{n}a^{2n+1}}{2n+1}\Big(1-e^{-\frac{h^{2}}{2}}\sum_{m=0}^{n}\frac{h^{2m}}{2^{m}m!}\Big).
\ee
Explicitly, the first few terms in the expansion of \rf{4.40}   in powers of $\xi$ are  thus (cf. \rf{413}) 
\be
\la{4.43}
\frac{\vev{\mc W^{3}}}{\vev{\mc W}^{3}} \simeq  1+3\, \xi +5\, \xi ^2+\tfrac{73}{15}\,\xi ^3
+\tfrac{113}{35}\, \xi ^4+\tfrac{508}{315}\, \xi ^5+\tfrac{33521}{51975}\, \xi ^6 
+\cdots.
\ee
Similar expansion can be found for all  $n$; as  we show in Appendix \ref{anew}, we have
\ba
{\vev{\mc W^{n}}\ov \vev{\mc W}^{n}} = 1 &+\frac{n\,(n-1)}{2}\,\xi+\frac{n\,(n-1)\, (3n-5)\,(n+2)}{24}\,\xi^{2}\notag \\
&+\frac{n\,(n-1)\, (15n^{4}+30n^{3}-75n^{2}-610 n+1064)}{720}\,\xi^{3}+
\mc O(\xi^{4}).\la{443}
\ea

\subsection{Correlator of loops in fundamental and anti-fundamental representations}
\la{s4.3}

Let us consider  now a correlator  of  one  Wilson loop in  $k$-fundamental   and another in  $k$-anti-fundamental representation of $U(N)$. In the matrix model description  it is given by  (cf. \rf{266}) 
\be
\vev{\mc W^{(k,-k)}}\equiv  \vev{ \W^{(k)} \, \W^{(-k)} }  =  \vev{ \tr U^{k}\,\tr U^{-k} } \ , \qquad \qquad 
U=  e^{\, \frac{g}{\sqrt 2}\,a} \ . \la{41}
\ee
We will  focus on the $k=1$ case as  (like  in the case of $k$-fundamental -- $k$-fundamental correlator
discussed above) 
the   dependence on $k$ can be  recovered   by the rescaling $g\to k g $  or 
 $\lambda\to k^{2}\lambda$. 
 Instead of  $\vev{\W^2}$ in \rf{4.2}   here one finds 
 \cite{Okuyama:2018aij}\foot{The peculiar first term in the r.h.s. of (\ref{42}) is due to would-be term in $\vev{\mc W^{(k, k')}}$ proportional to 
a certain Laguerre  $L_j^{(i)}\big(-(k+k')^{2}\frac{\lambda}{4N}\big)$
contribution 
 that happens to be $\lambda$ independent for $k+k'=0$.}
\ba\la{42}
\langle \mc W^{(1,-1)}\rangle = N +  \Big[e^{\frac{\lambda}{8N}}\,L_{N-1}^{(1)}\Big(-\frac{\lambda}{4N}\Big)\Big]^{2}
-e^{\frac{\lambda}{4N}}\,\sum_{i=0}^{N-1}\sum_{j=0}^{N-1}
(-1)^{i-j}\,\frac{j!}{i!}\Big(\frac{\lambda}{4N}\Big)^{i-j}\,\Big[L_{j}^{(i-j)}\Big(-\frac{\lambda}{4N}\Big)\Big]^{2}.
\ea
Its weak coupling expansion reads (cf. \rf{4.3}) 
\ba
\la{4.46}
\vev{\mc W^{(1,-1)}} = N^2 + (N^2-1)\Big[
\tfrac{1}{4} \lambda +\tfrac{5 }{192 } \lambda 
^2+\tfrac{7  }{4608}\lambda ^3 
  +\tfrac{(7N^2  +2) }{122880 N^2} \lambda ^4
+\tfrac{11 (N^2+1)}{7372800 
N^2}  \lambda ^5   
+\cdots\Big].
\ea
The first two terms in the $1/N$ expansion  are  as in \rf{4.4},\rf{47}:
\ba
\vev{\mc W^{(1,-1)}} &=  \sum^\infty_{p=0} {1\ov N^{2p-2}}\,  \vev{\mc W^{(1,-1)}}_{p} \ , \qquad \qquad 
\vev{\mc W^{(1,-1)}}_{0} = \frac{4}{\lambda}\,\big[I_{1}(\sqrt\lambda)\big]^{2}, \la{777} \\
\vev{\mc W^{(1,-1)}}_{1} &= -\frac{1}{2} \,\lambda\big[  I_0(\sqrt{\lambda })\big]^2+\frac{7}{12} 
\sqrt{\lambda }\, I_0(\sqrt{\lambda }) I_1(\sqrt{\lambda })+\frac{1}{6} 
(-1+3 \lambda ) \big[I_1(\sqrt{\lambda })\big]^2.\la{447}
\ea
Expanding these first two terms  at large $\lambda$ gives the analog of \rf{4.5},\rf{4.11} 
\be
\vev{\mc W^{(1,-1)}} = W_{1}^{2}\,\Big(1+\frac{\xi}{6} +\cdots\Big)\ ,    \qquad \qquad  
 W_1= \frac{\sqrt T}{{2\pi}\gs}\,e^{2\pi\,T} \ , \qquad  \xi = \pi { g^2_\str \ov T} \ . \la{448}
\ee
 To make an efficient ansatz  for higher order terms it is useful  to use  as in \rf{4.7}  the 
 representation in terms of the $_1F_2$  hypergeometric function 
\ba
\vev{\mc W^{(1,-1)}}_{0} &= {}_1F_2\Big(\tfrac{3}{2};2,3;\lambda \Big),
 \qquad\no \\
\vev{\mc W^{(1,-1)}}_{1}
& = -\tfrac{1}{4} \lambda  \, _1F_2\Big(\tfrac{1}{2};2,3;\lambda 
\Big)-\tfrac{1}{192} \lambda ^2 \, _1F_2\Big(\tfrac{3}{2};3,4;\lambda 
\Big)+\tfrac{1}{2304}\lambda ^3 \, _1F_2\Big(\tfrac{5}{2};4,5;\lambda \Big).
\ea
By some trial and error it is then possible to determine the higher genus contributions, e.g.,\footnote{As we mentioned previously, 
this is an efficient procedure equivalent to the rigorous analysis based on the  topological recursion \cite{Eynard:2004mh,Eynard:2008we}.}
\ba
\vev{\mc W^{(1,-1)}}_{2} =& -\tfrac{\lambda ^4}{61440} \, _1F_2\Big(\tfrac{5}{2};5,6;\lambda 
\Big)-\tfrac{\lambda ^5}{7372800} \, _1F_2\Big(\tfrac{7}{2};6,7;\lambda 
\Big)+\tfrac{\lambda ^6}{78643200} \, _1F_2\Big(\tfrac{9}{2};7,8;\lambda 
\Big)\lp
+\tfrac{\lambda ^7}{7927234560} \, _1F_2\Big(\tfrac{11}{2};8,9;\lambda 
\Big), \la{451}\\
\vev{\mc W^{(1,-1)}}_{3} = &-\tfrac{13 \lambda ^6}{1238630400} \, _1F_2\Big(\tfrac{7}{2};7,8;\lambda \Big)-
\tfrac{71 \lambda ^7}{208089907200} \, _1F_2\Big(\tfrac{9}{2};8,9;\lambda 
\Big)\lp
+\tfrac{389 \lambda ^8}{119859786547200} \, 
_1F_2\Big(\tfrac{11}{2};9,10;\lambda \Big)+\tfrac{4499 
\lambda ^9}{28766348771328000} \, _1F_2\Big(\tfrac{13}{2};10,11;\lambda \Big)\lp
+
\tfrac{169 \lambda ^{10}}{140896402145280000} \, _1F_2\Big(\tfrac{15}{2};11,12;\lambda 
\Big)+\tfrac{13 \lambda ^{11}}{5207531023289548800} \, 
_1F_2\Big(\tfrac{17}{2};12,13;\lambda \Big). \la{452}
\ea
Their weak-coupling expansions 
\ba
\vev{\mc W^{(1,-1)}}_{2} &= -\tfrac{\lambda ^4}{61440}-\tfrac{11 \lambda ^5}{7372800}-\tfrac{13 
\lambda ^6}{235929600}-\tfrac{13 \lambda ^7}{15854469120}+\tfrac{187 
\lambda ^8}{22830435532800}+\cdots, \\
\vev{\mc W^{(1,-1)}}_{3} &= -\tfrac{13 \lambda ^6}{1238630400}-\tfrac{83 \lambda 
^7}{83235962880}-\tfrac{289 \lambda ^8}{7491236659200} -\tfrac{12331 \lambda ^9}{17259809262796800}
+\cdots,
\ea
  agree with the large $N$ expansion of (\ref{4.46}) (as we checked up to $\mc O(\lambda^{30})$).
  Converting the hypergeometric functions into Bessel functions  gives 
  (cf. \rf{4.10}) 
\ba
\vev{\mc W^{(1,-1)}}_{2} = \tfrac{1}{11520} \Big[ -55 \lambda ^2 \big[I_0(\sqrt{\lambda })\big]^2&-4 \sqrt{\lambda } 
(24+11 \lambda ) I_0(\sqrt{\lambda }) I_1(\sqrt{\lambda })\no\\ & +(192+332 
\lambda +65 \lambda ^2) \big[I_1(\sqrt{\lambda })\big]^2\Big].
\ea
Expanding at  large $\lambda$, we obtain   higher  order terms in \rf{448} (cf. \rf{4.11}) 
and observe that they exponentiate 
\be
\la{4.56}
\vev{\mc W^{(1,-1)}} \simeq W_{1}^{2}\Big(
1+\tfrac{1 }{6}\xi +\tfrac{1}{72}\xi^2+\tfrac{1}{1296}\xi ^3+\tfrac{1}{31104}\xi ^4+\tfrac{1}{933120}\xi ^5+\tfrac{1}{33592320}\xi ^6+\cdots
\Big) \simeq   W_{1}^{2}\,  e^{\xi\ov6} \ . 
\ee
Comparing  this with the  sum of the  leading strong coupling terms  in $\vev{\W}$  given by $e^{{\xi\ov 12}}$  in \rf{1.5} 
we conclude that  in contrast to the nontrivial result for $\vev{W^2}$ in \rf{4.13} 
here one finds a  simple   factorization relation (valid  again  up to subleading terms in $1/T$)
\be
\la{4.57}
\vev{\mc W^{(1,-1)}} \simeq  \vev{\mc W}^{2}   \ . 
\ee 
Like \rf{4.13} this  can be proved to all orders 
 in $\xi$  using  the Toda recursion relations (cf. section  \ref{4.2.1}). To this aim,
 let us  define as in \rf{417},\rf{4.26} 
\be\la{458}
\vev{\mc W^{(1,-1)}}_{\rm conn}= e_{N}\Big(\sqrt{\textstyle \frac{\lambda}{2N}}, -\sqrt{\textstyle \frac{\lambda}{2N}}\Big) \equiv \overline\sigma(N,\lambda) = N^{2}\,\overline\sigma_{0}(\lambda)+\overline\sigma_{1}(\lambda)+\frac{1}{N^{2}}\overline\sigma_{2}(\lambda)+\cdots\ . 
\ee
The second recursion relation in (\ref{4.18}) reads (cf. \rf{4.28},\rf{4.22})
\ba
& \overline\sigma\Big(N+1, \lambda\frac{N+1}{N}\Big) +\overline\sigma\Big(N-1, \lambda\frac{N-1}{N}\Big) = 2\, \overline\sigma(N,\lambda)-\frac{\lambda^{2}}{16N^{4}}\, \big[w(N,\lambda)\big]^{2}, \notag \\
& w(N,\lambda) = \sqrt\frac{2}{\pi}\,e^{\sqrt\lambda}\,\sum_{n=0}^{\infty} \frac{1}{96^{n}\,n!}\,(\lambda^{-3/4}N)^{1-2n}+\cdots\ . 
\ea
Making an ansatz as in  \rf{4.29}  
\be
\overline\sigma(N,\lambda) = e^{2\sqrt\lambda}\,\overline G(z)+\cdots \ , \qquad \qquad z=\lambda^{-3/4}N \ , 
\ee
we find, expanding in large $N$  
\be
z^{-4/3}N^{-2/3}\overline G(z)+ z^{-8/3} N^{-4/3}\Big[\tfrac{1}{73728\pi} 
z^{2}e^{\frac{1}{48z^{2}}}+\tfrac{1}{12} \big((1-6z^{2})\overline G(z)+6z^{3}\overline G'(z)\big)\Big]+\mc O(N^{-2}) = 0.
\ee
In contrast to  the differential equation in \rf{4.30} here at leading  order 
in large $N$  we  get  simply   the constraint   
\be
\overline G(x)=0\ ,
\ee
implying the vanishing of the connected part \rf{458}  of $\vev{\mc W^{(1,-1)}}$  and  thus proving \rf{4.57}.


\section*{Acknowledgments}

We are grateful to Simone Giombi for a collaboration  at  an early stage of this project   and  many useful remarks  and suggestions. 
We also  thank  Nadav Drukker, Marcos Mari\~{n}o, Albrecht Klemm, Francesco Galvagno, and Marco Billo' for useful   communications   and 
discussions on various aspects of this work.  M.B. acknowledges the support of the INFN grant GSS (Gauge Theories, Strings and Supergravity). 
A.A.T. acknowledges the support of the 
 STFC grants ST/P000762/1  and ST/T000791/1.

\appendix

\section{On   $g_{\rm s}^2/T $ term   in   $\langle \mc W \rangle$ 
from  supergravity approximation }
\la{app:KK}
\def \s  {\sigma}

As discussed  in the Introduction,  the  form $T^{{1\ov 2}  - p}$ of the   string tension dependence  of the leading strong-coupling terms     in the 
$1/N $   expansion \rf{1.3}   of $\langle W \rangle$  has a  string-theory  explanation \ci{Giombi:2020mhz}   based on  the 
dependence of the ratio  of the  string  fluctuation determinants  (evaluated 
on a genus $p$  surface) on the AdS radius. 

At the same time,   since  in  the large $T$  limit the contributions of massive string modes in the virtual exchanges 
may be  expected to be   suppressed, 
one may  hope   \ci{Drukker:2000rr}, by analogy with a related discussion in \ci{Berenstein:1998ij},    
to give  an alternative explanation  
of this   dependence 
based on  including only the  massless  (supergravity) modes in  computing  string loop  
corrections to $\langle W \rangle$. 
If  such a  ``supergravity''  approach   could  be  shown   to work
  this  would  allow one to compute, e.g.,   the leading ``one-handle'' $g_{\rm s}^2/ T $
correction in \rf{1.3},\rf{1.4} 
\be
\la{A.1}
\vev{\mc W} = \frac{\sqrt T}{2\pi\, \gs}\, e^{2\pi T} \,\Big\{1+\frac{\pi}{12}\frac{\gs^{2}}{T}\big[1  +  \mc O(T^{-1})\big]
 +\mc O\Big(\big({\gs^{2}\ov T}\big)^{2}\Big) \Big\}\, ,
\ee
including its 
  coefficient.  
  As we will 
  explain below,  such a computation  does not   appear  to be straightforward  as  the  specific    $1\ov T$  dependence of the   $\gs^2$ term on the string tension
 should be a  consequence of a subtle   supersymmetry-related  cancellations of more dominant (for $T \gg 1$) terms. Also,  specific coefficients will depend  
 on  a particular choice of the   ``string''  UV cutoff ($\Lambda \sim {1\ov \sqrt{\a'}} \sim \sqrt T$, see, e.g.,  \ci{Metsaev:1987ju}). 
 
  One  may represent the  contribution of a thin handle   attached to a disc   by the sum of 
 massless  exchanges, each given   by the 
    two massless vertex operators  $V$ (integrated over the disc)  connected by the corresponding target space 
    ``massless'' propagator.  For example, in the flat target space  case  for  the 
    dilaton exchange  in  the bosonic string theory  in $D$ dimensions  we would have\foot{Note that the factor of string tension $T$ 
    in  $V$   is important for correct normalization of the dimensionless 
    dilaton  vertex   when it is combined   with the  massless string effective  action  as implied, e.g.,   by the 
     thin handle resummation of the  string loop expansion  (see  \ci{Tseytlin:1990vf}
    and a discussion in  \ci{Giombi:2020mhz}).}
    \be \la{a1} 
    V(x)  
   \sim    T  \int d^2 \s \sqrt g \Big [  \tfrac{1}{2}      \del^a  x^m \del_a  x_m 
     +  \tfrac{D-2}{ 4 T}  \tfrac{ 1 }{ 4 \pi}  R^{(2)}   \Big]\ \delta^{(D)} \big(x- x(\s)\big)\ .  \ee
     For large $T$    this   should be   evaluated  near  the relevant  minimal surface (flat disc  for the circular Wilson loop 
      in the   flat space case). 
The  relevant exchange contribution  will be     proportional 
to   \be \la{a2}
 X \sim  T^2 \int d^D x \int d^D x' \,  V(x)\,     G(x-x')\,  V(x')  \ , \qquad 
 \qquad  G(x-x')  \sim {1\ov |x-x'|^{D-2}} \ , \ee
 where $G(x-x')$    is the massless Green's function in $D$ dimensions.   The coefficient of the massless  scalar 
    kinetic term in the tree-level  string effective action is 
$ {1 \ov   \gs^2 (\sqrt{ \a' })^{D-2}} \sim  {1\ov  \gs^2 } T^{D/2 -1} $  so that 
 the  inverse of this  factor is to be included into  $X$. 
As a result, we will get 
\be   \la{a3} 
X\sim  \gs^2  T^{3-D/2} \int d^2 \s \sqrt {g(\s)} \int d^2 \s' \sqrt{ g(\s')} \  {1\ov | x(\s) -x(\s') |^{D-2}} \ , \ee
   where  $x^m(\s)$  represents 
    the minimal surface.  Since the integrals are   dominated   by the  short-distance region $\s \sim \s'$ 
where $x^i\sim \s^i$ ($i=1,2$)  (the  $D-2$   coordinates  $x^r$
 transverse to the disc   vanish on the classical solution)
    we thus find 
\be  \la{a4}  
 X\sim    \gs^2\,   T^{3-D/2}  \Lambda^{D-4} \sim   \gs^2  T  \ , \qquad \qquad  \Lambda \sim  T^{1/2}  \ , \qquad 
  T\gg 1 \ . \ee
  Here $\Lambda \gg 1 $   is  
 a  UV cutoff   that in the string theory context   should    have the interpretation of 
 a modular integral cutoff  set  up by the string  tension. 
  Then, up to subleading terms in $\Lambda$ dropped in \rf{a4}, 
    $X \sim  \gs^2  T$    universally   for any  target space  dimension $D$.

 Since   this  argument   involves   just the short-distance region, the result should  not be  sensitive to the target-space geometry.   Indeed, the same  expression 
    is found    by starting with the $D=10$   theory in  AdS$_5 \times S^5$  
 and compactifying on $S^5$, i.e.  considering  as in  \ci{Berenstein:1998ij}   the  5d dilaton    with dimension 
 $\Delta= 4 + k$
where $k$ is KK momentum.  In this  case $\delta^{(D)} \big(x- x(\s)\big)$ in \rf{a1}   is replaced by 
$[K(x) ]^\Delta  Y^I_k (y)$  where $K$ is the  bulk to boundary   propagator  in  AdS$_5$  and $Y^I_k$  is  $S^5$  spherical harmonic.  $G$ in \rf{a2} is  replaced by the  AdS$_5$  bulk-to-bulk  propagator.  Taking   into account the $k$-dependent 
normalization factors (see   \ci{Berenstein:1998ij}), summing over $k$ and extracting the leading UV divergent part of the resulting  analog of \rf{a3}    we end up   with same result   $X \sim  \gs^2  T$    as in \rf{a4}. This is 
 different from the expected  $ \gs^2 /  T$ ratio in \rf{A.1}.\foot{A  potential  problem  in  a similar  argument 
originally suggested    in \ci{Drukker:2000rr}   appears to be with  the contribution of summation over the  KK modes that 
   gives  
   $ T^{5/2}$ factor rather than $T^{1/2}$ assumed there.
   Indeed,  the kinetic term of the KK dilaton
      has  a  prefactor $B_k \sim [ 2^{k-1}  (k+1) (k+2)]^{-1}$ 
   that   then enters  in inverse power in the propagator. Also, the summation  over  quantum numbers of 
   spherical  harmonics with fixed $J^2= k (k+4)$ gives $\sum_I   Y^I_k Y^I_k \sim  2^{-k} (k+2) (k+3)$ 
   (see    \ci{Berenstein:1998ij} for details), 
   so that at the end  we get  $\sum _k   (k+1) (k+2)^2 (k+3)$ which diverges  as 
   $\sum^\Lambda_k   k^4  \sim \Lambda^5 \sim  T^{5/2}$ as appropriate for a 5-space. 
   We thank S. Giombi  for   a discussion of  this argument.
   }
   As already mentioned, details of  compactification   should  not actually matter as  the highest   divergence 
   depends on  the  power of the   UV singularity of  the $D=10$  massless 
   propagator   and is thus universal.
   In particular,  the same result  should  be found also in the AdS$_4\times CP^3$ case.

It is possible  that    once one  adds   together   similar exchanges  of all $D=10$  supergravity modes, 
the leading UV singularity will be reduced by 4  powers of  the cutoff $\Lambda$   due to supersymmetry  cancellations. In this case  one will  end up with  the following  analog of \rf{a4} (here $D=10$)
\be  \la{a5}  
 X\sim    \gs^2\,   T^{3-D/2} \Big(    0 \times  \Lambda^{D-4}   +   
 ...   +    0 \times  \Lambda^{D-7}   +   \Lambda^{D-8}\Big)\Big|_{ \Lambda \sim  T^{1/2}  }  \sim {\gs^2\ov T} \ . 
 \ee 
 Confirming this remains an open problem. 
 

\section{Remarks on   strong coupling expansion of  $\langle \W \rangle$ in SYM }
\la{app:B}

\subsection{Large $N$ expansion  in terms of Bessel functions}
\la{app:Bessel}

The  computation 
 of the explicit form of the   $\lambda$ dependent    coefficients  in 
  the $1/N$ expansion of the circular Wilson loop    correlator  $\langle \W \rangle$  in  
  the $\N=4$ SYM theory 
 first appeared in Appendix  A of \cite{Drukker:2000rr}  starting with a matrix model ansatz.
  A convenient algorithm to  find 
     these  coefficients     to any order in $1/N$  is   
discussed in \cite{Okuyama:2006ir} and leads to the following  compact 
representation (cf. \rf{266}  and footnote \ref{f0})
\be
\frac{1}{N}\vev{\mc W} = \frac{1}{N}\vev{\tr e^{{g\ov \sqrt 2}a}}  = \frac{2}{\sqrt\lambda}\,\mathop{\text{Res}}_{x=0} \Big[
\, e^{\frac{\lambda}{4N}\,H(\frac{\sqrt\lambda}{4N}\,x)}\sum_{n=0}^{\infty}\frac{I_{n}(\sqrt\lambda)}{x^{n}}\Big]\ ,
\qquad \quad  H(x) \equiv  \frac{1}{2}\Big(\coth x-\frac{1}{x}\Big).
\ee
Expanding $H$ around $x=0$ and taking the residue gives the following explicit expansion in terms of Bessel functions ($I_{n}\equiv I_{n}(\sqrt\lambda)$)
\ba
\la{B.2}
\frac{1}{N}\vev{\mc W} =  \frac{2 I_1}{\sqrt{\lambda }}& +\frac{\lambda  I_2}{48 
N^2}+\frac{1}{N^{4}}\Big(\frac{\lambda ^{5/2} I_3}{9216}-\frac{\lambda ^2 
I_4}{11520}\Big)+\frac{1}{N^{6}}\Big(\frac{\lambda ^4 I_4}{2654208}-\frac{\lambda 
^{7/2} I_5}{1105920}+\frac{\lambda ^3 
I_6}{1935360}\Big)\lp
+\frac{1}{N^{8}}\Big(\frac{\lambda ^{11/2} 
I_5}{1019215872}-\frac{\lambda ^5 I_6}{212336640}+\frac{\lambda 
^{9/2} I_7}{137625600}-\frac{\lambda ^4 
I_8}{309657600}\Big)+\cdots.
\ea
Keeping  only the   leading term  at large $\l$   at each order in $1/N$   we  observe  the exponentiation   \rf{1.5}
originally  found in  \cite{Drukker:2000rr}
\ba
\vev{\mc W}&\simeq \sqrt\frac{2}{\pi}\,{N\ov \lambda^{3/4}}\,e^{\sqrt\lambda}\,\Big(
1+\frac{\lambda^{3/2}}{96N^{2}}+\frac{\lambda^{3}}{18432N^{4}}+\frac{\lambda^{9/2}}{5308416N^{6}}+
\cdots
\Big) = \sqrt\frac{2}{\pi}\,{N\ov \lambda^{3/4}}\,e^{\sqrt\lambda}\ e^{\, \frac{\lambda^{3/2}}{96N^{2}}   } \ .\la{b3} 
\ea

\def \WW  {V}\def \aa  {{\rm a}}  \def \bb {{\rm b}}

\subsection{On the origin of the $N/\lambda^{3/4}$ prefactor  in  $\langle \W \rangle$ }

Let us explain  the origin of the  leading   strong-coupling   prefactor $N/\lambda^{3/4}$  
in \rf{b3}
 without resorting to the exact  Laguerre  representation \rf{1.4} of  $\langle \W \rangle$.
Let us  start with   a generic (one-cut)    matrix model  with potential $\WW$  and coupling $\rm g$
\be
\mc Z = \int \prod^{N-1}_{i=0}\frac{dm_{i}}{2\pi}\,\Delta^{2}(m)\,\exp\big[{-\frac{1}{\rm g}\,\sum_{j=0}^{N-1}\WW(m_{j})}\big]\ .
\ee
Let $t=N\rm g$ be the  analog of 't Hooft coupling. The planar resolvent for the one-cut distributions on $[\aa,\bb]$ is 
\be\la{b5}
\omega_{0}(z) = \frac{1}{N}\vev{\tr \frac{1}{z-M}}_{_{\rm planar}} = \frac{1}{t}\Big[\WW'(z)-\sqrt{(z-\aa)(z-\bb)}\,P(z)\Big],
\ee
where $P(z)$ is  a polynomial chosen so that  to reproduce the correct asymptotics of $\omega_{0}(z)$ at large $z$. For an even potential $V$  this  gives
$\omega_{0}(z) = \frac{1}{z}+\mc O({1\ov z^{3}})$.
The  analog of the Wilson loop expectation value is   given by 
\be\la{b6}
\vev{\mc W}_{_{\rm planar}}  = N\,\oint \frac{dz}{2\pi i }\,  e^{z}\, \omega_{0}(z)\ ,
\ee
where the contour encircles the cut $[-\aa,\aa]=[-\sqrt{2t},\sqrt{2t}]$.

The    case relevant for the SYM    theory is 
 the Gaussian  matrix model with  $\WW(x) = \frac{1}{2}x^{2}$  where  $\omega_{0}(z) = \frac{1}{t}(z-\sqrt{z^{2}-2t})$, \ $P(z)=1$   and 
 thus
\be\la{b7}
\vev{\mc W}_{_{\rm planar}}  = \frac{N}{t}\,\int_{-\sqrt{2t}}^{\sqrt{2t}} \frac{dx}{\pi} e^{x}\,\sqrt{2t-x^{2}} = \frac{2N}{\pi}\int_{-1}^{1}dx\,e^{\sqrt{2t}\,x}\sqrt{1-x^{2}} = N\,\sqrt\frac{2}{t}\,I_{1}(\sqrt{2t})\ .
\ee
In the  standard notation (cf. \rf{2.5},\rf{266})   we have 
 $\frac{1}{\rm g} =  \frac{2N}{\lambda}$, so that $t=N{\rm g}=\lambda/2$ and we 
recover the well known result of \ci{Erickson:2000af}.

Let us work out directly the large $t$ expansion of the intermediate expression in \rf{b7}.
By saddle point analysis
\be
\Big(\sqrt{2t}\,x+\frac{1}{2}\log(1-x^{2})\Big)' = 0 \quad  \to \quad x^{*} = \frac{\sqrt{1+8t}-1}{2\sqrt{2t}}\ .
\ee
Setting $x=x^{*}+\delta x$, expanding  in $\delta x$ and taking the large $t$ limit  gives 
$\exp[\sqrt{2t}\,x+\frac{1}{2}\log(1-x^{2})] = \exp[\sqrt{2t}-\frac{1}{4}\log t+ \cdots]-(2t+\cdots)\,(\delta x)^{2}+\cdots
$
and thus 
\be
\vev{\mc W}_{_{\rm planar}} \  \propto  \ N\,t^{-1/4}t^{-1/2}\,e^{\sqrt{2t}}\ \propto\  N\,t^{-3/4}\,e^{\sqrt{2t}}+\cdots,
\ee
where $t^{-1/4}$ comes from the  matrix model 
``action'' evaluated at $x^{*}$ while an additional $t^{-1/2}$ comes from integration over the quadratic fluctuations.
Similar analysis can be  repeated at  subleading order in $1/N$. The derivation is less transparent, but the same saddle point 
argument  gives the next term in the form $t^{3/4}/N$, 
as expected from the exact solution \rf{1.4}. 


As an aside, it  may be of  interest to generalize the above discussion 
 to the case of the matrix model with 
  a monomial potential
$\WW(x) = \frac{1}{2n}x^{2n}$.  Then the resolvent is given by  \rf{b5} with 
the following  polynomial  $P(z)\equiv P_n(z)$ (e.g., for $n=2,3$) 
\ba
P_2(z) = z^{2}+\frac{\aa^{2}}{2},\ \ \ \aa = \Big(\frac{8t}{3}\Big)^{1/4}\, ;   \qquad \qquad 
 P_3(z) = z^{4}+\frac{\aa^{2}}{2}\,z^{2}+\frac{3\aa^{4}}{8},\ \ \  \aa = \Big(\frac{16t}{5}\Big)^{1/6}.
\ea
Then in the quartic potential ($n=2$) case we  find for \rf{b6}
\be
\vev{\mc W}_{_{\rm planar}} = \frac{N}{t}\int_{-\aa}^{\aa}\frac{dx}{\pi}\,e^{x}\,\sqrt{\aa^{2}-x^{2}}\Big(z^{2}+\frac{\aa^{2}}{2}\Big) = \frac{3\aa^{2}N}{2t}\Big[\aa\,I_{1}(\aa)-2\,I_{2}(\aa)\Big]\ . 
\ee
The large $\aa\sim t^{1/4}$ expansion  gives 
\be
\vev{\mc W}_{_{\rm planar}} = \frac{3}{2\sqrt{2\pi}}{N\, t^{-1}\,\aa^{5/2}}\,e^{\aa}+\cdots\ .
\ee 
The prefactor  of  $e^{\aa}$ 
 thus scales is $\sim t^{-3/8}N$. In the sextic ($n=3$)  potential 
case  one finds a  similar result with  the prefactor $\sim t^{-1/4}N$.
For general $n$, 
it is easy to check that the details of the polynomial $P_n(z)$ are not important and each of its terms
 contributes at the same order  at large $t$;  as a result
 ($c$ is a numerical constant)  
\be
\vev{\mc W}_{_{\rm planar}}\  \propto\ N\,t^{-\frac{3}{4n}}\,\exp\big(c\,t^{\frac{1}{2n}}\big)+\cdots \ . 
\ee

\def \mcF {\varGamma}

\subsection{``D3-brane'' limit from  Toda recursion}
\la{app:Dbrane-Toda}
Let us consider the case of the  Wilson loop in  $k$-fundamental representation 
in the large $N$, large $ \l$   limit with 
\be
\kkappa = \frac{k\,\sqrt\lambda}{4N}=\text{fixed}\ .
\ee
Let us  apply  the Toda recursion (\ref{4.21}) to  derive the corresponding  expression for $\vev{\W}$. 
 From the matrix model point of view $k$ can be set to 1 since it  always appears with $\lambda$
 in the combination $k\sqrt\lambda$. Replacing $\lambda $ with $ (4N\kkappa)^{2}$ in (\ref{4.21}) and writing 
\be\la{b15} 
w\big(N,(4N\kkappa)^{2}\big) = N^{-1/2}\,e^{-\Gamma(N, \kkappa)}, \qquad \Gamma(N,\kkappa) = 
N\,\mcF_{0}(\kkappa) +\mcF_{1}(\kkappa) + \cdots,
\ee
we find
\be
(N+1)^{-1/2}\,e^{-\Gamma(N+1, \frac{N\kkappa}{\sqrt{N(N+1)}})}
+(N-1)^{-1/2}\,e^{-\Gamma(N-1, \frac{N\kkappa}{\sqrt{N(N-1)}})}
= 2\,(1+2\kkappa^{2})\,N^{-1/2}\,e^{-\Gamma(N,\kkappa)}.
\ee
Rearranging this as  
\be
\big(\tfrac{N+1}{N}\big)^{-1/2}\,e^{-\Gamma(N+1, \,\kkappa\,\sqrt\frac{N}{N+1})+\Gamma(N,\kkappa)}
+\big(\tfrac{N-1}{N}\big)^{-1/2}\,e^{-\Gamma(N-1, \,\kkappa\sqrt\frac{N}{N-1})+\Gamma(N,\kkappa)}
= 2\,(1+2\kkappa^{2}),
\ee
and expanding at large $N$ gives the following equation for the leading order  ``action'' $\mcF_{0}(\kkappa)$ in \rf{b15}
\ba
\mcF_{0}'(\kkappa) &= \frac{2}{\kkappa}\,\mcF_{0}(\kkappa)+\frac{1}{\kkappa}\log\big(1+2\kkappa^{2}\pm 2\,\kkappa\sqrt{1+\kkappa^{2}}\big)\ .
\ea
The equation  with the + sign is solved by the  expression  coinciding   with 
 the D3-brane action  evaluated  on the   corresponding semiclassical  solution
  \cite{Drukker:2005kx} (see also \ci{Hartnoll:2006is})
\be
\la{C.6}
\mcF_{0}(\kkappa) = -2\,\big(\kkappa\,\sqrt{1+\kkappa^{2}}+\text{arcsinh}\, \kkappa\big)\ .
\ee 
Including higher orders in  $1/N$ is straightforward.
 For instance, the next correction  in \rf{b15} 
 is obtained from 
\be
\mcF_{1}{\phantom{}'}(\kkappa) = \frac{3+4\kkappa^{2}}{2\,\kkappa(1+\kkappa^{2})}\quad\to\quad
\mcF_{1}(\kkappa) = \frac{1}{2}\log\big(\kkappa^{3}\,\sqrt{1+\kkappa^{2}}\big)\ ,
\ee
in agreement with  \cite{Kawamoto:2008gp}.


\section{String semiclassical limit  $J\sim \sql  \gg 1$ of $\vev{\W\, \OO_J}$ }
\la{sec:semi}

 On the string theory side,  taking   the   semiclassical  limit 
\be
\la{2.43}
\J=\frac{J}{\sqrt\lambda}=\text{fixed} \ , \qquad \qquad  \l \gg 1 \ , 
\ee
one finds that the  leading correction to the correlator  $\vev{\mc W\,\mc O_{J}}$
is described by a classical string solution 
 \cite{Zarembo:2002ph,Zarembo:2016bbk}. 
One may  consider the same limit also directly in the matrix model result for the correlator  \rf{2.24},\rf{225}.
This requires  the expansion of $I_{J}(\sqrt\lambda)$ in the limit (\ref{2.43})  which can be 
found by starting from the Debye expansion of the Bessel $\rJ$  function\footnote{See for instance, \url{https://dlmf.nist.gov/10.19}}
\ba
\rJ_{J}(J/\cosh\alpha) &= \frac{e^{-J(\alpha-\tanh\alpha)}}{\sqrt{2\pi J \tanh\alpha}}\sum_{k=0}^{\infty}\frac{U_{k}(\coth\alpha)}{J^{k}},
\ea
where
\be
U_{k+1}(x) = \frac{1}{2}x^{2}(1-x^{2})U_{k}'(x)+\frac{1}{8}\int_{0}^{x}dy\,(1-5y^{2})\,U_{k}(y), \qquad U_{0}(x)=1,
\ee
and analytically continuing to $\sech\alpha=i\,\J= i {J\ov \sql}$. 
This leads to 
\ba
\la{2.46}
\frac{\vev{\mc W\,\mc O_{J}}}{\vev{\mc W}} = &\big({\pi\ov 2}\big)^{\frac{J}{2}-1}\,\frac{\lambda\,\J}{4\,(1+\J^{2})^{1/4}}\,e^{\sqrt\lambda\,f(\J)}\,\Big[
1+\frac{h_{0}(\J)}{\sqrt\lambda}+\cdots\lp\qquad 
+\frac{1}{N^{2}}\,\frac{\lambda^{2}\J^{2}\big(1+2\J\,(\J+\sqrt{1+\J^{2}})\big)}{96}\,\Big(1+\frac{h_{1}(\J)}{\sqrt\lambda}+\cdots\Big)+\mc O\Big(\frac{1}{N^{4}}\Big)\Big],
\ea
where
\ba
f(\J)  &=  \sqrt{1+\J^2}-1 -  \J \log (\J+\sqrt{1+\J^2}) \ , \qquad h_{0}(\J) = \frac{3-2 \J^2+9 (1+\J^2)^{3/2}}{24 (1+\J^2)^{3/2}},\notag \\
h_{1}(\J) &= \frac{24+48 \J+3 \J^2+96 \J^3-26 \J^4+48 \J^5-3 \sqrt{1+\J^2} (8+16 \
\J+21 \J^2+8 \J^3+13 \J^4)}{24 \J^2 (1+\J^2)^{3/2}}.
\ea
This generalizes the  leading  exponential factor $e^{\sqrt\lambda\,f(\J)}$ found  in \ci{Zarembo:2002ph} to   subleading terms in $1/\sql$ and $1/N$.

\iffa 
The above functions admit the following expansions for $\J\to 0$
\be
f(\J) = -\frac{\J}{2}\log 2-\frac{\J^{2}}{2}+\cdots, \quad
h_{0}(\J) = \frac{1}{2}-\frac{13\,\J^{2}}{48}+\cdots, \quad
h_{1}(\J) = -3+2\,\J+\frac{29\,\J^{2}}{48}+\cdots,
\ee
and for $\J\to \infty$
\be
f(\J) = (1-\frac{3}{2}\log 2-\log\J\,)\J-1+\cdots, \quad
h_{0}(\J) =\frac{3}{8}-\frac{1}{12\J}+\cdots, \quad
h_{1}(\J) = \frac{3}{8}-\frac{25}{12\J}+\cdots,
\ee
\fi

\section{$1/N$ expansion of  
$\vev{\mc W^{n}}$ \la{anew} }

Let us consider  the correlators 
$\vev{\mc W^{n}}$  with $n>2$. Expanded in  large $N$,   the connected part 
$\vev{\mc W^{n}}_{\rm conn}$ starts at order $N^{2-2n}$, \ie one has the relations
\ba
\la{4.63}
& \frac{\vev{\mc W^{2}}}{N^{2}}-\Big(\frac{\vev{\mc W}}{N}\Big)^{2} = \mc O\Big(\frac{1}{N^{2}}\Big),\qquad 
\frac{\vev{\mc W^{3}}}{N^{3}}-3\,\frac{\vev{\mc W^{2}}}{N^{2}}\,\frac{\vev{\mc W}}{N}+2\,\Big(\frac{\vev{\mc W}}{N}\Big)^{3} = \mc O\Big(\frac{1}{N^{4}}\Big),
\notag \\
& \frac{\vev{\mc W^{4}}}{N^{4}} -4\,\frac{\vev{\mc W^{3}}}{N^{3}}\,\frac{\vev{\mc W}}{N} -3\,\Big(\frac{\vev{\mc W^{2}}}{N^{2}}\Big)^{2}
+12\,\frac{\vev{\mc W^{2}}}{N^{2}}\,\Big(\frac{\vev{\mc W}}{N}\Big)^{2}-6\,\Big(\frac{\vev{\mc W}}{N}\Big)^{4}
 = \mc O\Big(\frac{1}{N^{6}}\Big),\ \ \  {etc.}
\ea
These relations can be easily checked  using   
weak coupling expansions derived from the matrix model; like in \rf{4.3} we get 
\ba
& \vev{\mc W^{3}} = \textstyle N^{3}\Big[1+\Big(\frac{3}{8}+\frac{3}{4 N^2}\Big) \lambda +\frac{24+49 N^2+8 N^4
}{128 N^4}\lambda^{2}+\frac{462+248 N^2+19 N^4}{3072 
N^4} \lambda ^3
+\frac{6480+21373 N^2+4750 N^4+202 N^6}{491520 N^6} \lambda ^4\lp \textstyle +
\frac{449370+390763 N^2+44440 N^4+1162 N^6}{58982400 
N^6} \lambda ^5
+\frac{7960680+34567361 N^2+12437558 N^4+821534 N^6+14172 N^8}{19818086400 N^8}\lambda ^6+\cdots\Big], \notag \\
& \vev{\mc W^{4}} =\tet  N^{4}\Big[
1+\Big(\frac{1}{2}+\frac{3}{2 N^2}\Big) \lambda +\frac{(90+91 N^2+11 N^4) 
}{96 N^4}\lambda ^2 +\frac{(576+1854 N^2+605 N^4+37 N^6)}{2304 N^6} \lambda 
^3 \lp \tet 
+\frac{(62640+51999 N^2+7955 N^4+286 N^6) }{184320 N^6}\lambda 
^4  \tet +\frac{(460800+1813410 N^2+620777 N^4+52910 N^6+1223 
N^8)5}{11059200 N^8} \lambda ^5
+\cdots\Big] ,
\ea
that indeed satisfy (\ref{4.63}). 
 From  those relations we see that starting with  the  order $1/N^{6}$ expansion of 
 $\vev{\mc W^{n}}$ for  $n=1,2,3,4$, we can determine 
the order $1/N^{6}$ corrections in a closed 
 form for all higher $n>4$.
 
  For the $n=1$  case the $1/N$ 
 expansion in terms of Bessel functions 
was  given in  
Appendix~\ref{app:Bessel}. For $n=2$ we can use the results obtained  in section \ref{4}. 
The $1/N^{4}$ correction  in the  $n=3$  case 
is easily found by matching  the weak coupling expansion
 and this  also fixes the same-order correction in $n=4$ case. 
 As a result, we find
($I_{n}\equiv I_{n}(\sqrt\lambda)$):\footnote{
One can use recursion relations to bring all Bessel functions to $I_{0}$ and $I_{1}$ at the price of introducing polynomials in $\lambda$. In some cases, 
simpler expressions may be obtained in terms of higher index  Bessel functions. }
{\small 
\ba
\la{4.65}
&\frac{1}{N} \vev{\mc W} =\tet  \frac{2}{\sqrt\lambda}\,I_{1}+\frac{1}{N^{2}}\,\frac{\lambda}{48}\,I_{2}
+\frac{1}{N^{4}}\Big[\frac{\lambda^{5/2}}{9216}\,I_{3}-\frac{\lambda^{2}}{11520}\,I_{4}\Big]
+\frac{1}{N^{6}}\,\Big(
\frac{19 \lambda ^{7/2}}{6635520}\,I_{5}+\frac{48 \lambda ^3+35 
\lambda ^4}{92897280}\,I_{6}\Big)\notag+
\mc O\Big(\frac{1}{N^{8}}\Big), \notag \\
&\frac{1}{N^{2}} \vev{\mc W^{2}} =\tet \Big[\frac{2}{\sqrt\lambda}\,I_{1}\Big]^{2}
+\frac{\sqrt\lambda}{2N^{2}}\Big[
I_{0}\,I_{1}+\frac{1}{6}\,I_{1}\,I_{2}
\Big]
+\frac{1}{N^{4}}\Big[
\frac{37\lambda^{2}}{2304}\,I_{0}^{2}-\frac{\sqrt\lambda(24+131\lambda)}{2880}\,I_{0}\,I_{1}+\frac{192+332\lambda+185\lambda^{2}}{11520}\,I_{1}^{2}
\Big]\lp \tet +\frac{1}{N^{6}}\Big[
-\frac{\lambda ^2 (62+37 \lambda )}{23040}\,I_{0}^{2}+\frac{\sqrt{\lambda } (23040+56160 \lambda +40920 
\lambda ^2+6209 \lambda ^3)}{5806080}\,I_{0}\,I_{1}
-\frac{92160+111168 \lambda +85440 \lambda ^2+24857 
\lambda ^3}{11612160}\,I_{1}^{2}
\Big]+\mc O\Big(\frac{1}{N^{8}}\Big), \notag \\
&\frac{1}{N^{3}} \vev{\mc W^{3}} = \tet \Big[\frac{2}{\sqrt\lambda}\,I_{1}\Big]^{3}+\frac{1}{N^{2}}\,\Big(
\frac{13}{4}\,I_{1}^{2}\,I_{0}-\frac{1}{2\sqrt\lambda}\,I_{1}^{3}
\Big)+\frac{1}{N^{4}}\,\Big[
\frac{193}{384}\lambda^{3/2}I_{0}^{2}I_{1}-\frac{6+79\lambda}{240}\,I_{0}I_{1}^{2}+\frac{192+592\lambda+845\lambda^{2}}{3840\sqrt\lambda}I_{1}^{3}
\Big]\notag \\
&\tet\qquad  +\frac{1}{N^{6}}\Big[
\frac{2557 \lambda ^3 }{110592}\,I_{0}^{3}-\frac{\lambda 
^{3/2} (1776+7865 \lambda )}{92160}\,I_{0}^{2}\,I_{1}\tet \lp \tet \qquad +\frac{92160+474624 \lambda +878688 
\lambda ^2+572537 \lambda ^3}{7741440}\,I_{0}\,I_{1}^{2}-\frac{23040+46944 \lambda +64396 \lambda ^2+52073 
\lambda ^3}{967680 \sqrt{\lambda }}\,I_{1}^{3}
\Big]+\mc O\Big(\frac{1}{N^{8}}\Big),\notag \\
&\frac{1}{N^{4}} \vev{\mc W^{4}} =\tet  \Big[\frac{2}{\sqrt\lambda}\,I_{1}\Big]^{4}+\frac{1}{N^{2}}\,\Big(
\frac{38}{3\,\sqrt\lambda}\,
\,I_{1}^{3}\,I_{0}-\frac{4}{3\,\sqrt\lambda}\,I_{1}^{4}
\Big)+\frac{1}{N^{4}}\Big[
\frac{421}{96} \lambda\,  I_{0}^{2}I_{1}^{2}
-\frac{8+187 \lambda}{120 \sqrt{\lambda }}\,I_{0}\,I_{1}^{3}+\frac{192+852 \lambda +1985 \lambda ^2}{1440 \lambda }\,I_{1}^{4}
\Big]\notag \\
&\qquad \tet  +\frac{1}{N^{6}}\Big[
\frac{10567 \lambda ^{5/2} }{13824}\,I_{0}^{3}\,I_{1}-\frac{\lambda  (1032+9641 \lambda )}{11520}\,I_{0}^{2}\,I_{1}^{2}
+\frac{46080+386496 \lambda 
+1171740 \lambda ^2+1630489 \lambda ^3}{1451520 \sqrt{\lambda 
}}\,I_{0}\,I_{1}^{3}\notag \\
&\qquad \tet  -\frac{92160+264384 \lambda +520728 \lambda ^2+671447 \lambda ^3}{1451520 \lambda }\,I_{1}^{4}
\Big]+\mc O\Big(\frac{1}{N^{8}}\Big).
\ea
}
Applying repeatedly  the  relations like (\ref{4.63}) to 
determine the same expressions for $\frac{1}{N^{n}}\vev{\mc W^{n}}$ with  $n>4$, we obtain the following general result
{\small \ba
&\frac{1}{N^{n}}\vev{\mc W^{n}} =  \Big[\tfrac{2}{\sqrt\lambda}\,I_{1}\Big]^{n}
-\frac{1}{N^{2}}\,\tfrac{n\,2^{n-4}}{3}\,\lambda^{1-\frac{n}{2}}\,I_{1}^{n-1}\,\Big[\tfrac{5-6n}{2}\, \l^{1/2} \,I_{0}+I_{1}\Big]
\lp  +\frac{1}{N^{4}}\,\frac{1}{45}\,n\,2^{n-11}\,\lambda^{1-\frac{n}{2}}\,I_{1}^{n-2}\,\Big[
5(n-1)(36n^{2}-24n-59)\lambda^{2}I_{0}^{2}\notag \\
& -4\sqrt\lambda(24+(30n^{2}+35n-59)\lambda)I_{0}I_{1}+(192+4 (65n-47)\lambda+5(48n^{2}-108n+61)\lambda^{2})I_{1}^{2}
\Big]\notag \\
& +\frac{1}{N^{6}}\,\tfrac{1}{2835}\,n\,2^{n-16}\,\lambda^{1-\frac{n}{2}}\,I_{1}^{n-3}\,\Big[
35\lambda^{7/2}(n-1)(n-2)(216n^{3}+108n^{2}-738n-2033)I_{0}^{3}\notag \\
&\tet  +\lambda^{2}(n-1)(-2016(19+6n)-42(-4066+557n+600n^{2}+180n^{3})\lambda)\,I_{0}^{2}\,I_{1}\notag \\
&+\sqrt\lambda(92160+1152(-113+112n+21n^{2})\lambda+12(16398-25641 n +6671 n^{2}+2730 n^{3})\lambda^{2}\notag \\
& +7(-56278+99213 n -42570 n^{2}-4680 n^{3}+4320 n^{4})\lambda^{3})\,I_{0}\,I_{1}^{2}\notag \\
&+\big(-184320-1152(-73+133n)\lambda-8(3538-13839 n+11375 n^{2}) \lambda^{2}\notag \\
& -14(1901+1185 n-5940 n^{2}+2880 n^{3})\lambda^{3}\big)\,I_{1}^{3}
\Big]+\mc O\Big(\frac{1}{N^{8}}\Big).
\ea
}
Expanding   then in  large $\lambda$, we  find 
\ba
\vev{\mc W^{n}} =  &\vev{\mc W}^{n}\,\Big[\mathsf{R}_{n}(\xi)+\mc O(T^{-1})\Big] \ , \qquad \qquad 
\xi = \frac{\pi\gs^{2}}{T} = \frac{\lambda^{3/2}}{8N^{2}}\ ,  \la{dd5}  \\
\mathsf{R}_{n}(\xi) = 1 &+\tfrac{n\,(n-1)}{2}\,\xi+\tfrac{\,n\,(n-1)(3n-5)\,(n+2)}{24}\,\xi^{2}
+\tfrac{\,n\,(n-1)(15n^{4}+30n^{3}-75n^{2}-610 n+1064)}{720}\,\xi^{3}+
\mc O(\xi^{4}).\no \la{d55}
\ea
Setting,  in particular $n=3$,  we see  agreement with the first terms of (\ref{4.43}).

\section{$1/N$ expansion of $\vev{\W}$ for
  $\ha$-BPS Wilson  loop in ABJM}
\la{ABJM:simple:genus}

The localization computation in  \cite{Klemm:2012ii} proved that the expectation value of the $\frac{1}{2}$-BPS circular Wilson  loop  in ABJM  theory   
to all orders in $1/N$ expansion  at  fixed  level $k$  (i.e. in the M-theory limit) is  
\be
\la{F.6}
\vev{\mc W} = \frac{1}{2}\csc\big(\frac{2\pi}{k}\big)\,\frac{\text{Ai}\big[C^{-1/3}\big(N-\frac{k}{24}-\frac{7}{3k}\big)\big]}{\text{Ai}\big[
C^{-1/3}\big(N-\frac{k}{24}-\frac{1}{3k}\big)\big]},\qquad \ \ \ \ \ \  C = \frac{2}{\pi^{2}\,k}\ .
\ee
Let us set  $k=\frac{N}{\lambda}$  as in  \rf{1.2}
and  consider  the limit of $N\to \infty$.\foot{In the M-theory limit, i.e. expanding (\ref{F.6}) 
in  $1/N$  while  keeping 
$k$ fixed  we get 
\be
\notag
\vev{\W} =  \frac{1}{2}\text{csc}\big(\frac{2\pi}{k}\big)\, e^{\pi\sqrt{N/k}}\Big[1-\frac{\pi(32+k^2)}{24 \,\sqrt{2}\, 
k^{3/2}}\frac{1}{\sqrt N}+\Big(\frac{1}{2 k}+\frac{ \pi ^2(32+k^2)^2}{2304 k^3}\Big)\frac{1}{N}+\cdots\Big].
\ee
Note that a similar large $N$,  fixed  $k$  expansion of  the free energy of the ABJM theory on the 3-sphere  considered in \cite{Bhattacharyya:2012ye} contains an additional $\log N$ term. 
}
Using that the  Airy function may be replaced by its asymptotic expansion
\be\la{dd2}
\text{Ai}(x) \sim \frac{e^{-\frac{2}{3}\,x^{3/2}}}{2\sqrt\pi\,x^{1/4}}\ \sum_{n=0}^{\infty}\frac{(-{3\ov 4})^{n}\,\Gamma(n+\frac{5}{6})\Gamma(n+\frac{1}{6})}
{2\pi\,n!\,x^{3n/2}}\ , 
\ee
we find  for the resulting exponential factor  in \rf{F.6} 
\ba
&\exp\Big\{-\frac{\pi\sqrt N}{144\sqrt{3\,\lambda}}\Big[\Big(24N-\frac{N}{\lambda}-\frac{56\lambda}{N}\Big)^{3/2}-\Big(24N-\frac{N}{\lambda}-\frac{8\lambda}{N}\big)^{3/2}\Big)\Big]
\Big\}\ \stackrel{N\to\infty}{\to} \  e^{\pi\sqrt{2(\lambda-\frac{1}{24})}}.
\ea
Keeping only  the  leading large $\l$  terms (or 
doing  the shift   $\lambda\to \lambda+\frac{1}{24}$  \cite{Drukker:2010nc})  this   gives  the 
$e^{\pi\sqrt{2\lambda}}$  factor in \rf{1.6}.\foot{Let us 
 note  that the expression  (\ref{F.6}) is expected to be valid up to terms which are exponentially suppressed at large $N$
\cite{Klemm:2012ii}. 
Such 
 terms  may not  
 be a priori  negligible in the type IIA  string theory limit with fixed $\lambda = N/k$.
  Nevertheless, if one is interested  only in the leading large $\lambda$  corrections 
 it seems reasonable to neglect  these exponential corrections.}

The pre-exponential part of the ratio of  the Airy functions in \rf{F.6} 
 is  $1+\frac{\lambda^{2}}{2(\lambda-{1\ov 24})N^{2}}+\cdots$. At large $\lambda$ it  is 1 
 up to subleading contribution 
$\sim \lambda/N^{2}$ (instead of  the leading $\lambda^{2}/N^{2}\sim {1/k^2} \sim {g^2_\str/ T}$ coming from 
the expansion of $\text{csc}\big(\frac{2\pi}{k}\big)$ in \rf{F.6}). 
This leads to the  simple   expression \rf{1.7}  for the sum of the   leading large  $\l$  terms in $\vev{\mc W}$ 
(cf. \rf{1.6}) 
\be
\la{F.10}
\langle \mc W\rangle \simeq \frac{1}{2}\,\csc\big(\frac{2\pi\lambda}{N}\big)\,e^{\pi\sqrt{2\lambda}} 
= \Big(\frac{N}{4\pi\,\lambda}+\frac{\pi\lambda}{6N}+\frac{7\pi^{3}\lambda^{3}}{90N^{3}}+
\frac{31\pi^{5}\lambda^{5}}{945N^{5}} +\cdots \Big)\,e^{\pi\sqrt{2\lambda}}.
\ee
\iffa with the following explicit first terms
\be
\langle \mc W\rangle = \Big(\frac{N}{4\pi\,\lambda}+\frac{\pi\lambda}{6N}+\frac{7\pi^{3}\lambda^{3}}{90N^{3}}+\frac{31\pi^{5}\lambda^{5}}{945N^{5}}
+\frac{127\pi^{7}\lambda^{7}}{9450N^{7}}+\cdots\Big)\,e^{\pi\sqrt{2\lambda}}.
\ee \fi 
The log of \rf{F.10}  has the following expansion  ($2\pi\frac{\lambda}{N} = \sqrt\frac{\pi}{2}\frac{\gs}{\sqrt T}$, see \rf{1.2})
\be
\la{w8}
\log\vev{\mc W}  \simeq  2 \pi T 
- \ha \log\big(  { 2 \pi  g^2_\str \ov T} \big)
 + \frac{\pi}{12}\,\frac{\gs^{2}}{T}+\frac{\pi^{2}}{720}\,\frac{\gs^{4}}{T^{2}}+\cdots\ .
\ee
Compared to  the SYM  case  where  the analogous  expansion  representing the
  leading-order  terms at strong coupling   
stops at $\gs^2/T$  and thus gives  a   simple exponentiation in \rf{1.5}, 
this  
 does not happen in the ABJM case. Other differences emerge even at planar level when subleading corrections
 in large $T$ are considered.   While in SYM we have  $-\frac{3}{16\pi T}$ in (\ref{x1}), in the ABJM the 
 analogous term in  (\ref{w8}) is $-\frac{\pi}{48\,T}$,  i.e. the coefficient of the $1/T$  correction  in  the planar 
 part of $\vev{\W}$ in \rf{1.6},\rf{F.6},\rf{F.10}
  has an opposite  power of $\pi$.\footnote{Contrary to what happens in the SYM case, at higher order in $\gs^{2}/T$, the $1/T^{n}$ corrections  to $\log\vev{\mc W}$ are rational combinations of different powers of $\pi$, see  Eq.~(4.127) in \cite{Klemm:2012ii}.}
%

Let us also note that  the structure of 
(\ref{w8}) is essentially similar  to the one that  appears when one replaces the circular $\frac{1}{2}$-BPS loop by the latitude loops  considered in \cite{Bianchi:2018bke}.
Also, one can consider  the  $\mathbb{Z}_{r}$ abelian orbifolds of the ABJM 
theory  with reduced  amount of supersymmetry  where the expectation value of the  $\frac{1}{2}$-BPS loop  was 
computed in  \cite{Ouyang:2015hta}. The exact expression  for $ \vev{\mc W} $ 
   differs due to the dependence on the integer $r$.
Nevertheless, in the large $N$ limit at fixed $\l=N/k$ we 
again obtain  a simple prefactor 
$(2r)^{-1}  \csc[2\pi/(r k)]$ (with the ratio of   the prefactors in the Airy functions being again 1 to the  leading order).

\section{Correlators  of coincident  $\ha$-BPS  Wilson loops in ABJM\la{5}}

Here  we  shall study the strong coupling expansion 
 of the expectation value of coincident circular  Wilson loops in the ABJM theory.
We shall  first 
present perturbative results at weak coupling  for  finite $N$, then consider  the   $1/N$  expansion  with coefficients that are 
exact functions of the coupling $\l$  and  then   consider the 
 strong coupling  limit.

\subsection{Weak coupling expansion} 

Using  the notation of  \cite{Kapustin:2009kz}  the ABJM  matrix model partition function  may be written as 
\ba\la{cc1}
Z &= \int \Big(\prod_{i}e^{-\frac{N}{2\alpha}\mu_{i}^{2}-\frac{N}{2\widehat{\alpha}}\nu_{i}^{2}}
d\mu_{i}d\nu_{i}\Big)\,\Delta(\mu)^{2}\,\Delta(\nu)^{2}\notag \\
&\times  \exp\Big\{\sum_{i<j}\Big[
2\log\Big(\frac{2\sinh\frac{\mu_{i}-\mu_{j}}{2}}{\frac{\mu_{i}-\mu_{j}}{2}}\Big)+2\log\Big(
\frac{2\sinh\frac{\nu_{i}-\nu_{j}}{2}}{\frac{\nu_{i}-\nu_{j}}{2}}
\Big)
\Big]-2\sum_{i,j}\log\Big(2\cosh\frac{\mu_{i}-\nu_{j}}{2}\Big)
\Big\},
\ea
with the two  couplings  
associated with the two factors in the gauge group $U(N)\times U(N)$
being 
\be
\alpha = -\widehat\alpha = 2\,\pi\,i\,\frac{N}{k} = 2\,\pi\,i\,\lambda.
\ee
Setting $M = \text{diag}(\mu_{1}, \dots, \mu_{N})$ and $\widehat M = \text{diag}(\nu_{1}, \dots, \nu_{N})$ and using the $U(N|N)$ matrix block notation 
$U=\text{diag}(M, -N)$, the  matrix model counterpart   of the 
$\frac{1}{2}$-BPS Wilson loop reads \cite{Drukker:2009hy}
\be
\la{w1}
\mc W = \text{Str}\,U = \tr e^{M}+\tr e^{\widehat M} \ . 
\ee 
As in the SYM case (see footnote \ref{f0})  here  we do   not include the 
 $1/N$ prefactors in  the definition of $\mc W$.
 By computing the perturbative in small  $\alpha$ expansion
of $\mc W$   we find 
\ba
\la{D.11}
 \frac{1}{2N}\,\vev{\mc W} =
1 &+\Big(-\frac{1}{12}-\frac{1}{24 N^2}\Big) \alpha^2+\Big(-\frac{7}{960}+\frac{7}{5760 N^4}+\frac{11}{576 N^2}\Big) \alpha^4\no\\
&+\Big(-\frac{173}{80640}-\frac{31}{967680 N^6}-\frac{29}{8640 
N^4}+\frac{97}{23040 N^2}\Big) \alpha ^6+\cdots,
\ea
where the dependence on $N$ at each order in $\alpha$ is exact. Similarly, we get 
\ba
\la{D.17}
\frac{1}{(2N)^{2}}
\vev{\mc W^{2}} = 
1 &+\Big(-\frac{1}{6}+\frac{1}{6 N^2}\Big) \alpha^2
+\Big(-\frac{11}{1440}-\frac{13}{480 N^4}+\frac{5}{144 N^2}\Big) \alpha^4\lp
+\Big(-\frac{31}{10080}+\frac{61}{12096 N^6}-\frac{1}{90 
N^4}+\frac{79}{8640 N^2}\Big) \alpha ^6+\cdots.
\ea
At leading order we have  the expected large $N$ (planar) factorization $\vev{\mc W^{2}} = \vev{\mc W}^{2}+\mc O(N^{0})$ with corrections to it being 
\ba
 \vev{\mc W^{2}}-\vev{\mc W}^{2} = 
\alpha ^2&-\Big(\frac{1}{24}+\frac{1}{8 N^2}\Big) \alpha^4
+
\Big(\frac{19}{1440}-\frac{1}{96 N^2}+\frac{1}{48 N^4}\Big) \alpha ^6+ \cdots  \  .\ea 
In general, we can write the $1/N$ expansion of the connected part of the correlator in the form 
\ba
\la{w2}
& \vev{\mc W^{2}}_{\rm conn} = \vev{\mc W^{2}}-\vev{\mc W}^{2} = \Sigma^{(0)}+\frac{1}{N^{2}}\,\Sigma^{(1)} + \cdots \ , 
\ea
where  at weak coupling  
\ba
\la{D.20}
& \Sigma^{(0)}(\alpha) = 
\alpha ^2-\frac{\alpha^{4}}{24}+\frac{19\,\alpha^{6}}{1440}+\frac{2347\,\alpha^{8}}{322560}+\cdots, \qquad 
\Sigma^{(1)}(\alpha) = 
-\frac{\alpha^{4}}{8}-\frac{\alpha^{6}}{96}-\frac{323\,\alpha^{8}}{23040}+\cdots.
\ea
We  shall now apply the algebraic curve solution of the ABJM  matrix model
 in order to obtain  the closed expressions for these two  functions  valid for all  values of  the coupling  $\alpha=2\pi i \,\lambda$
 and then   consider their  expansion at strong coupling $\l \gg 1$.
 

\subsection{Algebraic curve solution  and strong  coupling expansion  } 

As discussed in  detail in \cite{Marino:2009jd,Drukker:2010nc,Klemm:2012ii}, the ABJM model may be solved after  considering it as a restriction of the  
lens space model $L(2,1)$ with generic left and right gauge group ranks and couplings  (see also \cite{Marino:2002fk,Aganagic:2002wv}). Let us denote by 
$z$ the large $N$ continuum limit of the eigenvalues $\mu_{i}$ and $\nu_{i}$ in \rf{cc1}. At leading order, the eigenvalues condense on two cuts 
$\mc C_{1} = (a^{-1}, a)$, $\mc C_{2} = (-b^{-1}, -b)$ in the $Z=e^{z}$ plane. Using mirror symmetry,
the position of the branch points may be expressed in terms of the coupling $\l$  using  the following implicit parametrization \cite{Marino:2009jd}
\ba
\la{w3}
a(\kappa) &= \tfrac{1}{2}\Big[2+i\kappa+\sqrt{\kappa\,(4i-\kappa)}\Big], \qquad 
b(\kappa) = \tfrac{1}{2}\Big[2-i\kappa+\sqrt{-\kappa\,(4i+\kappa)}\Big], \\
\lambda(\kappa) &= \frac{\kappa}{8\pi}\ _{3}F_{2}\Big(\tfrac{1}{2}, \tfrac{1}{2}, \tfrac{1}{2}; 1,\tfrac{3}{2}; -\tfrac{1}{16}\,\kappa^{2}\Big)\ . \la{ww1}
\ea
 Integrating the eigenvalue densities 
along the two cuts, the explicit expression of the $\frac{1}{2}$-BPS loop expectation value is reduced to a residue at infinity
(cf. \rf{b6})
\be\la{ww4}
\frac{1}{2N}\vev{\mc W} = \oint_{\infty}\frac{dZ}{4\pi i}\, \omega(Z) \ , \qquad \qquad \omega(Z) = \gcs\, \vev{\tr\frac{Z+U}{Z-U}},\qquad \gcs \equiv  \frac{2\pi i}{k} \ , 
\ee
where $\omega(Z)$ is  the resolvent of the ABJM matrix model. 
At the planar level, one can use the explicit expression of $\omega(Z)$ in  \cite{Halmagyi:2003ze} 
to obtain the  large $N$  part of  \rf{D.11} ($\kappa=\kappa(\alpha)$, $\alpha=2\pi i \,\lambda$)
\be
\frac{1}{2N}\vev{\mc W}_{\rm planar } = \frac{i\kappa}{4N\,\gcs} = 1-\frac{\alpha^{2}}{12}-\frac{7\alpha^{4}}{960}-\frac{173\alpha^{6}}{80640}+\cdots\ , 
\ee
At strong coupling,  we find  from \rf{ww1}  that (cf. \rf{1.7}) 
 \ba  \lambda(\kappa) &= \frac{1}{2\pi^{2}}\log^{2}\kappa+\frac{1}{24}+\cdots \ , \ \ \ \ \qquad   \l\gg 1 \ , \la{ww2} \\
\vev{\mc W}_{\rm planar} &=   \frac{i}{2\gcs}e^{\sqrt{2\pi^{2}\lambda}} +\cdots = 
{N\ov 4\pi\lambda}\,e^{\pi\,\sqrt{2\lambda}}
+\cdots\  . 
 \la{f14}
\ea
The same approach can be applied to the calculation of  \rf{w2} 
\be
\frac{1}{(2N)^{2}}\big [\vev{\mc W^{2}} - \vev{\mc W}^{2}\big] =   \oint_{\infty} \frac{dZ}{4\pi i}\oint_{\infty} \frac{dZ'}{4\pi i}\omega_{2}(Z,Z')  \ ,
\qquad 
 \ \ 
\omega_{2} =\frac{1}{N^{2}}\,\omega_{2,0}+\frac{1}{N^{4}}\,\omega_{2,1}+\cdots\ , 
 \la{ww6} 
\ee
where each integration is around both cuts and $\omega_{2}$ 
is the 
connected part of the  two-point resolvent. The large $N$ leading order term $\omega_{2,0}$ 
was  computed in a generic two-cut hermitian matrix model  \cite{Akemann:1996zr} and provides the first term $\Sigma^{(0)}$ in (\ref{w2}).
Evaluating  the double residue at infinity  gives 
\be
\la{D.54}
\Sigma^{(0)} = \frac{1}{16}\Big[ \frac{(a+b)^{2}(1+ab)^{2}}{a^{2}b^{2}}
 -\frac{4(1+ab)^{2}}{ab}\frac{\mathbb E}{\mathbb K}\Big]\ ,
\ee
where $\mathbb E$ and $\mathbb K$ are complete elliptic integrals with  the squared elliptic modulus
\be
\la{D.34}
\mathsf{m} = 1-\Big(\frac{a+b}{1+ab}\Big)^{2}.
\ee
Using (\ref{w3})  we observe that  at weak coupling the expression \rf{D.54}  matches 
 perfectly the expansion in (\ref{D.20}). 
Expanded  at strong coupling,  (\ref{D.54})   gives 
 \be
 \la{D.56}
 \Sigma^{(0)} = 
 -\frac{\kappa ^2}{8 \log \kappa }+\frac{1-2 \log \kappa +2 \log 
^2\kappa}{2 \log ^2\kappa}- \frac{8-3 \log \kappa-12 \log 
^2\kappa+8 \log ^3\kappa}{4 \kappa 
^2 \log ^3\kappa }+\cdots\ . 
\ee
 Fig.~\ref{fig:conn}  gives the plot of  $\Sigma^{(0)} $  in (\ref{D.54}) as a   function  of $\lambda$
  and  the comparison  with the weak and strong coupling expansions.
\begin{figure}
\begin{center}
\includegraphics[width=0.7\textwidth]{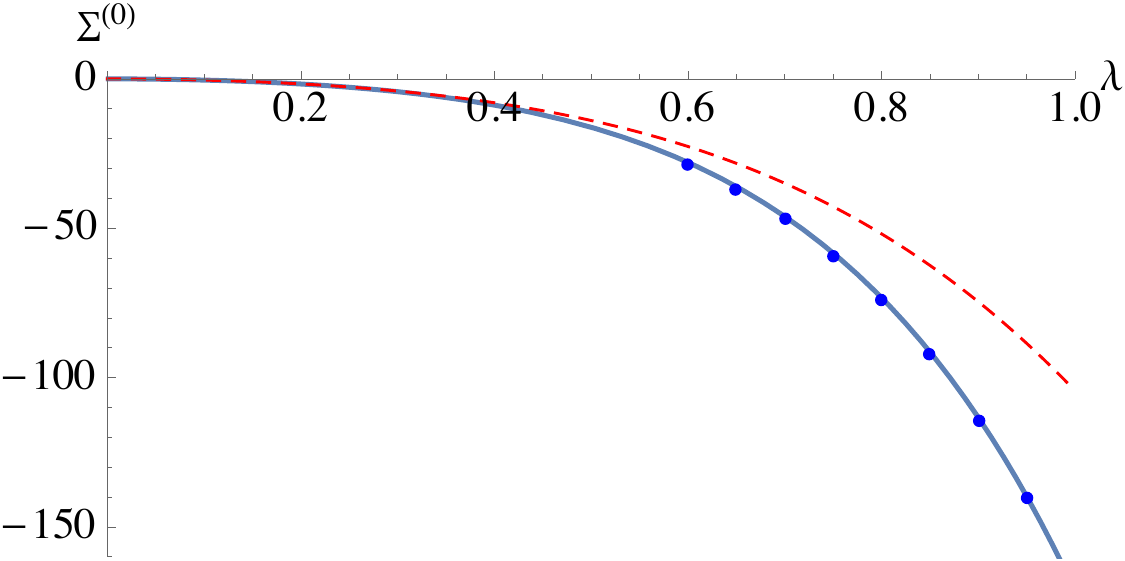}
\end{center}
\caption{
\la{fig:conn}\small 
Plot of the function $\Sigma^{(0)}$ in (\ref{D.54}), \ie the planar contribution to $\vev{\mc W^{2}}_{\rm conn}$ (blue line). 
The red thin dashed line  represents the contribution  of  the first two terms of  the
weak coupling expansion (\ref{D.20}). The blue dots  represent  the  leading  term  in the 
strong coupling expansion $-\frac{1}{8\pi\,\sqrt{2\lambda'}}\,e^{2\pi\,\sqrt{2\lambda'}}$, 
where $\lambda' = \lambda-\frac{1}{24}$. The agreement is very good already at the moderate values  of the coupling $\lambda\sim 1$.
}
\end{figure}
Expressing  $\kappa$ in terms of $\lambda$ using \rf{ww2} we obtain  for $\l \gg 1$
\ba
\vev{\mc W^{2}} &= N^{2}\,(4\pi\lambda)^{-2}e^{2\pi\,\sqrt{2\lambda}}\,\Big(1+2\times \frac{2\pi^{2}\lambda^{2}}{3N^{2}}+\cdots\Big)+(2N)^{2}\frac{1}{N^{2}}\Big(
-\frac{1}{32\pi\,\sqrt{2\lambda}}\,e^{2\pi\,\sqrt{2\lambda}}\Big)+\cdots\notag \\
&= W_{1}^{2}\Big[1+\frac{4\pi^{2}\lambda^{2}}{3N^{2}}-\frac{\sqrt 2 \pi \lambda^{3/2}}{N^{2}}+\mc O(N^{-4})\Big] \ , \la{ww5}
\ea
where $W_{1}$ is the leading-order  planar strong  coupling  part   in $\vev{\mc W}$  in \rf{1.6},\rf{F.10},\rf{f14}. 
In terms of the dual string theory  parameters in \rf{1.2})   it reads 
\be\la{www}
W_{1} = \frac{N}{4\pi\lambda}\,e^{\pi\,\sqrt{2\lambda}} 
=\frac{1}{\sqrt{2\pi}}  \frac{\sqrt T}{\gs}\, e^{2\pi T} \ . 
\ee
The  leading correction $\frac{4\pi^{2}\lambda^{2}}{3N^{2}} = \frac{\pi}{6}\frac{\gs^{2}}{T}$
in  \rf{ww5}  is just twice the correction  in $\vev{\W}$  in \rf{F.10},\rf{1.6}   corresponding  to the  factorized contribution
 $\vev{\mc W}^{2}$  while 
 the  connected contribution is  thus subleading ($\lambda^{3/2}$  vs.  $\lambda^{2}$)  at large $\l$. 
  We conclude that  to leading order  at strong coupling  $\vev{\mc W^{2}} $ factorizes (cf. \rf{1.19}) 
  \be
\la{w4}
\vev{\mc W^{2}} = W_{1}^{2}\,\Big(1+\frac{\pi}{6}\,\frac{\gs^{2}}{T}+\cdots\Big)= \vev{\mc W}^2  +... \ , 
\ee
   i.e. the  connected  contribution \rf{w2}   is   subleading  at large $\l\sim T^2 $ at  order  $1/N^2 \sim  g^2_\str$. 
    
It is tempting to  conjecture  that this factorization   continues to be  true  also at higher orders in $1/N$  
(as that happened in the SYM  case for the  fundamental -- anti-fundamental Wilson 
loop correlator  (\ref{4.56})).
A   test of  this conjecture   requires a  much more involved 
calculation of $\Sigma^{(1)}$  term in  \rf{w2}  presented in the next subsection.
Since $W_1^2 \sim N^2$
this requires  computing the $1/N^4\sim g^4_\str $ term   in  the brackets in  \rf{ww5}. 

\subsubsection{
 $\vev{\mc W^{2}}$}

The next to leading order correction to    the two-point resolvent  in \rf{ww6}   and thus to  $\vev{\mc W^{2}}$
can be computed by working out the 
$1/N$ expansion of the loop equations of the ABJM matrix model (see, e.g.,  \cite{Klemm:2012ii}).
The exact result  for  $ \Sigma^{(1)}  $  in \rf{D.54} valid for all  values  of the coupling $\a$ is 
quite involved    
{\small 
\ba
\la{D.61}
& \Sigma^{(1)} = 
8\,\alpha^{2}\,\Big[
-\frac{(a+b)^2 (1+a b)^2 (1+4 a b+b^2+a^2 (1+b^2))}{192 a (-1+a^2)^2 
b (-1+b^2)^2}\lp
+\frac{\mb E}{\mb K}\,\frac{1}{192 a^2 (-1+a^2)^4 b^2 (-1+b^2)^4}\Big(
(1+a b)^2 \big[b^4+b^6+a^{10} (b^4+b^6)+3 a^2 b^2 (6-7 b^2-7 b^4+6 b^6)\lp
+3 
a^8 b^2 (6-7 b^2-7 b^4+6 b^6)+a^4 (1+b^2) (1-22 b^2+44 b^4-22 
b^6+b^8)\lp
+a^6 (1+b^2) (1-22 b^2+44 b^4-22 b^6+b^8)+2 a (b+b^3-2 
b^5+b^7+b^9)+2 a^9 (b+b^3-2 b^5+b^7+b^9)\lp
+2 a^3 (b+12 b^3-34 b^5+12 
b^7+b^9)+2 a^7 (b+12 b^3-34 b^5+12 b^7+b^9)\lp
-4 a^5 (b+17 b^3-42 b^5+17 
b^7+b^9)\big]
\Big)+\Big(\frac{\mb E}{\mb K}\Big)^{2}\,\frac{1}{192 a^2 (-1+a^2)^4 b^2 (-1+b^2)^4}\Big(
(1+a b)^4 (a^2+a^6\lp
-9 a (-1+a^2)^2 (1+a^2) b+(1-32 a^2+54 a^4-32 
a^6+a^8) b^2+9 a (-1+a^2)^2 (1+a^2) b^3\lp
+6 a^2 (9-16 a^2+9 a^4) b^4+9 
a (-1+a^2)^2 (1+a^2) b^5+(1-32 a^2+54 a^4-32 a^6+a^8) b^6\lp
-9 a 
(-1+a^2)^2 (1+a^2) b^7+a^2 (1+a^4) b^8)
\Big) +\Big(\frac{\mb E}{\mb K}\Big)^{3}\,\frac{1}{192 a^2 (-1+a^2)^4 b^2 (-1+b^2)^4}\Big(
-(1+a b)^6 \big[5 (b^2+b^4) \lp+5 a^6 (b^2+b^4)+4 a^3 b (3+4 b^2+3 b^4)-4 a b 
(4-3 b^2+4 b^4)-4 a^5 b (4-3 b^2+4 b^4)\lp
+5 a^2 (1-2 b^2-2 b^4+b^6)+5 
a^4 (1-2 b^2-2 b^4+b^6)\big]\Big)
+\Big(\frac{\mb E}{\mb K}\Big)^{4}\,\frac{(a-b)^2 (-1+a b)^2 (1+a b)^8}{32 a^2 (-1+a^2)^4 b^2 (-1+b^2)^4}
\Big].
\ea
}
Expanded  at weak coupling  \rf{D.61}   is in agreement  with \rf{D.20}. 
At strong coupling (i.e. large $\kappa$ in \rf{ww1},\rf{ww2}), we obtain for  the leading term
\ba
\Sigma^{(1)} = -\frac{\kappa^{2}\,\log^{3}\kappa}{24\pi^{2}}+\cdots.
\ea
This gives an additional $1/N^4$ correction to the brackets  in \rf{ww5}, i.e.   
\be
\vev{\mc W^{2}} = W_1^2 \Big[1 + \OO\big( {1\ov N^2}\big)   -\frac{4\sqrt 2 \pi^{3}}{3}\frac{\lambda^{7/2}}{N^{4}}  + ... \Big] \ ,
\ee
which is  indeed subleading compared to the similar term $(\gs^{2}/T)^{2} \sim {\l^4/ N^4}$  in   the square of 
 $\vev{\mc W}$ in  
 \rf{F.10}, i.e. 
\be  
\vev{\mc W}^2 = W_1^2 \Big[1 + {\pi \ov 6} {  \gs^{2}\ov T}     +  {\pi^2 \ov 60}  \Big( {  \gs^{2}\ov T} \Big)^2 + ... \Big] \  . 
\ee
We conclude that 
there is no $(\gs^{2}/T)^{2}$ correction to (\ref{w4}),
\ie we have $\vev{\mc W^{2}} = \vev{\mc W}^{2}$  to this order.

\subsubsection{
 $\vev{\mc W^{3}}$}

It is interesting to consider also  the first correction to the  correlator $\vev{\mc W^{3}}$ of the three  coincident Wilson loops
at the leading order at strong coupling. 
 $\vev{\mc W^{3}}$ may be   again decomposed into  factorized and  connected contributions.
  From the usual scaling arguments,  the $\gs^{2}/T$ correction  may come 
only from $\vev{\mc W}^{3}$ and $\vev{\mc W}  \vev{\mc W^{2}}_{\rm conn}$   while  corrections to 
 $\vev{\mc W^{3}}_{\rm conn}$ start at  order $(\gs^{2}/T)^{2}$.
From the above   result  \rf{ww5} for  $\vev{\mc W^{2}}$ (implying  that  $\vev{\mc W^{2}}_{\rm conn}$ is subleading) 
 it follows that
 the $\gs^{2}/T$ term in  $\vev{\mc W^{3}}$    is   precisely three times that in   $\vev{\mc W}$,  i.e.   comes  only from 
 $\vev{\mc W}^{3}$. 
 At the next  $(\gs^{2}/T)^{2}$  order we may have contribution only from $\vev{\mc W^{3}}_{\rm conn}$, 
 since  
 according to the  result of the previous subsection 
 there is no such leading  term in  $\vev{\mc W^{2}}_{\rm conn}$.

The weak coupling expansion of $\vev{\mc W^{3}}_{\rm conn}$ 
computed from the matrix model  turns out to be 
\ba
\vev{\mc W^{3}}_{\rm conn} = \frac{6}{N}\,\alpha^{2} &+\Big(\frac{1}{2 
N}+\frac{9}{4 N^3}\Big)\, \alpha ^4+\Big(-\frac{27}{160 N}-\frac{17}{96 N^3}+\frac{81}{320 N^5}\Big)\,
\alpha ^6\lp
+\Big(-\frac{467}{13440 N}+\frac{247}{3840 N^3}-\frac{1}{960 
N^5}+\frac{243}{17920 N^7}\Big)\, \alpha ^8+\cdots
\ea
Hence,  defining  the  coefficients in the $1/N$  expansion as  (\cf (\ref{w2}))
\be
\vev{\mc W^{3}}_{\rm conn} = \frac{1}{N}\,C^{(0)}+\frac{1}{N^{3}}\,C^{(1)}+\cdots,
\ee
we have 
\be
\la{D.66}
C^{(0)} =  6\,\alpha^{2}+\frac{\alpha^{4}}{2}
-\frac{27\,\alpha^{6}}{160}-\frac{467\,\alpha^{8}}{13440}+\cdots.
\ee
From the loop equations of the ABJM theory \cite{Klemm:2012ii} we can determine  the exact expression for the function $C^{(0)}$ 
by computing the planar three-point resolvent (cf.  \rf{ww6}) 
{\small 
\ba
\la{D.67}
& C^{(0)}= 4\,\alpha\,\Big[
\frac{(a+b)^2 (1+a b)^2}{64 a^3 (-1+a^2)^2 b^3 (-1+b^2)^2}
\Big (b-2 b^3+b^5+a^6 b (-1+b^2)^2-a (1+b^6)\no \\
&\qquad \qquad \qquad \qquad  +2 a^3 
(1+b^6)-a^5 (1+b^6)\Big)+\frac{(a-b) 
(-1+a b) (1+a b)^6}{32 a^2 (-1+a^2)^2 b^2 (-1+b^2)^2}\,\Big(\frac{\mb E}{\mb K}\Big)^{3}\\
&-\frac{3 (a-b) (-1+a b) (1+a b)^4 \big[1+4 a b+b^2+a^2 (1+b^2)\big]
}{64 a^2 (-1+a^2)^2 b^2 (-1+b^2)^2}\,\Big(\frac{\mb E}{\mb K}\Big)^{2}
+\frac{3 
(a-b) (a+b)^2 (-1+a b) (1+a b)^4}{32 a^2 (-1+a^2)^2 b^2 
(-1+b^2)^2}\,\frac{\mb E}{\mb K}
\Big].\no 
\ea
}
This reproduces the weak coupling expansion  (\ref{D.66}). 
At  strong coupling one finds 
\be
\la{D.69}
C^{(0)} =  \frac{(3\log\kappa-1)(12\log^{2}\kappa+\pi^{2})}{384\,\pi\,\log^{3}\kappa}\,\kappa^{3}+\cdots\  \to \  \frac{3\kappa^{3}}{32\pi}+\cdots.
\ee
As a result,  
\ba
\vev{\mc W^{3}}_{\rm conn} &= \frac{1}{N}
\frac{3}{32\pi}\,e^{3\pi\,\sqrt{2\lambda}}+\cdots
= W_{1}^{3}\,\frac{6\pi^{2}\lambda^{3}}{N^{4}}+\cdots.
\ea
This contribution   is subleading  compared to the  one  $\sim  W_{1}^{3}  {g^4_\str\ov T^3} \sim W_{1}^{3} {\l^4\ov N^4}  $ 
  from factorized parts of the  correlator $\vev{\mc W^{3}}$. 
We  conclude that 
 $\vev{\mc W^{3}}=\vev{\mc W}^{3}$ at order $(\frac{g_\str^{2}}{T})^{2}$, 
  i.e.    confirming     \rf{1.19}.


\iffa 
\section{Differential operators $\mathscr{D}_{J}$ for odd $J$}
\la{app:odd}

As we discussed in section 2, it is possible to write  $\vev{\mc W\,\mathsf{O}_{J}} = \mathscr{D}_{J}\vev{\mc W}$, where $\mathscr D_{J}$
is a differential operator polynomial in $\partial\equiv \partial_{\lambda}$ with coefficients that depend on $\lambda$ and $N$.
We already gave in the main text the explicit expression of $\mathscr D_{J}$ for even $J\le 8$. In this short Appendix, we collect the 
expressions for odd $J\le 9$ (the case $J=1$ is just a multiplication, see footnote \ref{fo14})

\ba
& \mathscr{D}_{1} = \frac{1}{2}\sqrt{\textstyle \frac{\lambda}{2N}}, \notag \\
& \mathscr{D}_{3} = -\frac{1}{4}\sqrt\frac{\lambda  N}{2}\,(1-8\partial-16\partial^{2}), \notag \\
& \mathscr{D}_{5} = \frac{1}{8}\sqrt\frac{\lambda }{2N}\Big[ N^{2}-1-6 (\lambda 
+4 N^2) \partial-48 (-4+\lambda ) N^2 \partial^{2}+768 \lambda  
N^2 \partial^{3}+256 \lambda ^2 N^2 \partial^{4}\Big], \notag \\
& \mathscr{D}_{7} = \frac{1}{32N}\sqrt{\textstyle \frac{\lambda}{2N}}\Big[
\lambda -2 N^4 +16 N^2 (-5+3 \lambda +6 N^2) \partial+32 N^2 (-35 \lambda -60 N^2+6 \lambda  N^2) \partial^{2}\notag \\
& -640 N^2 (
\lambda ^2-24 N^2+12 \lambda  N^2) \partial^{3}-2560 
(-36+\lambda ) \lambda  N^4 \partial^{4}+61440 \lambda ^2 N^4 \partial^{5}+8192 \lambda ^3 N^4 \partial^{6}
\Big], \notag \\
& \mathscr{D}_{9} = -\frac{1}{64}\sqrt{\textstyle \frac{\lambda}{2N}}\Big[
42+5 \lambda -20 N^2-2 N^4+20 (-3 \lambda +20 N^2+6 
\lambda  N^2+8 N^4) \partial\lp
-80 (3 \lambda ^2+62 \lambda  N^2+72 
N^4-4 \lambda  N^4) \partial^{2}-3840 N^2 (-21 \lambda +\lambda 
^2-28 N^2+6 \lambda  N^2) \partial^{3} \lp
-7680 N^2 (-14 \lambda 
^2+112 N^2-84 \lambda  N^2+\lambda ^2 N^2) \partial^{4}+21504 
\lambda  N^2 (\lambda ^2-320 N^2+20 \lambda  N^2) \partial^{5}\lp
+57344 (-120+\lambda ) \lambda ^2 N^4 \partial^{6}-1835008 
\lambda ^3 N^4 \partial^{7}-131072 \lambda ^4 N^4 
\partial^{8}
\Big].
\ea
\fi 

\small

\bibliography{BT-Biblio}

\providecommand{\href}[2]{#2}\begingroup\raggedright\begin{thebibliography}{10}

\bibitem{Erickson:2000af}
J.~K. Erickson, G.~W. Semenoff and K.~Zarembo, \emph{{Wilson loops in N=4
  supersymmetric Yang-Mills theory}},
  \href{http://dx.doi.org/10.1016/S0550-3213(00)00300-X}{\emph{Nucl. Phys.}
  {\bf B582} (2000) 155--175}, [\href{http://arxiv.org/abs/hep-th/0003055}{{\tt
  hep-th/0003055}}].

\bibitem{Drukker:2000rr}
N.~Drukker and D.~J. Gross, \emph{{An Exact prediction of N=4 SUSYM theory for
  string theory}}, \href{http://dx.doi.org/10.1063/1.1372177}{\emph{J. Math.
  Phys.} {\bf 42} (2001) 2896--2914},
  [\href{http://arxiv.org/abs/hep-th/0010274}{{\tt hep-th/0010274}}].

\bibitem{Pestun:2007rz}
V.~Pestun, \emph{{Localization of gauge theory on a four-sphere and
  supersymmetric Wilson loops}},
  \href{http://dx.doi.org/10.1007/s00220-012-1485-0}{\emph{Commun. Math. Phys.}
  {\bf 313} (2012) 71--129}, [\href{http://arxiv.org/abs/0712.2824}{{\tt
  0712.2824}}].

\bibitem{Zarembo:2016bbk}
K.~Zarembo, \emph{{Localization and AdS/CFT Correspondence}},
  \href{http://dx.doi.org/10.1088/1751-8121/aa585b}{\emph{J. Phys.} {\bf A50}
  (2017) 443011}, [\href{http://arxiv.org/abs/1608.02963}{{\tt 1608.02963}}].

\bibitem{Kapustin:2009kz}
A.~Kapustin, B.~Willett and I.~Yaakov, \emph{{Exact Results for Wilson Loops in
  Superconformal Chern-Simons Theories with Matter}},
  \href{http://dx.doi.org/10.1007/JHEP03(2010)089}{\emph{JHEP} {\bf 03} (2010)
  089}, [\href{http://arxiv.org/abs/0909.4559}{{\tt 0909.4559}}].

\bibitem{Marino:2009jd}
M.~Mari\~no and P.~Putrov, \emph{{Exact Results in ABJM Theory from Topological
  Strings}}, \href{http://dx.doi.org/10.1007/JHEP06(2010)011}{\emph{JHEP} {\bf
  06} (2010) 011}, [\href{http://arxiv.org/abs/0912.3074}{{\tt 0912.3074}}].

\bibitem{Drukker:2010nc}
N.~Drukker, M.~Mari\~no and P.~Putrov, \emph{{From Weak to Strong Coupling in
  ABJM Theory}},
  \href{http://dx.doi.org/10.1007/s00220-011-1253-6}{\emph{Commun. Math. Phys.}
  {\bf 306} (2011) 511--563}, [\href{http://arxiv.org/abs/1007.3837}{{\tt
  1007.3837}}].

\bibitem{Giombi:2020mhz}
S.~Giombi and A.~A. Tseytlin, \emph{{Strong coupling expansion of circular
  Wilson loops and string theories in AdS$_5 \times {\rm S}^5$ and AdS$_4
  \times {\rm CP}^3$}},
  \href{http://dx.doi.org/10.1007/JHEP10(2020)130}{\emph{JHEP} {\bf 10} (2020)
  130}, [\href{http://arxiv.org/abs/2007.08512}{{\tt 2007.08512}}].

\bibitem{Maldacena:1997re}
J.~M. Maldacena, \emph{{The Large N limit of superconformal field theories and
  supergravity}},
  \href{http://dx.doi.org/10.1023/A:1026654312961}{\emph{Int.J.Theor.Phys.}
  {\bf 38} (1999) 1113--1133}, [\href{http://arxiv.org/abs/hep-th/9711200}{{\tt
  hep-th/9711200}}].

\bibitem{Aharony:2008ug}
O.~Aharony, O.~Bergman, D.~L. Jafferis and J.~Maldacena,
  \emph{{${\mathcal{N}}\!=6$ Superconformal Chern-Simons-Matter Theories,
  M2-Branes and Their Gravity Duals}},
  \href{http://dx.doi.org/10.1088/1126-6708/2008/10/091}{\emph{JHEP} {\bf 10}
  (2008) 091}, [\href{http://arxiv.org/abs/0806.1218}{{\tt 0806.1218}}].

\bibitem{Berenstein:1998ij}
D.~E. Berenstein, R.~Corrado, W.~Fischler and J.~M. Maldacena, \emph{{The
  Operator product expansion for Wilson loops and surfaces in the large N
  limit}}, \href{http://dx.doi.org/10.1103/PhysRevD.59.105023}{\emph{Phys.
  Rev.} {\bf D59} (1999) 105023},
  [\href{http://arxiv.org/abs/hep-th/9809188}{{\tt hep-th/9809188}}].

\bibitem{Drukker:1999zq}
N.~Drukker, D.~J. Gross and H.~Ooguri, \emph{{Wilson loops and minimal
  surfaces}}, \href{http://dx.doi.org/10.1103/PhysRevD.60.125006}{\emph{Phys.
  Rev.} {\bf D60} (1999) 125006},
  [\href{http://arxiv.org/abs/hep-th/9904191}{{\tt hep-th/9904191}}].

\bibitem{Drukker:2000ep}
N.~Drukker, D.~J. Gross and A.~A. Tseytlin, \emph{{Green-Schwarz string in
  AdS(5) x S5: Semiclassical partition function}},
  \href{http://dx.doi.org/10.1088/1126-6708/2000/04/021}{\emph{JHEP} {\bf 04}
  (2000) 021}, [\href{http://arxiv.org/abs/hep-th/0001204}{{\tt
  hep-th/0001204}}].

\bibitem{Klemm:2012ii}
A.~Klemm, M.~Mari\~no, M.~Schiereck and M.~Soroush,
  \emph{{Aharony-Bergman-Jafferis--Maldacena Wilson Loops in the Fermi Gas
  Approach}}, \href{http://dx.doi.org/10.5560/ZNA.2012-0118}{\emph{Z.
  Naturforsch. A} {\bf 68} (2013) 178--209},
  [\href{http://arxiv.org/abs/1207.0611}{{\tt 1207.0611}}].

\bibitem{Zarembo:2002an}
K.~Zarembo, \emph{{Supersymmetric Wilson loops}},
  \href{http://dx.doi.org/10.1016/S0550-3213(02)00693-4}{\emph{Nucl. Phys.}
  {\bf B643} (2002) 157--171}, [\href{http://arxiv.org/abs/hep-th/0205160}{{\tt
  hep-th/0205160}}].

\bibitem{Drukker:2006ga}
N.~Drukker, \emph{{1/4 BPS circular loops, unstable world-sheet instantons and
  the matrix model}},
  \href{http://dx.doi.org/10.1088/1126-6708/2006/09/004}{\emph{JHEP} {\bf 09}
  (2006) 004}, [\href{http://arxiv.org/abs/hep-th/0605151}{{\tt
  hep-th/0605151}}].

\bibitem{Drukker:2007yx}
N.~Drukker, S.~Giombi, R.~Ricci and D.~Trancanelli, \emph{{Wilson Loops: from
  Four-Dimensional SYM to Two-Dimensional YM}},
  \href{http://dx.doi.org/10.1103/PhysRevD.77.047901}{\emph{Phys. Rev.} {\bf
  D77} (2008) 047901}, [\href{http://arxiv.org/abs/0707.2699}{{\tt
  0707.2699}}].

\bibitem{Drukker:2007dw}
N.~Drukker, S.~Giombi, R.~Ricci and D.~Trancanelli, \emph{{More Supersymmetric
  Wilson Loops}},
  \href{http://dx.doi.org/10.1103/PhysRevD.76.107703}{\emph{Phys. Rev. D} {\bf
  76} (2007) 107703}, [\href{http://arxiv.org/abs/0704.2237}{{\tt 0704.2237}}].

\bibitem{Drukker:2007qr}
N.~Drukker, S.~Giombi, R.~Ricci and D.~Trancanelli, \emph{{Supersymmetric
  Wilson loops on $S^{3}$}},
  \href{http://dx.doi.org/10.1088/1126-6708/2008/05/017}{\emph{JHEP} {\bf 05}
  (2008) 017}, [\href{http://arxiv.org/abs/0711.3226}{{\tt 0711.3226}}].

\bibitem{Semenoff:2001xp}
G.~W. Semenoff and K.~Zarembo, \emph{{More exact predictions of SUSYM for
  string theory}},
  \href{http://dx.doi.org/10.1016/S0550-3213(01)00455-2}{\emph{Nucl. Phys.}
  {\bf B616} (2001) 34--46}, [\href{http://arxiv.org/abs/hep-th/0106015}{{\tt
  hep-th/0106015}}].

\bibitem{Pestun:2002mr}
V.~Pestun and K.~Zarembo, \emph{{Comparing strings in AdS(5) x S5 to planar
  diagrams: An Example}},
  \href{http://dx.doi.org/10.1103/PhysRevD.67.086007}{\emph{Phys. Rev.} {\bf
  D67} (2003) 086007}, [\href{http://arxiv.org/abs/hep-th/0212296}{{\tt
  hep-th/0212296}}].

\bibitem{Semenoff:2006am}
G.~W. Semenoff and D.~Young, \emph{{Exact 1/4 BPS Loop: Chiral primary
  correlator}},
  \href{http://dx.doi.org/10.1016/j.physletb.2006.10.047}{\emph{Phys. Lett.}
  {\bf B643} (2006) 195--204}, [\href{http://arxiv.org/abs/hep-th/0609158}{{\tt
  hep-th/0609158}}].

\bibitem{Giombi:2009ds}
S.~Giombi and V.~Pestun, \emph{{Correlators of local operators and 1/8 BPS
  Wilson loops on $S^2$ from 2d YM and matrix models}},
  \href{http://dx.doi.org/10.1007/JHEP10(2010)033}{\emph{JHEP} {\bf 10} (2010)
  033}, [\href{http://arxiv.org/abs/0906.1572}{{\tt 0906.1572}}].

\bibitem{Giombi:2012ep}
S.~Giombi and V.~Pestun, \emph{{Correlators of Wilson Loops and Local Operators
  from Multi-Matrix Models and Strings in AdS}},
  \href{http://dx.doi.org/10.1007/JHEP01(2013)101}{\emph{JHEP} {\bf 01} (2013)
  101}, [\href{http://arxiv.org/abs/1207.7083}{{\tt 1207.7083}}].

\bibitem{Bassetto:2009rt}
A.~Bassetto, L.~Griguolo, F.~Pucci, D.~Seminara, S.~Thambyahpillai and
  D.~Young, \emph{{Correlators of supersymmetric Wilson-loops, protected
  operators and matrix models in N=4 SYM}},
  \href{http://dx.doi.org/10.1088/1126-6708/2009/08/061}{\emph{JHEP} {\bf 08}
  (2009) 061}, [\href{http://arxiv.org/abs/0905.1943}{{\tt 0905.1943}}].

\bibitem{Bassetto:2009ms}
A.~Bassetto, L.~Griguolo, F.~Pucci, D.~Seminara, S.~Thambyahpillai and
  D.~Young, \emph{{Correlators of supersymmetric Wilson loops at weak and
  strong coupling}},
  \href{http://dx.doi.org/10.1007/JHEP03(2010)038}{\emph{JHEP} {\bf 03} (2010)
  038}, [\href{http://arxiv.org/abs/0912.5440}{{\tt 0912.5440}}].

\bibitem{Bonini:2014vta}
M.~Bonini, L.~Griguolo and M.~Preti, \emph{{Correlators of chiral primaries and
  1/8 BPS Wilson loops from perturbation theory}},
  \href{http://dx.doi.org/10.1007/JHEP09(2014)083}{\emph{JHEP} {\bf 09} (2014)
  083}, [\href{http://arxiv.org/abs/1405.2895}{{\tt 1405.2895}}].

\bibitem{Giombi:2006de}
S.~Giombi, R.~Ricci and D.~Trancanelli, \emph{{Operator product expansion of
  higher rank Wilson loops from D-branes and matrix models}},
  \href{http://dx.doi.org/10.1088/1126-6708/2006/10/045}{\emph{JHEP} {\bf 10}
  (2006) 045}, [\href{http://arxiv.org/abs/hep-th/0608077}{{\tt
  hep-th/0608077}}].

\bibitem{Gomis:2008qa}
J.~Gomis, S.~Matsuura, T.~Okuda and D.~Trancanelli, \emph{{Wilson loop
  correlators at strong coupling: From matrices to bubbling geometries}},
  \href{http://dx.doi.org/10.1088/1126-6708/2008/08/068}{\emph{JHEP} {\bf 08}
  (2008) 068}, [\href{http://arxiv.org/abs/0807.3330}{{\tt 0807.3330}}].

\bibitem{Aprile:2020uxk}
F.~Aprile, J.~Drummond, P.~Heslop, H.~Paul, F.~Sanfilippo, M.~Santagata et~al.,
  \emph{{Single particle operators and their correlators in free $ \mathcal{N}
  $ = 4 SYM}}, \href{http://dx.doi.org/10.1007/JHEP11(2020)072}{\emph{JHEP}
  {\bf 11} (2020) 072}, [\href{http://arxiv.org/abs/2007.09395}{{\tt
  2007.09395}}].

\bibitem{Drukker:2005kx}
N.~Drukker and B.~Fiol, \emph{{All-genus calculation of Wilson loops using
  D-branes}},
  \href{http://dx.doi.org/10.1088/1126-6708/2005/02/010}{\emph{JHEP} {\bf 02}
  (2005) 010}, [\href{http://arxiv.org/abs/hep-th/0501109}{{\tt
  hep-th/0501109}}].

\bibitem{Okuyama:2006ir}
K.~Okuyama, \emph{{'T Hooft Expansion of 1/2 BPS Wilson Loop}},
  \href{http://dx.doi.org/10.1088/1126-6708/2006/09/007}{\emph{JHEP} {\bf 09}
  (2006) 007}, [\href{http://arxiv.org/abs/hep-th/0607131}{{\tt
  hep-th/0607131}}].

\bibitem{Gross:1998gk}
D.~J. Gross and H.~Ooguri, \emph{{Aspects of Large $N$ Gauge Theory Dynamics as
  Seen by String Theory}},
  \href{http://dx.doi.org/10.1103/PhysRevD.58.106002}{\emph{Phys. Rev. D} {\bf
  58} (1998) 106002}, [\href{http://arxiv.org/abs/hep-th/9805129}{{\tt
  hep-th/9805129}}].

\bibitem{Zarembo:1999bu}
K.~Zarembo, \emph{{Wilson Loop Correlator in the AdS / CFT Correspondence}},
  \href{http://dx.doi.org/10.1016/S0370-2693(99)00717-0}{\emph{Phys. Lett. B}
  {\bf 459} (1999) 527--534}, [\href{http://arxiv.org/abs/hep-th/9904149}{{\tt
  hep-th/9904149}}].

\bibitem{Correa:2018lyl}
D.~H. Correa, P.~Pisani and A.~Rios~Fukelman, \emph{{Ladder Limit for
  Correlators of Wilson Loops}},
  \href{http://dx.doi.org/10.1007/JHEP05(2018)168}{\emph{JHEP} {\bf 05} (2018)
  168}, [\href{http://arxiv.org/abs/1803.02153}{{\tt 1803.02153}}].

\bibitem{Correa:2018pfn}
D.~Correa, P.~Pisani, A.~Rios~Fukelman and K.~Zarembo, \emph{{Dyson Equations
  for Correlators of Wilson Loops}},
  \href{http://dx.doi.org/10.1007/JHEP12(2018)100}{\emph{JHEP} {\bf 12} (2018)
  100}, [\href{http://arxiv.org/abs/1811.03552}{{\tt 1811.03552}}].

\bibitem{Dorn:2018srz}
H.~Dorn, \emph{{On Wilson Loops for Two Touching Circles with Opposite
  Orientation}}, \href{http://dx.doi.org/10.1088/1751-8121/ab0003}{\emph{J.
  Phys. A} {\bf 52} (2019) 095401},
  [\href{http://arxiv.org/abs/1811.00799}{{\tt 1811.00799}}].

\bibitem{Giombi:2009ms}
S.~Giombi, V.~Pestun and R.~Ricci, \emph{{Notes on Supersymmetric Wilson Loops
  on a Two-Sphere}},
  \href{http://dx.doi.org/10.1007/JHEP07(2010)088}{\emph{JHEP} {\bf 07} (2010)
  088}, [\href{http://arxiv.org/abs/0905.0665}{{\tt 0905.0665}}].

\bibitem{Sysoeva:2018xig}
E.~Sysoeva, \emph{{Wilson Loop and Its Correlators in the Limit of Large
  Coupling Constant}},
  \href{http://dx.doi.org/10.1016/j.nuclphysb.2018.09.015}{\emph{Nucl. Phys. B}
  {\bf 936} (2018) 383--399}, [\href{http://arxiv.org/abs/1803.00649}{{\tt
  1803.00649}}].

\bibitem{CanazasGaray:2018cpk}
A.~F. Canazas~Garay, A.~Faraggi and W.~M\"uck, \emph{{Antisymmetric Wilson
  loops in $ \mathcal{N}=4 $ SYM: from exact results to non-planar
  corrections}}, \href{http://dx.doi.org/10.1007/JHEP08(2018)149}{\emph{JHEP}
  {\bf 08} (2018) 149}, [\href{http://arxiv.org/abs/1807.04052}{{\tt
  1807.04052}}].

\bibitem{Okuyama:2018aij}
K.~Okuyama, \emph{{Connected correlator of 1/2 BPS Wilson loops in
  $\mathcal{N}=4$ SYM}},
  \href{http://dx.doi.org/10.1007/JHEP10(2018)037}{\emph{JHEP} {\bf 10} (2018)
  037}, [\href{http://arxiv.org/abs/1808.10161}{{\tt 1808.10161}}].

\bibitem{CanazasGaray:2019mgq}
A.~F. Canazas~Garay, A.~Faraggi and W.~M\"uck, \emph{{Note on generating
  functions and connected correlators of 1/2-BPS Wilson loops in
  $\mathcal{N}=4$ SYM theory}},
  \href{http://dx.doi.org/10.1007/JHEP08(2019)149}{\emph{JHEP} {\bf 08} (2019)
  149}, [\href{http://arxiv.org/abs/1906.03816}{{\tt 1906.03816}}].

\bibitem{Muck:2019hnz}
W.~M\"uck, \emph{{Combinatorics of Wilson loops in $ \mathcal{N} $ 4 SYM
  theory}}, \href{http://dx.doi.org/10.1007/JHEP11(2019)096}{\emph{JHEP} {\bf
  11} (2019) 096}, [\href{http://arxiv.org/abs/1908.11582}{{\tt 1908.11582}}].

\bibitem{Arutyunov:2001hs}
G.~Arutyunov, J.~Plefka and M.~Staudacher, \emph{{Limiting Geometries of Two
  Circular Maldacena-Wilson Loop Operators}},
  \href{http://dx.doi.org/10.1088/1126-6708/2001/12/014}{\emph{JHEP} {\bf 12}
  (2001) 014}, [\href{http://arxiv.org/abs/hep-th/0111290}{{\tt
  hep-th/0111290}}].

\bibitem{Drukker:2009hy}
N.~Drukker and D.~Trancanelli, \emph{{A Supermatrix Model for
  ${\mathcal{N}}\!=6$ Super Chern-Simons-Matter Theory}},
  \href{http://dx.doi.org/10.1007/JHEP02(2010)058}{\emph{JHEP} {\bf 02} (2010)
  058}, [\href{http://arxiv.org/abs/0912.3006}{{\tt 0912.3006}}].

\bibitem{Papathanasiou:2009zm}
G.~Papathanasiou and M.~Spradlin, \emph{{Two-Loop Spectroscopy of Short ABJM
  Operators}}, \href{http://dx.doi.org/10.1007/JHEP02(2010)072}{\emph{JHEP}
  {\bf 02} (2010) 072}, [\href{http://arxiv.org/abs/0911.2220}{{\tt
  0911.2220}}].

\bibitem{Fucito:2015ofa}
F.~Fucito, J.~F. Morales and R.~Poghossian, \emph{{Wilson loops and chiral
  correlators on squashed spheres}},
  \href{http://dx.doi.org/10.1007/JHEP11(2015)064}{\emph{JHEP} {\bf 11} (2015)
  064}, [\href{http://arxiv.org/abs/1507.05426}{{\tt 1507.05426}}].

\bibitem{Billo:2018oog}
M.~Billo, F.~Galvagno, P.~Gregori and A.~Lerda, \emph{{Correlators between
  Wilson loop and chiral operators in $ \mathcal{N}=2 $ conformal gauge
  theories}}, \href{http://dx.doi.org/10.1007/JHEP03(2018)193}{\emph{JHEP} {\bf
  03} (2018) 193}, [\href{http://arxiv.org/abs/1802.09813}{{\tt 1802.09813}}].

\bibitem{Maldacena:1998im}
J.~M. Maldacena, \emph{{Wilson loops in large N field theories}},
  \href{http://dx.doi.org/10.1103/PhysRevLett.80.4859}{\emph{Phys. Rev. Lett.}
  {\bf 80} (1998) 4859--4862}, [\href{http://arxiv.org/abs/hep-th/9803002}{{\tt
  hep-th/9803002}}].

\bibitem{Alday:2011pf}
L.~F. Alday and A.~A. Tseytlin, \emph{{On Strong-Coupling Correlation Functions
  of Circular Wilson Loops and Local Operators}},
  \href{http://dx.doi.org/10.1088/1751-8113/44/39/395401}{\emph{J. Phys. A}
  {\bf 44} (2011) 395401}, [\href{http://arxiv.org/abs/1105.1537}{{\tt
  1105.1537}}].

\bibitem{Giombi:2018hsx}
S.~Giombi and S.~Komatsu, \emph{{More Exact Results in the Wilson Loop Defect
  CFT: Bulk-Defect Ope, Nonplanar Corrections and Quantum Spectral Curve}},
  \href{http://dx.doi.org/10.1088/1751-8121/ab046c}{\emph{J. Phys.} {\bf A52}
  (2019) 125401}, [\href{http://arxiv.org/abs/1811.02369}{{\tt 1811.02369}}].

\bibitem{Kristjansen:2002bb}
C.~Kristjansen, J.~Plefka, G.~Semenoff and M.~Staudacher, \emph{{A New Double
  Scaling Limit of ${\mathcal{N}}\!=4$ Superyang-Mills Theory and PP Wave
  Strings}}, \href{http://dx.doi.org/10.1016/S0550-3213(02)00749-6}{\emph{Nucl.
  Phys. B} {\bf 643} (2002) 3--30},
  [\href{http://arxiv.org/abs/hep-th/0205033}{{\tt hep-th/0205033}}].

\bibitem{Okuyama:2006jc}
K.~Okuyama and G.~W. Semenoff, \emph{{Wilson Loops in ${\mathcal{N}}\!=4$ Sym
  and Fermion Droplets}},
  \href{http://dx.doi.org/10.1088/1126-6708/2006/06/057}{\emph{JHEP} {\bf 06}
  (2006) 057}, [\href{http://arxiv.org/abs/hep-th/0604209}{{\tt
  hep-th/0604209}}].

\bibitem{Gerchkovitz:2016gxx}
E.~Gerchkovitz, J.~Gomis, N.~Ishtiaque, A.~Karasik, Z.~Komargodski and S.~S.
  Pufu, \emph{{Correlation Functions of Coulomb Branch Operators}},
  \href{http://dx.doi.org/10.1007/JHEP01(2017)103}{\emph{JHEP} {\bf 01} (2017)
  103}, [\href{http://arxiv.org/abs/1602.05971}{{\tt 1602.05971}}].

\bibitem{Billo:2017glv}
M.~Billo, F.~Fucito, A.~Lerda, J.~F. Morales, {\relax Ya}.~S. Stanev and
  C.~Wen, \emph{{Two-point Correlators in N=2 Gauge Theories}},
  \href{http://dx.doi.org/10.1016/j.nuclphysb.2017.11.003}{\emph{Nucl. Phys.}
  {\bf B926} (2018) 427--466}, [\href{http://arxiv.org/abs/1705.02909}{{\tt
  1705.02909}}].

\bibitem{Hartnoll:2006is}
S.~A. Hartnoll and S.~Kumar, \emph{{Higher Rank Wilson Loops from a Matrix
  Model}}, \href{http://dx.doi.org/10.1088/1126-6708/2006/08/026}{\emph{JHEP}
  {\bf 08} (2006) 026}, [\href{http://arxiv.org/abs/hep-th/0605027}{{\tt
  hep-th/0605027}}].

\bibitem{Kawamoto:2008gp}
S.~Kawamoto, T.~Kuroki and A.~Miwa, \emph{{Boundary Condition for D-Brane from
  Wilson Loop, and Gravitational Interpretation of Eigenvalue in Matrix Model
  in AdS/CFT Correspondence}},
  \href{http://dx.doi.org/10.1103/PhysRevD.79.126010}{\emph{Phys. Rev. D} {\bf
  79} (2009) 126010}, [\href{http://arxiv.org/abs/0812.4229}{{\tt 0812.4229}}].

\bibitem{Buchbinder:2012vr}
E.~I. Buchbinder and A.~A. Tseytlin, \emph{{Correlation function of circular
  Wilson loop with two local operators and conformal invariance}},
  \href{http://dx.doi.org/10.1103/PhysRevD.87.026006}{\emph{Phys. Rev.} {\bf
  D87} (2013) 026006}, [\href{http://arxiv.org/abs/1208.5138}{{\tt
  1208.5138}}].

\bibitem{Aguilera-Damia:2017znn}
J.~Aguilera-Damia, D.~H. Correa, F.~Fucito, V.~I. Giraldo-Rivera, J.~F. Morales
  and L.~A. Pando~Zayas, \emph{{Strings in Bubbling Geometries and Dual Wilson
  Loop Correlators}},
  \href{http://dx.doi.org/10.1007/JHEP12(2017)109}{\emph{JHEP} {\bf 12} (2017)
  109}, [\href{http://arxiv.org/abs/1709.03569}{{\tt 1709.03569}}].

\bibitem{Akemann:2001st}
G.~Akemann and P.~Damgaard, \emph{{Wilson loops in $N$=4 supersymmetric
  Yang-Mills theory from random matrix theory}},
  \href{http://dx.doi.org/10.1016/S0370-2693(01)00675-X}{\emph{Phys. Lett. B}
  {\bf 513} (2001) 179}, [\href{http://arxiv.org/abs/hep-th/0101225}{{\tt
  hep-th/0101225}}]. [Erratum: Phys.Lett.B 524, 400--400 (2002)].

\bibitem{Gerasimov:1990is}
A.~Gerasimov, A.~Marshakov, A.~Mironov, A.~Morozov and A.~Orlov, \emph{{Matrix
  Models of 2-D Gravity and Toda Theory}},
  \href{http://dx.doi.org/10.1016/0550-3213(91)90482-D}{\emph{Nucl. Phys. B}
  {\bf 357} (1991) 565--618}.

\bibitem{Morozov:1994hh}
A.~Morozov, \emph{{Integrability and Matrix Models}},
  \href{http://dx.doi.org/10.1070/PU1994v037n01ABEH000001}{\emph{Phys. Usp.}
  {\bf 37} (1994) 1--55}, [\href{http://arxiv.org/abs/hep-th/9303139}{{\tt
  hep-th/9303139}}].

\bibitem{Morozov:1995pb}
A.~Morozov, \emph{{Matrix Models as Integrable Systems}},  in \emph{{Crm-Cap
  Summer School on Particles and Fields `94}}, pp.~127--210, 1, 1995.
\newblock \href{http://arxiv.org/abs/hep-th/9502091}{{\tt hep-th/9502091}}.

\bibitem{Mironov:2005qn}
A.~Mironov, \emph{{Matrix Models Vs. Matrix Integrals}},
  \href{http://dx.doi.org/10.1007/s11232-006-0007-7}{\emph{Theor. Math. Phys.}
  {\bf 146} (2006) 63--72}, [\href{http://arxiv.org/abs/hep-th/0506158}{{\tt
  hep-th/0506158}}].

\bibitem{Morozov:2009uy}
A.~Morozov and S.~Shakirov, \emph{{Exact 2-Point Function in Hermitian Matrix
  Model}}, \href{http://dx.doi.org/10.1088/1126-6708/2009/12/003}{\emph{JHEP}
  {\bf 12} (2009) 003}, [\href{http://arxiv.org/abs/0906.0036}{{\tt
  0906.0036}}].

\bibitem{Eynard:2004mh}
B.~Eynard, \emph{{Topological Expansion for the 1-Hermitian Matrix Model
  Correlation Functions}},
  \href{http://dx.doi.org/10.1088/1126-6708/2004/11/031}{\emph{JHEP} {\bf 11}
  (2004) 031}, [\href{http://arxiv.org/abs/hep-th/0407261}{{\tt
  hep-th/0407261}}].

\bibitem{Eynard:2008we}
B.~Eynard and N.~Orantin, \emph{{Algebraic Methods in Random Matrices and
  Enumerative Geometry}},  \href{http://arxiv.org/abs/0811.3531}{{\tt
  0811.3531}}.

\bibitem{Metsaev:1987ju}
R.~R. Metsaev and A.~A. Tseytlin, \emph{{On loop corrections to string theory
  effective actions}},
  \href{http://dx.doi.org/10.1016/0550-3213(88)90306-9}{\emph{Nucl. Phys.} {\bf
  B298} (1988) 109--132}.

\bibitem{Tseytlin:1990vf}
A.~A. Tseytlin, \emph{{On 'macroscopic string' approximation in string
  theory}}, \href{http://dx.doi.org/10.1016/0370-2693(90)90792-5}{\emph{Phys.
  Lett.} {\bf B251} (1990) 530--534}.

\bibitem{Zarembo:2002ph}
K.~Zarembo, \emph{{Open String Fluctuations in $\mathrm{AdS}_5$ $\times$ $ S^5$
  and Operators with Large R Charge}},
  \href{http://dx.doi.org/10.1103/PhysRevD.66.105021}{\emph{Phys. Rev. D} {\bf
  66} (2002) 105021}, [\href{http://arxiv.org/abs/hep-th/0209095}{{\tt
  hep-th/0209095}}].

\bibitem{Bhattacharyya:2012ye}
S.~Bhattacharyya, A.~Grassi, M.~Mari\~no and A.~Sen, \emph{{A One-Loop Test of
  Quantum Supergravity}},
  \href{http://dx.doi.org/10.1088/0264-9381/31/1/015012}{\emph{Class. Quant.
  Grav.} {\bf 31} (2014) 015012}, [\href{http://arxiv.org/abs/1210.6057}{{\tt
  1210.6057}}].

\bibitem{Bianchi:2018bke}
M.~S. Bianchi, L.~Griguolo, A.~Mauri, S.~Penati and D.~Seminara, \emph{{A
  Matrix Model for the Latitude Wilson Loop in ABJM Theory}},
  \href{http://dx.doi.org/10.1007/JHEP08(2018)060}{\emph{JHEP} {\bf 08} (2018)
  060}, [\href{http://arxiv.org/abs/1802.07742}{{\tt 1802.07742}}].

\bibitem{Ouyang:2015hta}
H.~Ouyang, J.-B. Wu and J.-j. Zhang, \emph{{Exact Results for Wilson Loops in
  Orbifold ABJM Theory}},
  \href{http://dx.doi.org/10.1088/1674-1137/40/8/083101}{\emph{Chin. Phys. C}
  {\bf 40} (2016) 083101}, [\href{http://arxiv.org/abs/1507.00442}{{\tt
  1507.00442}}].

\bibitem{Marino:2002fk}
M.~Mari\~no, \emph{{Chern-Simons Theory, Matrix Integrals, and Perturbative
  Three Manifold Invariants}},
  \href{http://dx.doi.org/10.1007/s00220-004-1194-4}{\emph{Commun. Math. Phys.}
  {\bf 253} (2004) 25--49}, [\href{http://arxiv.org/abs/hep-th/0207096}{{\tt
  hep-th/0207096}}].

\bibitem{Aganagic:2002wv}
M.~Aganagic, A.~Klemm, M.~Mari\~no and C.~Vafa, \emph{{Matrix Model as a Mirror
  of Chern-Simons Theory}},
  \href{http://dx.doi.org/10.1088/1126-6708/2004/02/010}{\emph{JHEP} {\bf 02}
  (2004) 010}, [\href{http://arxiv.org/abs/hep-th/0211098}{{\tt
  hep-th/0211098}}].

\bibitem{Halmagyi:2003ze}
N.~Halmagyi and V.~Yasnov, \emph{{The Spectral Curve of the Lens Space Matrix
  Model}}, \href{http://dx.doi.org/10.1088/1126-6708/2009/11/104}{\emph{JHEP}
  {\bf 11} (2009) 104}, [\href{http://arxiv.org/abs/hep-th/0311117}{{\tt
  hep-th/0311117}}].

\bibitem{Akemann:1996zr}
G.~Akemann, \emph{{Higher Genus Correlators for the Hermitian Matrix Model with
  Multiple Cuts}},
  \href{http://dx.doi.org/10.1016/S0550-3213(96)00542-1}{\emph{Nucl. Phys. B}
  {\bf 482} (1996) 403--430}, [\href{http://arxiv.org/abs/hep-th/9606004}{{\tt
  hep-th/9606004}}].

\end{thebibliography}\endgroup
\bibliographystyle{JHEP}
\end{document}